\documentclass[reprint,floatfix, aps,prc, twocolumn,notitlepage,
nofootinbib,nobibnotes,
amsmath,amssymb,preprintnumbers]{revtex4-1}
\usepackage{graphicx,xcolor}
\usepackage{physics}
\usepackage{enumitem}   
\usepackage{dcolumn}
\usepackage{hyperref}
\usepackage[normalem]{ulem}
\usepackage{mathtools}
\usepackage{graphicx}
\usepackage{bm}
\usepackage{subfigure} 
\usepackage{xcolor}
\usepackage{float}
\usepackage{placeins}
\usepackage{tabularx}
\usepackage{url}
\usepackage{latexsym}
\usepackage{epsf}
\usepackage{etoolbox}

\makeatletter
\appto\abstract{%
  \let\latexlist\list
  \def\list{\edef\keeprightskip{\the\rightskip}\latexlist}%
  \patchcmd\latexlist{\ignorespaces}{\rightskip\keeprightskip\ignorespaces}{}{}%
}
\makeatother
\def\trento{T$_{\rm R}$ENTo \hspace{0.07cm}}

\begin{document}
\title{Role of initial transverse momentum in a hybrid approach }
\author{Niklas G\"otz$^{1,2}$, Lucas Constantin$^{1}$  and Hannah Elfner$^{3,1,2,4}$}
\affiliation{$^1$Goethe University Frankfurt, Department of Physics, Institute for Theoretical Physics, Frankfurt, Germany}
\affiliation{$^2$Frankfurt Institute for Advanced Studies, Ruth-Moufang-Strasse 1, 60438
Frankfurt am Main, Germany}
\affiliation{$^3$GSI Helmholtzzentrum f\"ur Schwerionenforschung, Planckstr. 1, 64291
Darmstadt, Germany}
\affiliation{$^4$Helmholtz Research Academy Hesse for FAIR (HFHF), GSI Helmholtz Center,
Campus Frankfurt, Max-von-Laue-Straße 12, 60438 Frankfurt am Main, Germany}

\date{\today}
\begin{abstract}
\begin{description}
\item[Background] Significant theoretical uncertainties exist with respect to the initial condition of the hydrodynamic description of ultrarelativistic heavy-ion collisions. Several approaches exist, of which some contain initial momentum information. Its impact is commonly assumed to be small and final flow is seen as a linear response to the initial state eccentricity.
\item[Purpose] The purpose of this work is to study the effect of exchanging initial condition models in a modular hybrid approach. The focus lies on the event-by-event correlations of elliptic and triangular flow. 
\item[Method] This study is performed in the hybrid approach SMASH-vHLLE, composed of the hadronic transport approach SMASH and the (3+1)d viscous hydrodynamic code vHLLE. The initial condition models investigated are SMASH IC, \trento and IP-Glasma. Correlations are calculated on an event-by-event basis between the eccentricities and momentum anisotropies of the initial state as well as the momentum anisotropies in the final state, both for ultra-central and off-central collisions for AuAu collisions at $\sqrt{s_{NN}} = 200$ GeV. The response of the final state to initial state properties is studied.
 \item[Results] This work demonstrates that, although averaged values for the eccentricities of these models are very similar, substantial differences exist both in the distributions of eccentricities, the correlations amongst the initial state properties as well as in the correlations between initial state and final state properties.
 Notably, whereas initial state momentum anisotropy is shown to not affect the final state flow, the presence of radial flow affects the emergence of final state momentum anisotropies. 
\item[Conclusions] Inclusion of radial flow in the linear fit improves the prediction of final state flow from initial state properties. The presence of momentum in the initial state has an effect on the emergence of flow and is therefore a relevant part of initial state models, challenging the common understanding of final state momentum anisotropies being a linear response to initial state eccentricity only.
\end{description}
 
\end{abstract}

\maketitle
\section{Introduction}
One of the most important results of studying fundamental properties of matter by colliding atomic nuclei at relativistic velocities is the observation of collective behavior. Gold-gold collisions at the Relativistic Heavy Ion Collider (RHIC) at BNL showed signatures of a strongly interacting system with a small mean free path, as long range correlations and a significant elliptic flow have been observed \cite{PHENIX:2011yyh,STAR:2004jwm}.

The current understanding for this elliptic (as well as higher order) flows is the existence of initial state eccentricities $\epsilon_n$, originating from the geometry of the collision and initial state fluctuations. They result in pressure anisotropies during the evolution of the hot and dense matter. These, in turn, lead to the momentum of the particles emitted to depend on the azimuthal angle \cite{Elfner:2022iae}. The resulting flows $v_n$ are often assumed to be a linear response to the $\epsilon_n$ \cite{Noronha-Hostler:2015dbi}.

From the theoretical perspective, heavy-ion collisions have been successfully described in hybrid approaches. These are based on relativistic viscous fluid dynamics for the hot and dense stage and hadronic transport for the non-equilibrium rescattering \cite{Schafer:2021csj,Karpenko_2015,Petersen_2008,Wu_2022,Shen_2017,akamatsu2018dynamically,du20203+, nandi2020constraining}. They combine the successful description of hadronic transport at low energies, where hadronic interactions prevail, with the high energy description of relativistic hydrodynamic calculations. The hadronic transport typically serves as a hadronic afterburner, which has shown to successfully reproduce experimental observables \cite{Petersen_2014}.

Relativistic viscous hydrodynamic calculations provide direct access to the properties of the medium and a potential phase transition, as the equation of state and the transport coefficients are given as an input. Determining the functional form of the shear viscosity over entropy ratio of nuclear matter $\eta/s$ is of special interest, as one expects a minimum near the point of phase transition \cite{JETSCAPE:2020mzn,Ghiglieri:2018dib,Auvinen:2020mpc,Nijs:2020ors,Nijs:2020roc,greiner2011hagedorn,Gorenstein_2008,csernai2006strongly}. As the shear viscosity counteracts pressure gradients in the fluid during hydrodynamic evolution, its main effect is the reduction of the flow coefficients. Therefore, the flow coefficients are the main observables to study $\eta/s$ \cite{Elfner:2022iae}.  In order to determine the minimum of the shear viscosity as a function of the temperature, one commonly resorts to Bayesian inference, which optimizes the input parameters, including the ones for the parameterization of the shear viscosity, in order to describe experimental data \cite{Liyanage:2023nds, JETSCAPE:2020shq,JETSCAPE:2020mzn,Nijs:2020ors,Nijs:2020roc}.
Although this approach has been successful, it also became evident that such an inference is highly model dependent, as has been demonstrated for the choice of the particlization scheme \cite{JETSCAPE:2020shq}.

A similar impact can be expected for choosing different initial condition models. Due to the extremely short lifetime, the initial state of heavy ion collisions is experimentally not accessible, leading to a variety of initial state models to exist \cite{Kolb:2001qz,Bartels:2002cj,Kowalski:2003hm,Lin:2004en,Hirano:2005xf, Drescher:2006pi,Drescher_2007,Petersen:2008dd,Werner:2010aa,Accardi:2012qut,Schenke:2012wb,Rybczynski:2013yba,Moreland:2014oya,vanderSchee:2015rta,Schafer:2021csj,Garcia-Montero:2023gex}.  This theoretical uncertainty of the initial state model is potentially limiting the predictive strength of a Bayesian inference based on a singular initial state model. Indeed, the initial state model provides the geometry and fluctuations which are evolved into momentum anisotropy in the final state, from which the shear viscosity is primarily extracted \cite{Holopainen:2010gz, Song:2010mg}.

In this work, the effect of changing the initial state model in a hybrid approach is studied. We compare three initial conditions models for Au-Au collisions at $\sqrt{s_{NN}}$ = 200 GeV: the SMASH IC, which performs hadronic interactions until a constant hypersurface of proper time, the parametric model \trento \cite{Moreland:2014oya} and the CGC-based model IP-Glasma \cite{Schenke:2012wb}. A special focus is given to event-by-event correlations between initial state properties and final state flow observables, as well as to the impact of momentum space in the initial state.

This work is structured as follows: In Sec. \ref{sec:model}, the hybrid approach {SMASH-vHLLE-hybrid}, within which this study is performed, is briefly summarized, as well as the different initial conditions used in this study. In Sec. \ref{sec:results}, we show that, although averaged properties of the initial conditions are similar, systematic differences arise on an event-by-event bases in the correlations between the initial state and final state properties. To conclude, a brief summary and outlook can be found in Section \ref{sec:Conclusion}.

\section{Model Description}\label{sec:model}

The theoretical calculations presented in this work are performed using the {SMASH-vHLLE-hybrid} approach \cite{hybridurl}, which is publicly available and suited for the theoretical description of heavy-ion collisions between $\sqrt{s_{NN}}$ = 4.3 GeV and $\sqrt{s_{NN}}$ = 5.02 TeV. {SMASH-vHLLE-hybrid}  has been shown to reproduce experimental data well across a wide range of collision energies and conserves all charges ($B, Q, S$). It is especially successful in reproducing the longitudinal baryon dynamics at intermediate collision energies \cite{Schafer:2021csj}, which was the primary motivation behind its development.

In the following section, a short overview of its components is given (a more detailed description can be found in \cite{Schafer:2021csj}).

\subsection{Hybrid approach}
Hybrid approaches combine well established models for the different stages of a heavy ion collision. The hot and dense stage has been successfully described by (viscous) fluid dynamics. On the other hand, microscopic non-equilibrium transport approaches are commonly chosen to describe the cold and dilute stage.  

\subsubsection{Hadronic transport}
{SMASH} \cite{Weil_2016,dmytro_oliinychenko_2021_5796168,smashurl} effectively solves the relativistic Boltzmann equation by simulating the collision integral through formations, decays and scatterings of hadronic resonances, for which all hadrons listed by the PDG up to a mass of 2.35 GeV are included \cite{ParticleDataGroup:2018ovx}. For hadronic interactions at high energies, hard scatterings are carried out within {Pythia 8} \cite{Sj_strand_2008} and a soft string model is employed.

The relativistic Boltzmann equation becomes applicable in the late stage of the evolution, when the system has cooled down and expanded far enough to be out of equilibrium. In this last stage of the hybrid evolution, {SMASH} is employed for hadron rescattering. The hadrons  obtained from particlization on the Cooper-Frye hypersurface are propagated back to the earliest time and appear subsequently in the hadronic transport evolution and scatter or decay. The calculation is terminated when the medium becomes sufficiently dilute for any interactions to cease.

\subsubsection{Hydrodynamic evolution}
{vHLLE} \cite{Karpenko_2014} is a 3+1D viscous hydrodynamics code and used to model the evolution of the hot and dense fireball. It solves the hydrodynamic equations
\begin{equation}
    \partial_\nu T^{\mu \nu} =0 \hspace{0.25em}, 
\end{equation}
\begin{equation}
    \partial_\nu j^\nu_B=0 \hspace{1.75em} \partial_\nu j_Q^\nu=0 \hspace{1.75em} \partial_\nu j_S^\nu=0  .
\end{equation}
These equations represent the conservation of net-baryon, net-charge and net-strangeness number currents as well as the conservation of energy and momentum. The energy-momentum tensor is decomposed as 
\begin{equation}
    T^{\mu\nu}=\epsilon u^\mu u^\nu -\Delta^{\mu \nu}(p+\Pi)+\pi^{\mu\nu}
\end{equation}
with $\epsilon$ the local rest frame energy density, $p$ and $\Pi$ the equilibrium and bulk pressure and $\pi^{\mu\nu}$ is the shear stress tensor. These equations are solved in the second order Israel-Stewart framework \cite{Denicol_2014,Ryu_2015}.

At this step, particles have been converted into fluid elements which are evolved using a chiral mean field model equation of state \cite{Steinheimer:2011ea,Motornenko:2019arp,Most:2022wgo}. This evolution is performed until a switching energy density is reached, which is set in this work to a default value of $\epsilon_{\text{switch}}$=0.5 GeV/fm$^3$.  At this point, the freezeout hypersurface is constructed using the {CORNELIUS} subroutine \cite{Huovinen_2012} and the thermodynamical properties of the freezeout elements are calculated according to the {SMASH}  hadron resonance gas equation of state \cite{Schafer:2021csj} to prevent discontinuities. The shear viscosity is chosen according to \cite{Gotz:2022naz}, which defines an energy-density dependent value of the shear viscosity. This reduces the impact of the choice of the technical parameter $\epsilon_{sw}$, the value energy density at which the degrees of freedom in the description are changed to particles.
As the hybrid approach is 3D, the 2D initial state models need to be extended. This is performed in the same way as in \cite{Cimerman:2020iny}: the 2D profile is extended on a rapidity plateau, until it falls off with a Gaussian profile.

\subsubsection{Particle sampler}

As {SMASH} evolves particles and not fluid elements, particlization has to be performed, for which the {SMASH-vHLLE} hybrid approach applies the {SMASH-hadron-sampler} \cite{samplerurl}. The {SMASH-hadron-sampler} employs the grand-canonical ensemble in order to particlize each surface element independently. Hadrons are sampled according to a Poisson distribution with the mean at the thermal multiplicity, and momenta are sampled according to the Cooper-Frye formula \cite{cooper1975landau}. The corrections to the distribution function $\delta f_{\rm shear}$ associated to a finite shear viscosity are considered using the Grad's 14-moment ansatz, for which we assume that the correction is the same for all hadron species. Due to ($T,\mu_B$) dependence of the shear viscosity, the corrections also have this dependence implicitly. This procedure provides a particle list that can be evolved by {SMASH}. A more thorough introduction into the sampling procedure can be found in \cite{Karpenko_2015}.

\subsection{{Initial conditions}}

The input to the viscous relativistic hydrodynamic model is the initial condition, which describes the evolution of the system before the applicability of hydrodynamics. In the following, three different initial condition models are introduced: initial conditions extracted from SMASH, the \trento \cite{Moreland:2014oya} as well as the IP-Glasma \cite{Schenke:2012wb} model. Whereas the SMASH initial condition is the default choice in SMASH-vHLLE-hybrid, both \trento as well as IP-Glasma are amongst the most common choices of initial conditions for Bayesian inference. As \trento and IP-Glasma are 2D models, they are suited for high-energy collisions only, whereas SMASH provides full 3D initial conditions. We therefore restrict ourselves to a comparison at $\sqrt{s_{NN}}= 200$ GeV. Additionally, we compare our results with SMASH to the intermediate energy case of $\sqrt{s_{NN}}= 17.3$ GeV

\subsubsection{SMASH IC}

SMASH generates initial conditions by letting the particles of the colliding nuclei interact until the individual particles reach a proper time $\tau_0$, upon which they are removed from the simulation. This continues until the last particle has reached the hypersurface of constant proper time. This procedure is necessary, as SMASH is realised in Cartesian coordinates, whereas the hydrodynamic evolution is performed in Milne coordinates. $\tau_0$ corresponds to the passing time of the two nuclei
\begin{equation}
    \tau_0 = \frac{R_P + R_T}{\sqrt{\left(\frac{\sqrt{s_{NN}}}{2 m_N}\right)^2)}-1},
\end{equation}
where $R_P $ and $R_T$ are the radii of projectile and target, respectively. $\sqrt{s_{NN}}$ is the collision energy and $m_N$ is the nucleon mass. 
This follows the assumption that at this time local equilibrium is reached and the hydrodynamic description becomes applicable \cite{Oliinychenko:2015lva,Inghirami:2022afu}.
In order to initialise the hydrodynamic evolution with the iso-$\tau$ particle list generated from {SMASH}, smoothing has to be performed in order to prevent shock waves. For this purpose, a Gaussian smearing kernel \cite{Karpenko_2015} with the parameters taken from \cite{Schafer:2021csj} is applied. The values of the smearing parameters are essential for tuning the initial condition in order to agree with experimental data, as has been shown with similar approaches \cite{Auvinen:2013sba}. 
It is important to note that the SMASH IC gives access to the initial momentum of the fluid, as well as to charge, baryon and strange density. Additionally, it has been shown to capture baryon stopping well \cite{Schafer:2021csj}.

\subsubsection{\trento}
\trento is generalized, parametric ansatz for the entropy density deposition from the participant
nucleons as follows:
\begin{equation}
    T_R(p; T_A, T_B)= (T_A^p + T_B^p)^{\frac{1}{p}},
\end{equation}
where $T_{A,B}$ are the thickness profiles of the two incoming nuclei, and p is a dimensionless parameter which interpolates between the simplified functional forms of the initial entropy density profile in different initial state models, e.g. $p = 1$ corresponds to a Monte Carlo wounded nucleon model, whereas $p = 0$ is functionally similar to EKRT and IP-Glasma models. Other important parameters are the nucleon width and the shape of the nucleon fluctuations. In this work, we use the parameters extracted from Bayesian inference \cite{JETSCAPE:2020mzn}.
As \trento offers only a description of the entropy profile, information about the momentum space is lacking. It can therefore serve as a benchmark in comparison to the other models used in this work, which provide initial momentum information.
\trento has been extensively used for Bayesian inference studies, as its parametric setup allows to investigate various shapes of the density profile \cite{JETSCAPE:2022cob,JETSCAPE:2020mzn,Bernhard:2019bmu,Bernhard:2016tnd}

\subsubsection{IP-Glasma}
Another way to describe the initial state is motivated from the Color-Glass Codensate framework \cite{Gelis:2010nm}, an effective field theory representation of QCD. The CGC action can be written as follows \cite{Krasnitz:1998ns}
\begin{equation}
    S_{CGC}= \int d^4 x \left(-\frac{1}{4} F^a_{\mu\nu}F^{a\, \mu\nu}+J^{a\, \mu} A_\mu^a\right)
\end{equation}
where $J^{a\hspace{0.25em}\mu}$ is the source of soft gluons, which is the current represents hard partons, and $F^{a\hspace{0.25em}\mu\nu}$ is the non-Abelian field strength tensor. $a$ is the color index. 
The CGC can be used to generate initial collisions in form of the IP-Glasma model \cite{Gale:2013da,Schenke:2012hg, Schenke:2012wb}, which uses the IP-Sat approach \cite{Kowalski:2003hm,Bartels:2002cj} to incorporate the fluctuation of the initial color configuration of the two ultrarelativistic colliding nuclei. The color charges are sources for soft gluon fields at small $x$, which, due to their large occupation number, can be treated as classically. This means that their evolution obeys the Yang-Mills equation:
\begin{equation}
    [ D_\nu, F^{\mu\nu}]^a = J^{a\,\mu}
\end{equation}
with $D^a_\mu=\delta_\mu - i g A_\mu t^a$, where $t^a$ represents the color SU(3) matrices. 
The color current $J^{a \, \mu}=\delta^{\mu\pm}\rho^a_{A/B}(x^\mp,\mathbf{x}_\perp)$ is generated by nucleus A/B moving along the light-cone direction $x^+/x^-$ and $\rho^a$ represents the color charge distribution generated from IP-Sat.
The input to fluid dynamics at proper time $\tau_0$ is generated by solving the Classical Yang-Mills equations event-by-event, which produces the Glasma distributions.

\subsubsection{Configuration details}
For each setup presented, unless stated otherwise, 750 event-by-event viscous hydrodynamics events were generated, each initialised from a separate initial state event. The IP-Glasma events were generated by running the glasma evolution until $\tau = 0.5$ fm. From the resulting freezeout hypersurface 1000 events are sampled for the hadronic afterburner evolution, in order to average out final state effects.
Unless stated otherwise, the code versions {SMASH-vHLLE-hybrid:298ebfa}, {vHLLE:9c6655d}, {SMASH-2.2}, {vhlle-params:99ef7b4} and {SMASH-hadron-sampler-1.1} are used in this work. 
\FloatBarrier
\section{Results}\label{sec:results}
In the following, we compare results based on the following properties of the different initial state models:
\begin{itemize}
    \item The initial elliptic and triangular eccentricities of the energy densities generated by the initial condition model, $\epsilon_2$ and $\epsilon_3$, which are calculated by integrating on the grid
    \begin{equation}
        \epsilon_n = \frac{\int r^n \epsilon(\mathbf{r})e^{i n \phi}d\mathbf{r}}{\int r^n \epsilon(\mathbf{r})d\mathbf{r}}
    \end{equation}
    with the energy density in the grid cell at $\mathbf{r}$, $\epsilon(\mathbf{r})$
    \item The final state elliptic and triangular flow, calculated from the scalar product method \cite{STAR:2002hbo}, $v_2$ and $v_3$
    \item The final state transverse flow at midrapidity, $\langle p_T \rangle$
    \item In the case of the SMASH and IP-Glasma initial condition, the average initial state radial flow, $\langle p_T^{IC} \rangle$
    \item In the case of the SMASH initial condition, the initial state elliptic and triangular flow, calculated from the particles in the initial state using the scalar product method \cite{STAR:2002hbo}, $v_2^{IC}$ and $v_3^{IC}$
    \item In the case of the IP-Glasma initial condition, the initial state momentum anisotropy, $\epsilon_p$ as defined in \cite{Schenke_2020}
\end{itemize}
Flows and eccentricities were calculated using the software package SPARKX \cite{hendrik_roch_2023_10288639}.

\subsection{Averaged quantities}\label{sec:average}
\begin{figure}
    \centering
    \includegraphics{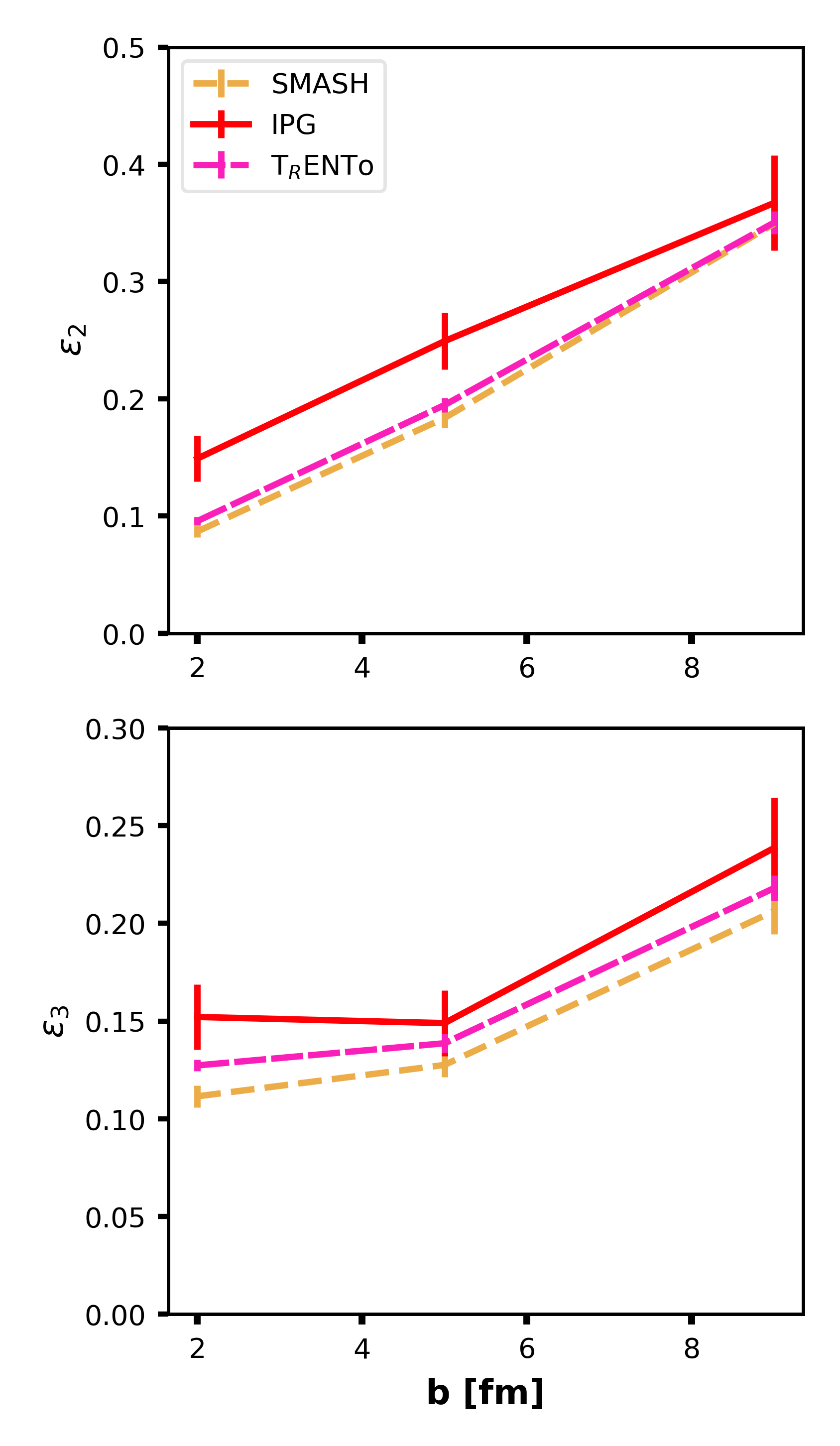}
    \caption{Initial state eccentricities for Au-Au collisions at $\sqrt{s_{NN}}$ = 200 GeV as a function of the impact parameter for the three models.}
    \label{fig:ecc_Au}
\end{figure}
\begin{figure}
    \centering
    \includegraphics{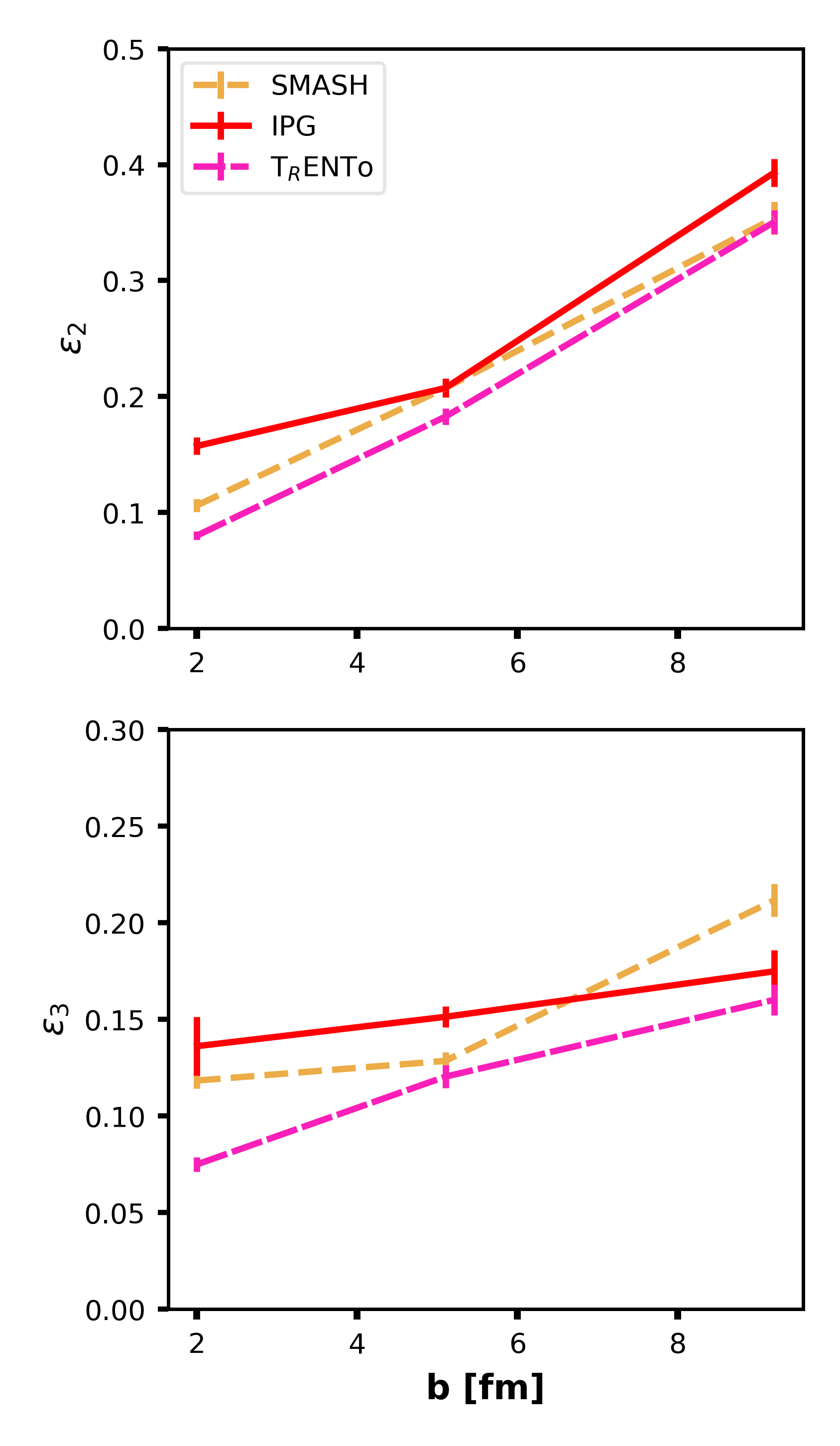}
    \caption{Initial state eccentricities for Pb-Pb collisions at $\sqrt{s_{NN}}$ = 5020 GeV as a function of the impact parameter for the three models.}
    \label{fig:ecc_Pb}
\end{figure}
\begin{figure}
    \centering
    \includegraphics{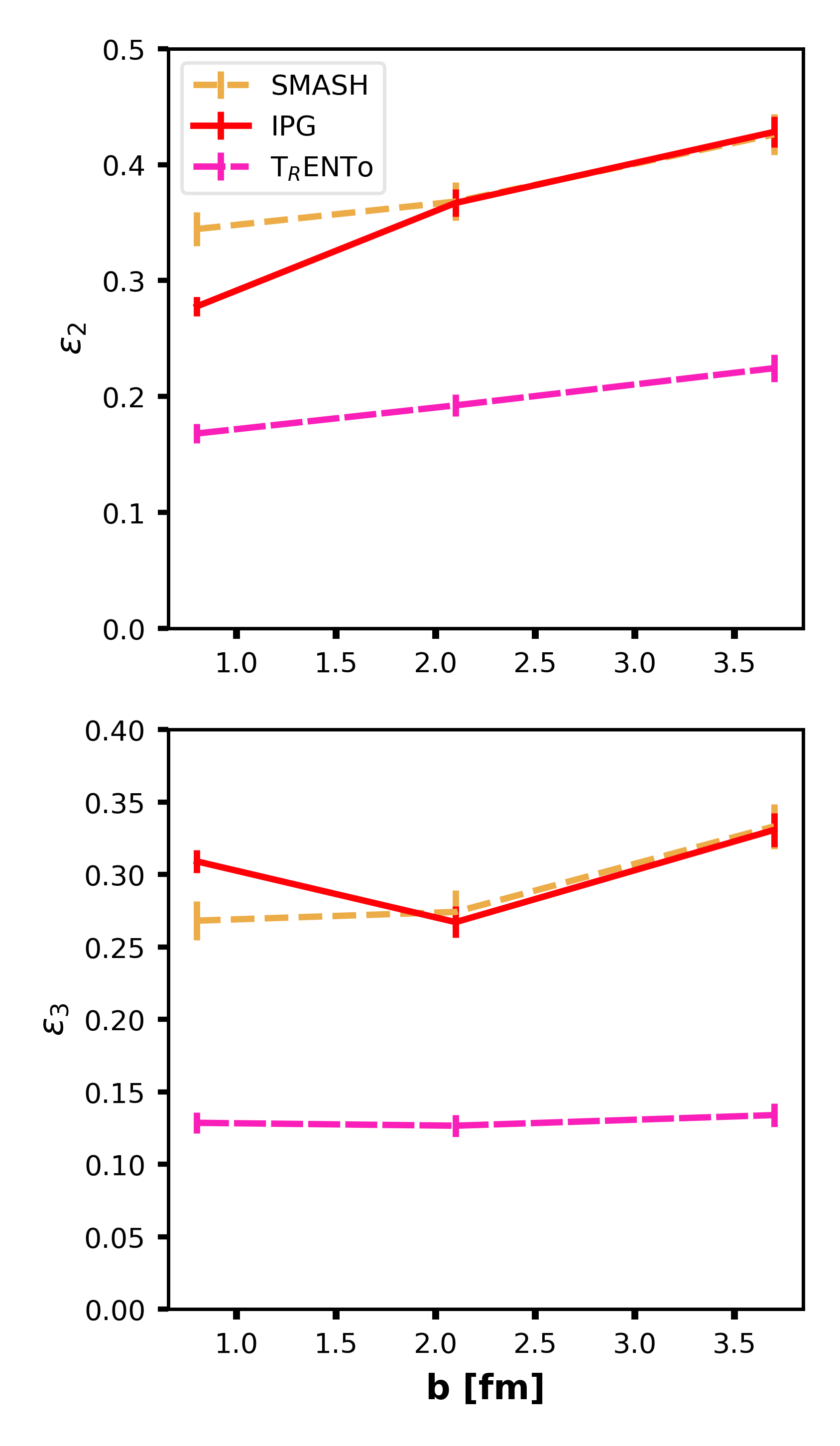}
    \caption{Initial state eccentricities for O-O collisions at $\sqrt{s_{NN}}$ = 7 TeV as a function of the impact parameter for the three models.}
    \label{fig:ecc_O}
\end{figure}
Figures \ref{fig:ecc_Au}, \ref{fig:ecc_Pb} and \ref{fig:ecc_O} show $\epsilon_2$ and $\epsilon_3$ as a function of the impact parameter for three experimentally relevant systems: Au-Au collisions at $\sqrt{s_{NN}}$ = 200 GeV, Pb-Pb collisions at $\sqrt{s_{NN}}$ = 5020 GeV and O-O collisions at $\sqrt{s_{NN}}$ = 7 TeV. As \trento does not provide oxygen configurations, they were taken from \cite{Alvioli:2009ab}. 
For both Au-Au collisions at $\sqrt{s_{NN}}$ = 200 GeV andd Pb-Pb collisions at $\sqrt{s_{NN}}$ = 5020 GeV, the three models give similar results with comparable increase as a function of the impact parameter. In both cases, IP-Glasma produces the highest values for the eccentricities, expect for $\epsilon_3$ in very peripheral collisions at high energies, where SMASH gives slightly higher values.
This changes substantially when looking at small nuclei in Fig. \ref{fig:ecc_O}. SMASH and IP-Glasma are still comparable, but \trento gives now much more spherical profiles with only small impact-parameter dependence.
\begin{figure}
    \centering
    \includegraphics{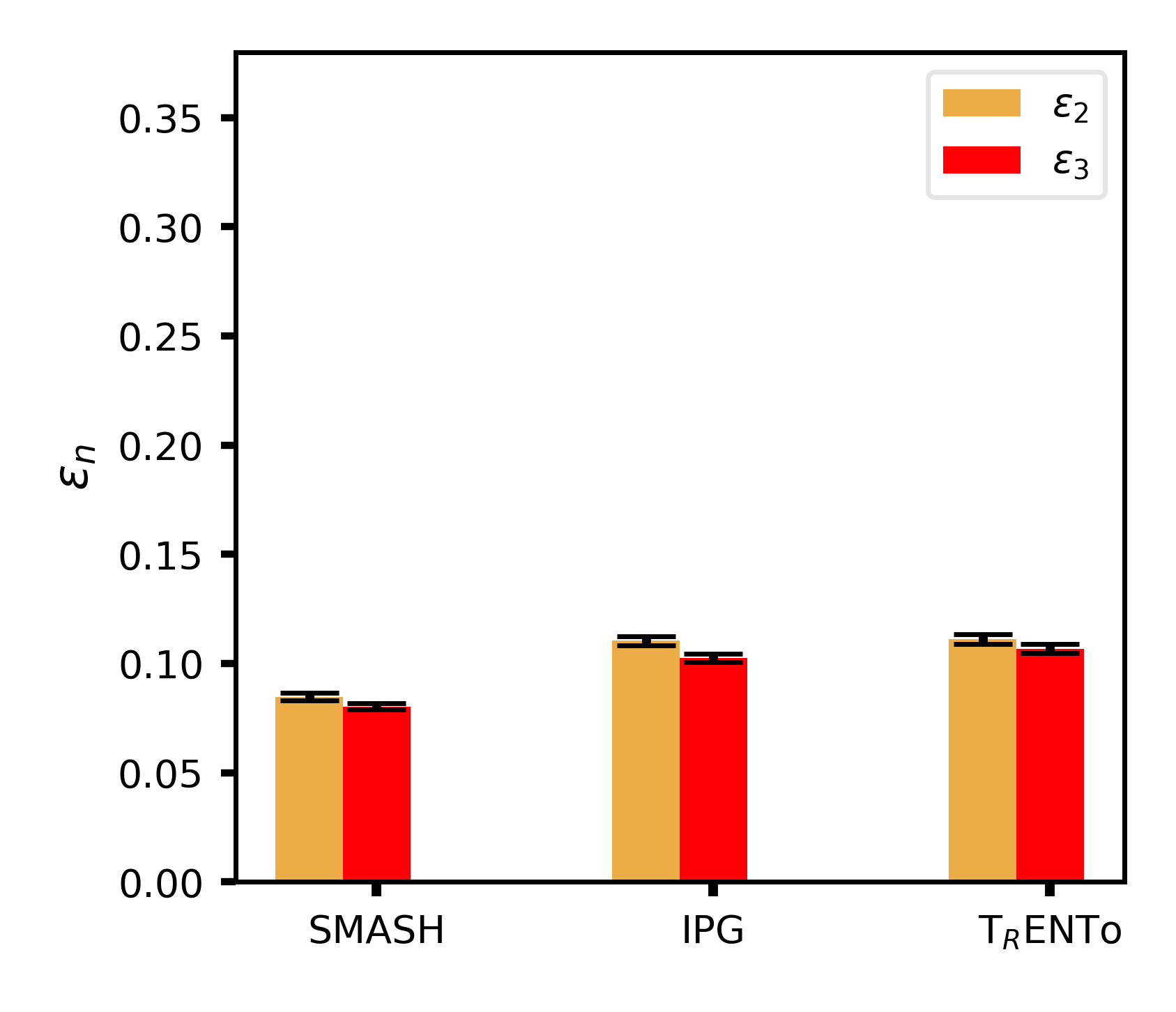}
    \includegraphics{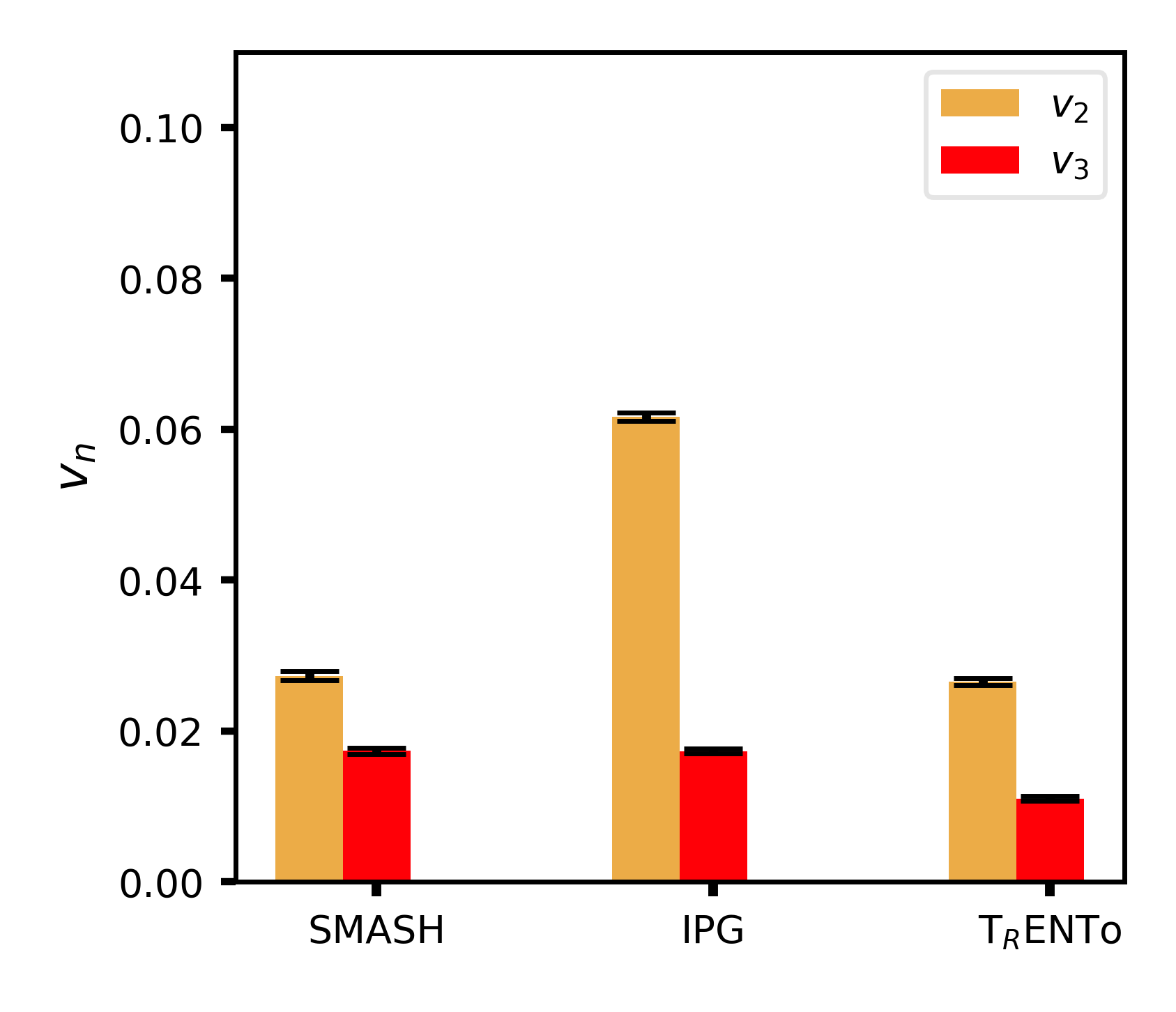}
    \caption{Initial state eccentricities and final state flows for all three models at 0-5\% centrality.}
    \label{fig:averaged_05}
\end{figure}
\begin{figure}
    \centering
    \includegraphics{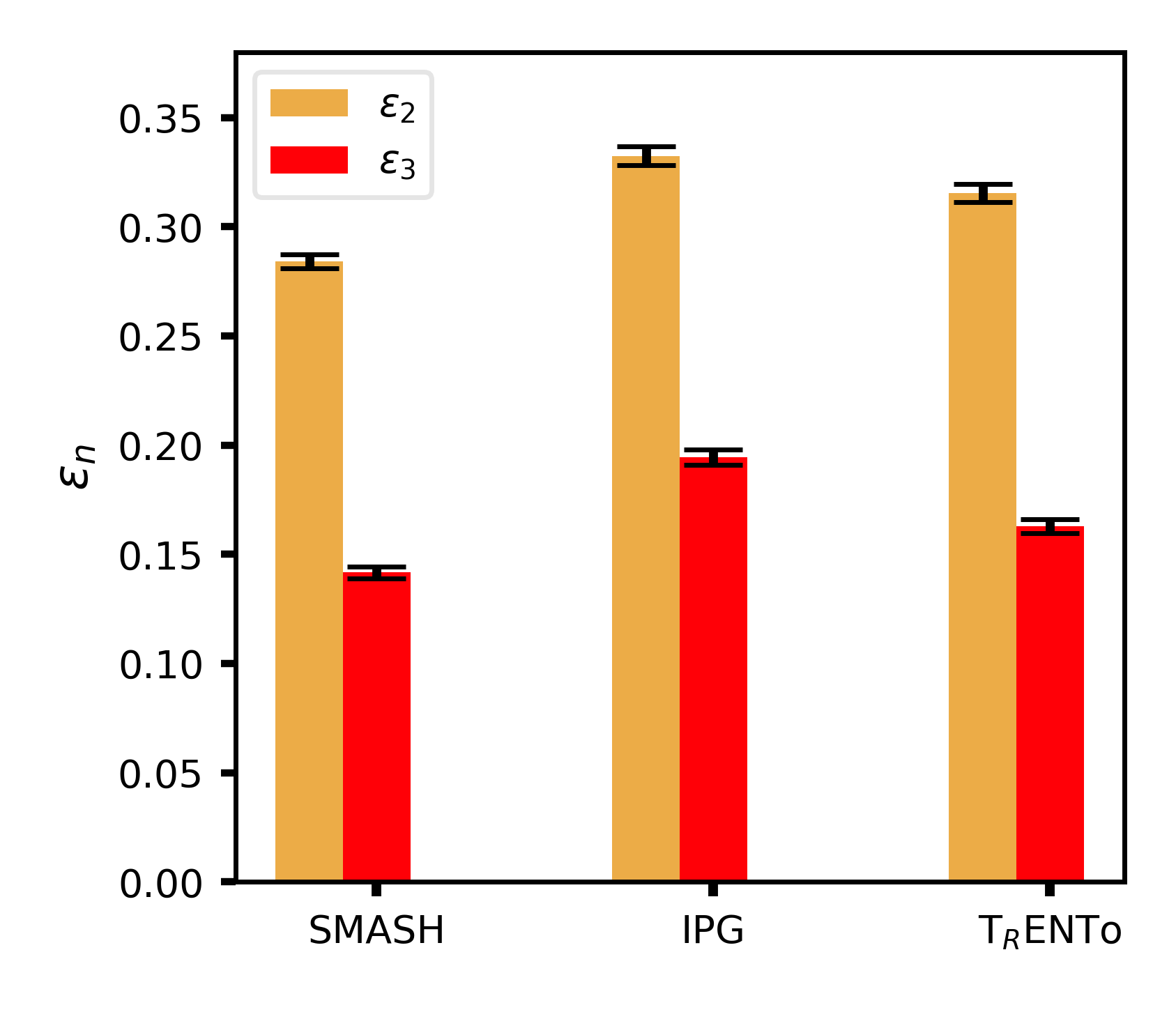}
    \includegraphics{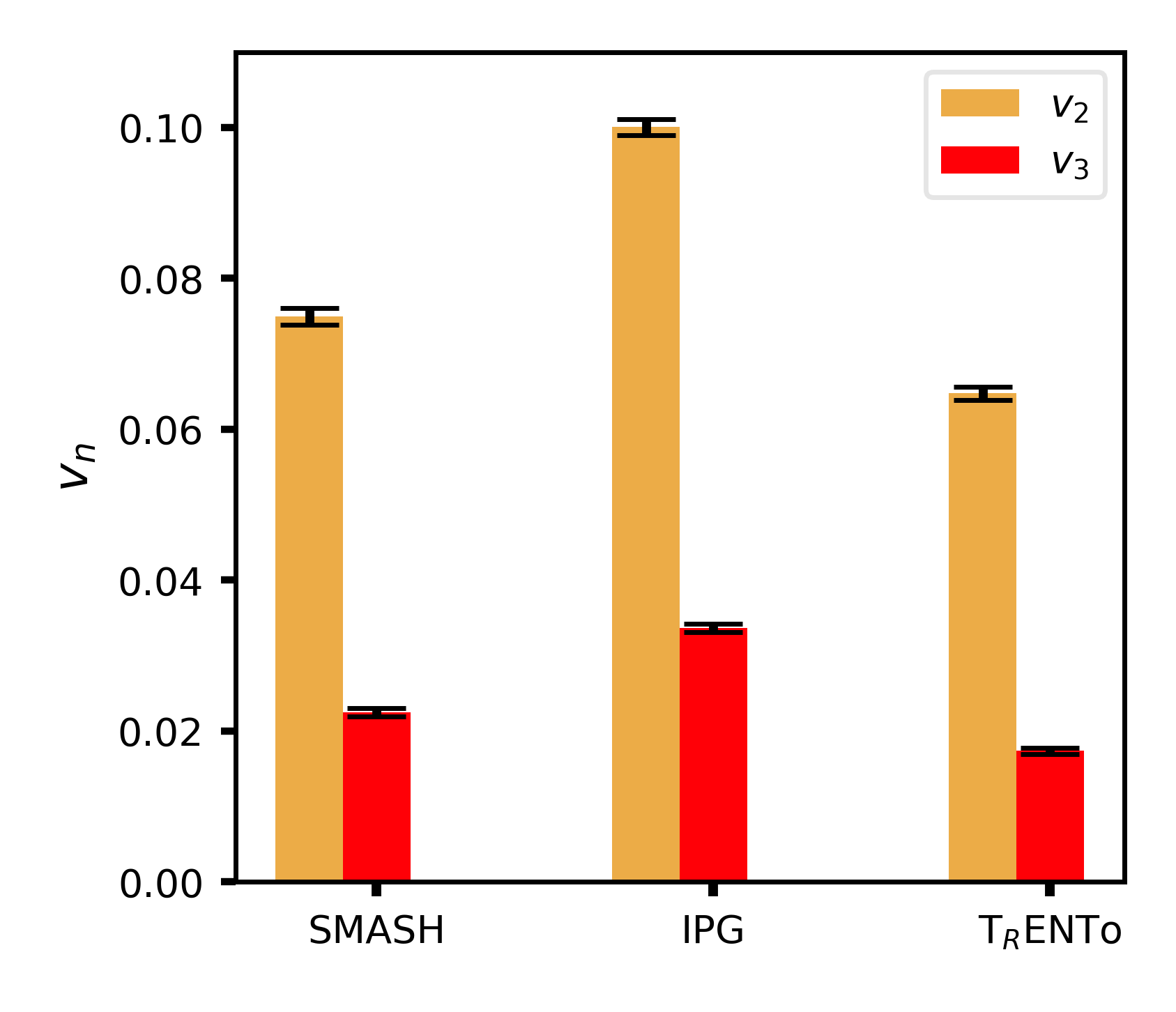}
    \caption{Initial state eccentricities and final state flows for all three models at 20-30\% centrality.}
    \label{fig:averaged_2030}
\end{figure}
\begin{figure}
    \centering
    \includegraphics{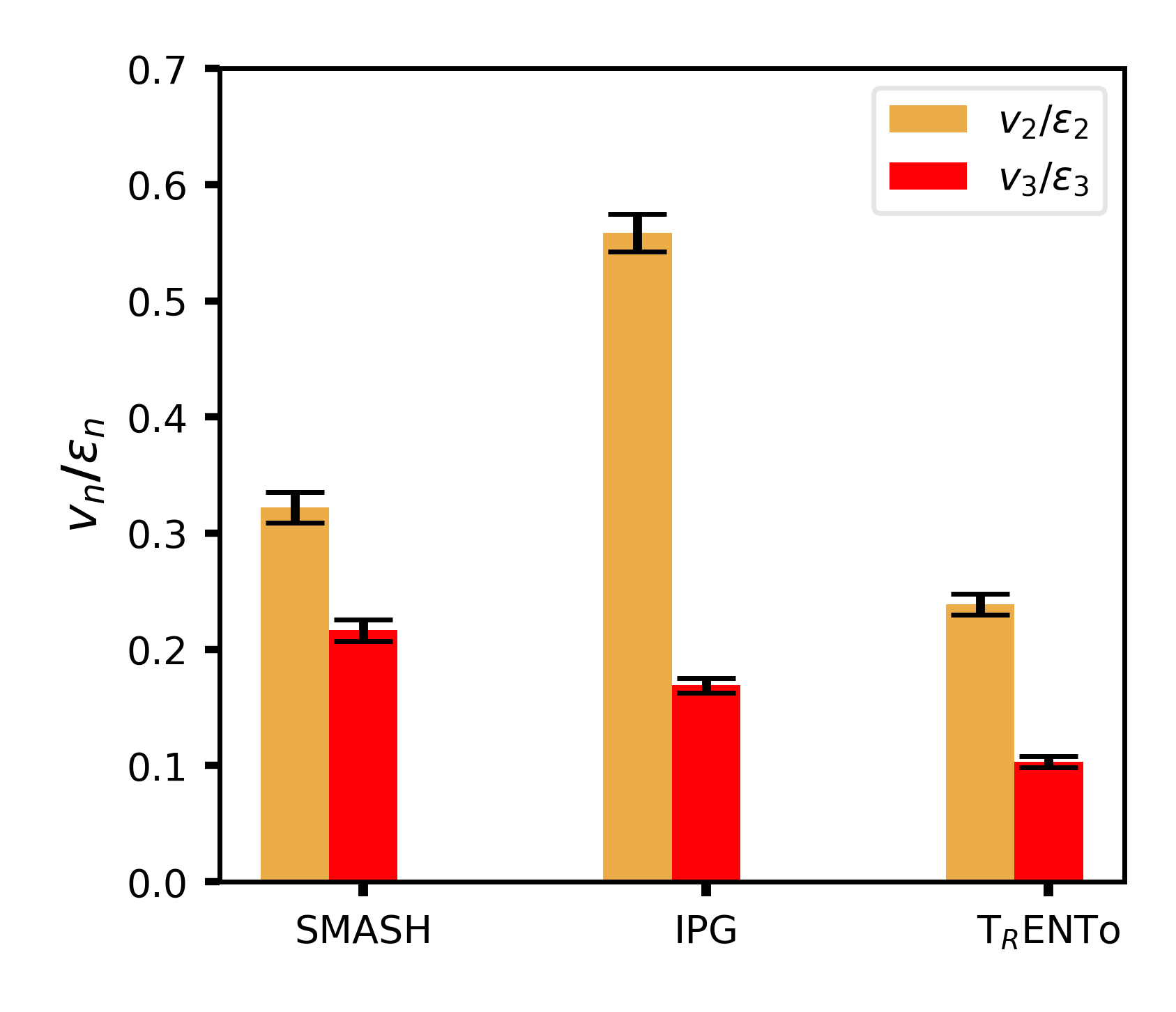}
    \includegraphics{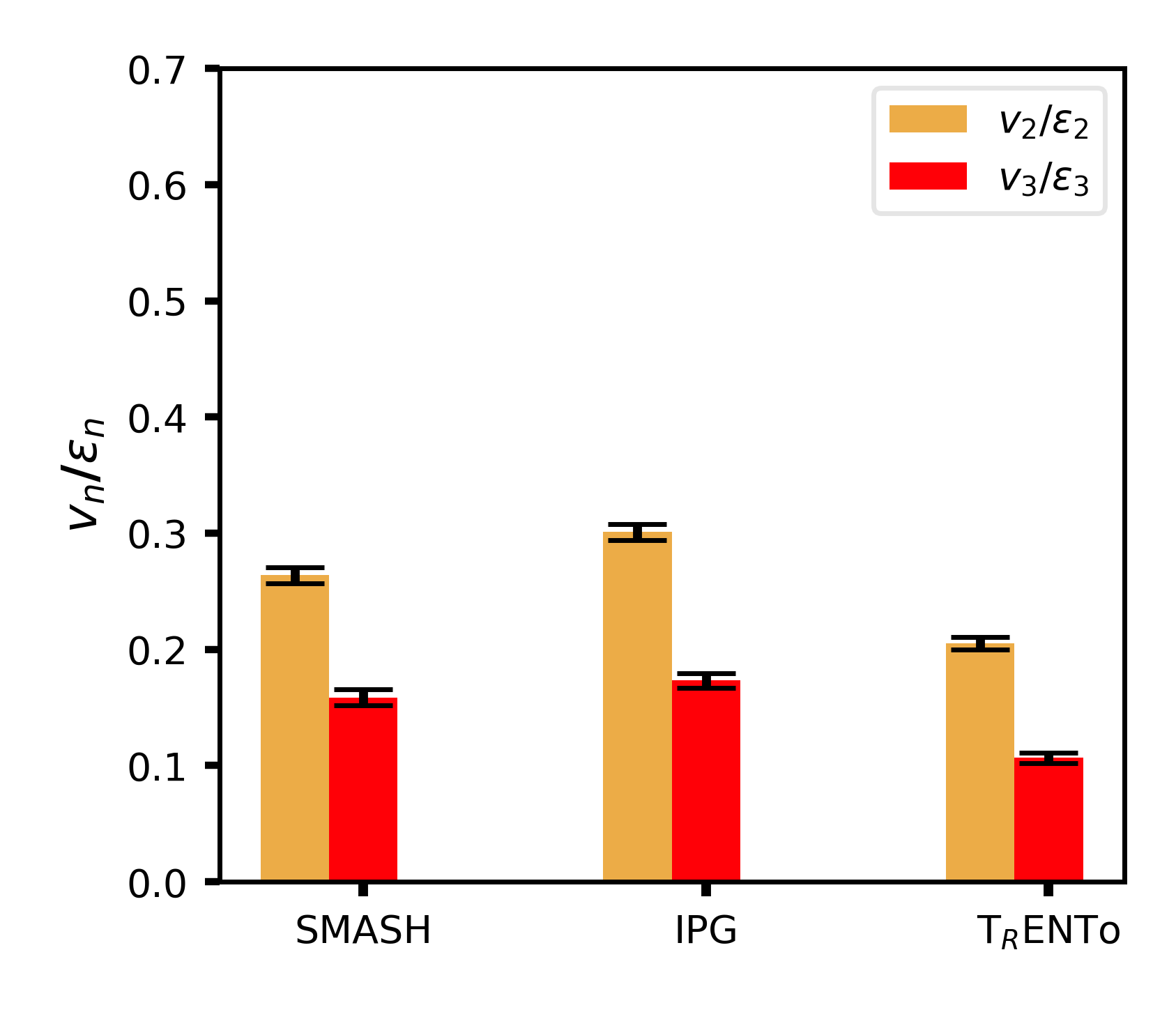}
    \caption{The response function $v_n$/$\epsilon_n$ for all the models at 0-5\% centrality (top) and 20-30\% (bottom).}
    \label{fig:ratio}
\end{figure}
In the following, we restrict ourselves to $\sqrt{s_{NN}}$ = 200 GeV in order to limit computational costs while at the same time achieving statistically significant statements. 
Fig. \ref{fig:averaged_05} and \ref{fig:averaged_2030} show the averaged values of $\epsilon_2$, $\epsilon_3$, $v_2$ and $v_3$ for the three models for Au-Au collisions at $\sqrt{s_{NN}}$ = 200 GeV at 0-5\% and 20-30\% centrality, respectively. As expected from Fig. \ref{fig:ecc_Au}, the eccentricities of the models have similar values in both centrality classes. The differences are slightly greater for the final state flow, which is in general greater when IP-Glasma is employed as an initial state model, especially for central collisions. Although this suggests that the models produce comparable results,  more intricate differences arise when looking at the response function in \ref{fig:ratio}. Although the only change in the three setups is the choice of the initial condition model, the response to initial state eccentricities differs, especially at low eccentricities. Deeper insights into this can be gained by looking into the distribution of the variables in an event-by-event basis.

\subsection{Event-by-event distributions}
\begin{figure}
    \centering
    \includegraphics{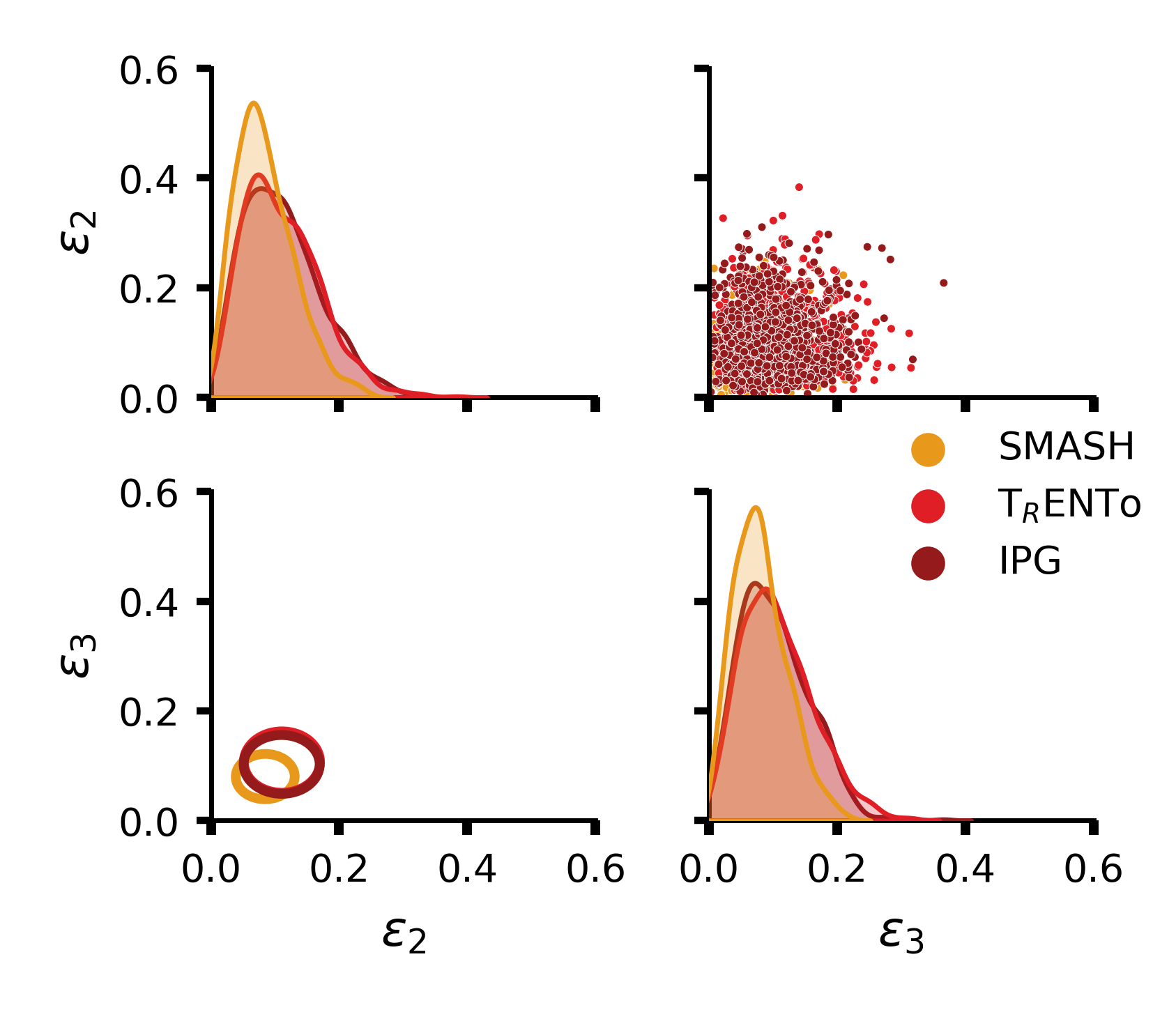}
    \includegraphics{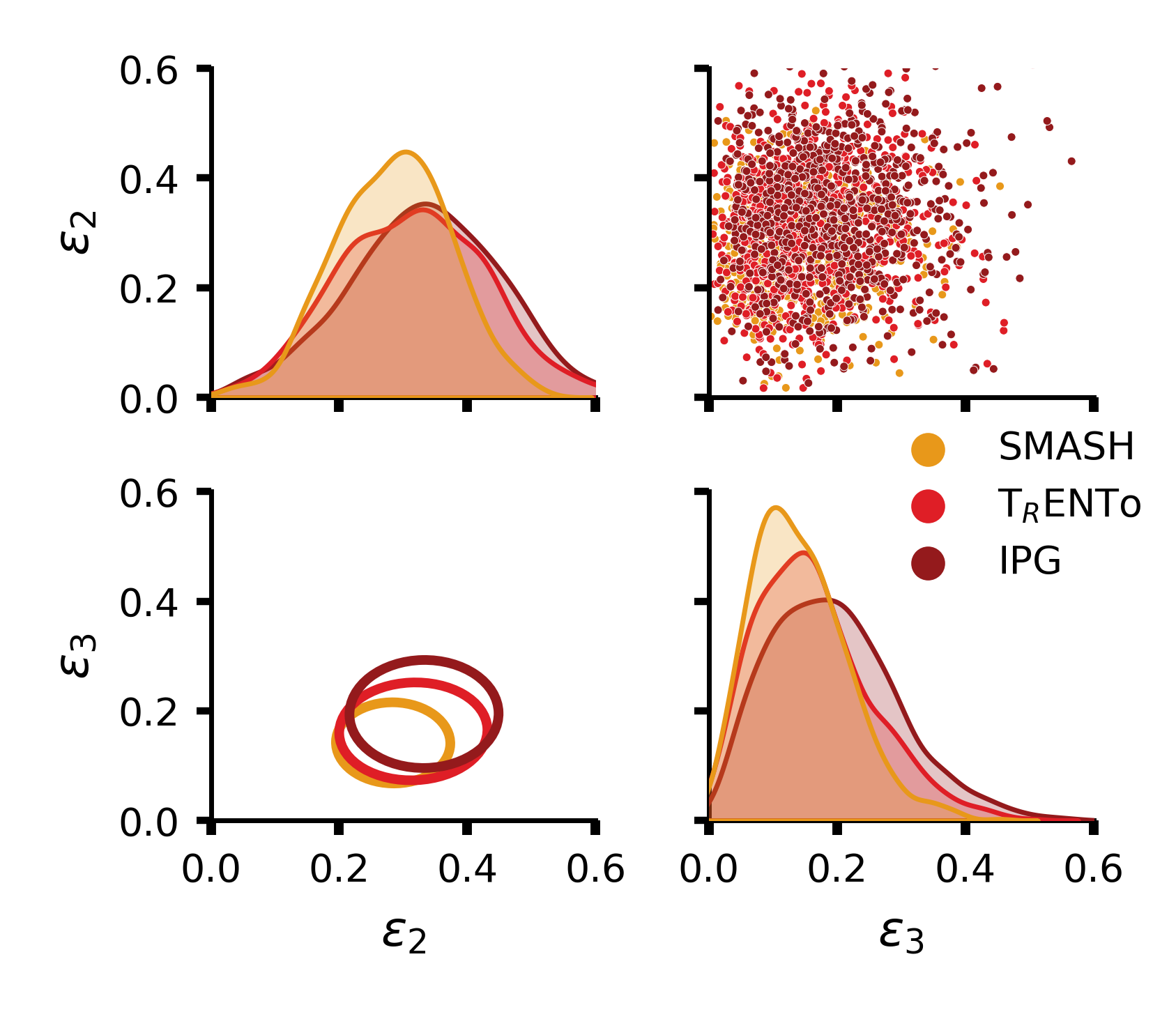}
    \caption{Distribution of the initial state eccentricities. On the diagonal, the normalized probability density distribution is shown. On the top right, a 2D scatter plot of the $\epsilon_2$-$\epsilon_3$ distribution. The lower left plot shows the relationship of the two quantities, with the center of the ellipse at the mean of $\epsilon_2$ and $\epsilon_3$, width and height are the variance of $\epsilon_2$ and $\epsilon_3$, respectively, and the angle shows the covariance.
    Data at 0-5\% centrality (top) and 20-30\% centrality (bottom).}
    \label{fig:eps_distr}
\end{figure}
\begin{figure}
    \centering
    \includegraphics{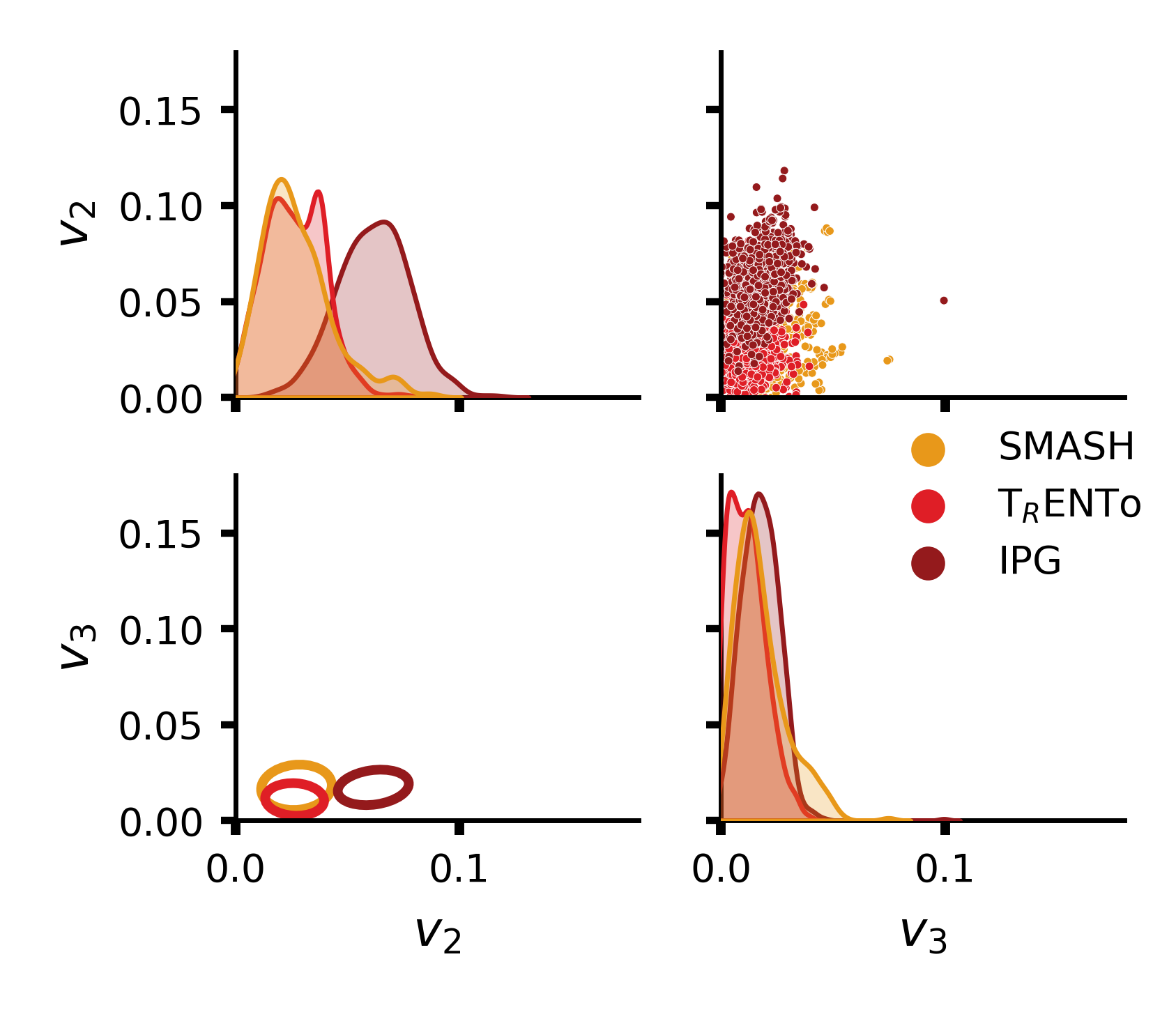}
    \includegraphics{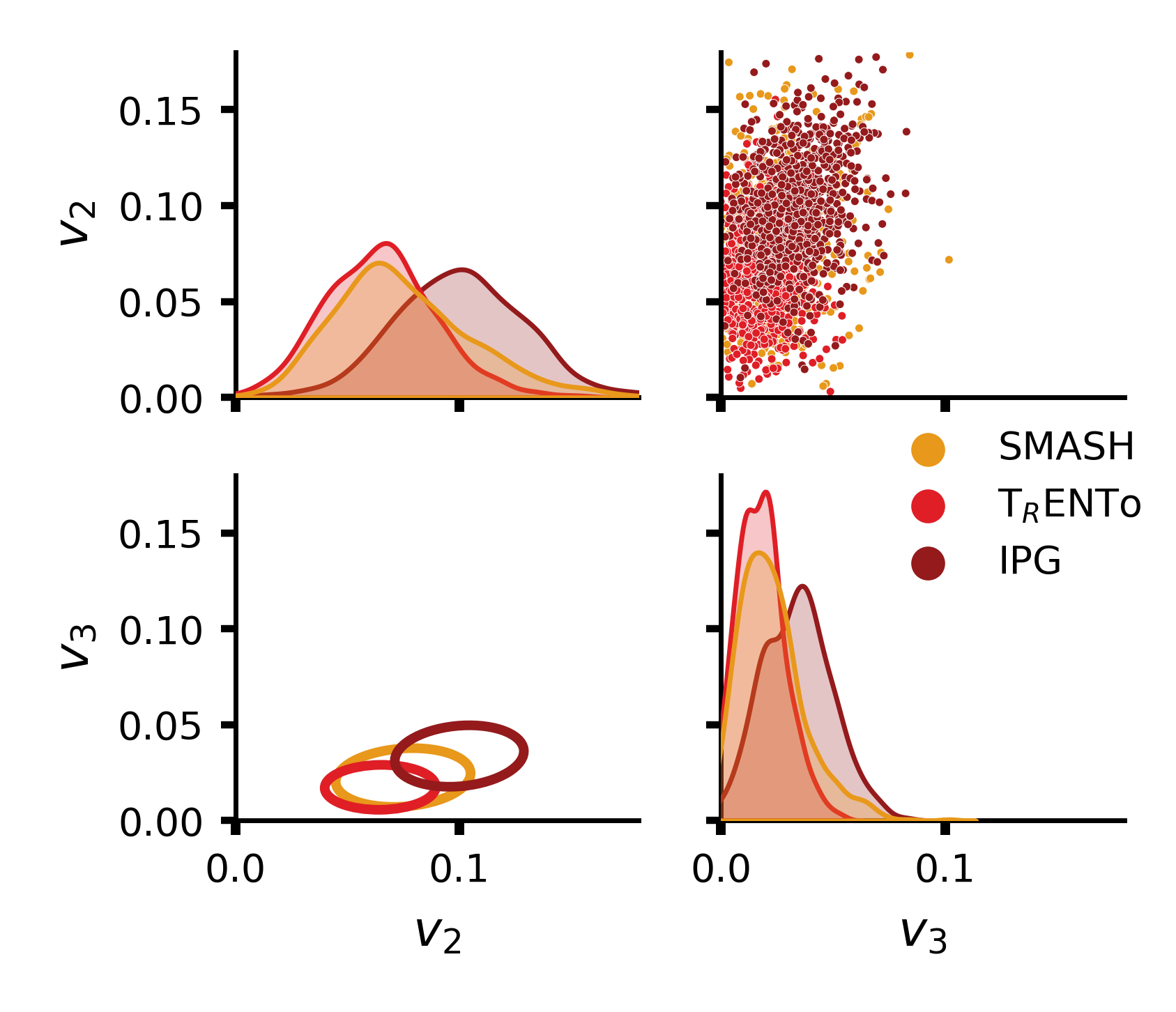}
    \caption{Distribution of the initial state eccentricities. The diagonal shows the normalized probability density distribution. On the top right, a 2D scatter plot of the $v_2$-$v_3$ distribution. The lower left plot shows the relationship of the two quantities, with the center of the ellipse at the mean of $v_2$ and $v_3$, width and height are the variance of $v_2$ and $v_3$, respectively, and the angle shows the covariance.
    Data at 0-5\% centrality (top) and 20-30\% centrality (bottom).}
    \label{fig:flow_distr}
\end{figure}

Fig.\ref{fig:eps_distr} and \ref{fig:flow_distr} show the eccentricity and flow distribution, respectively. In the case of eccentricities, we see that SMASH has a lot more peaked distribution of eccentricities, especially at central collisions, where the distribution for IP-Glasma and \trento is very similar. This changes for off-central collisions, where the distribution of SMASH stays more peaked for $\epsilon_2$, but becomes comparable to \trento for $\epsilon_3$. In both cases, IP-Glasma is spread out considerably wider. For both centralities and all three models, there is no significant $\epsilon_2$-$\epsilon_3$ correlation, and the spread in $\epsilon_2$ is greater than for $\epsilon_3$.
Although final state flows are often seen as a linear response to initial state eccentricities, the initial state distributions vary in their behaviour from the final state properties. For both centrality classes, the distributions for the \trento model are more peaked than the results with SMASH and IP-Glasma. Additionally, we now observe significant correlations, as can be seen in the orientations of the ellipses at the lower left corner. SMASH and IP-Glasma results show a positive correlation between $v_2$ and $v_3$, whereas \trento exhibits a negative correlation for central events.
This shows that the properties of final state flow cannot be exclusively reduced to the initial state eccentricties, even if other elements of the hybrid approach are fixed. In the following, we take a look at possible candidates to explain the deviations between the models.

\subsection{Pearson Correlations}\label{sec:corr}
\begin{figure*}
    \centering
    \includegraphics{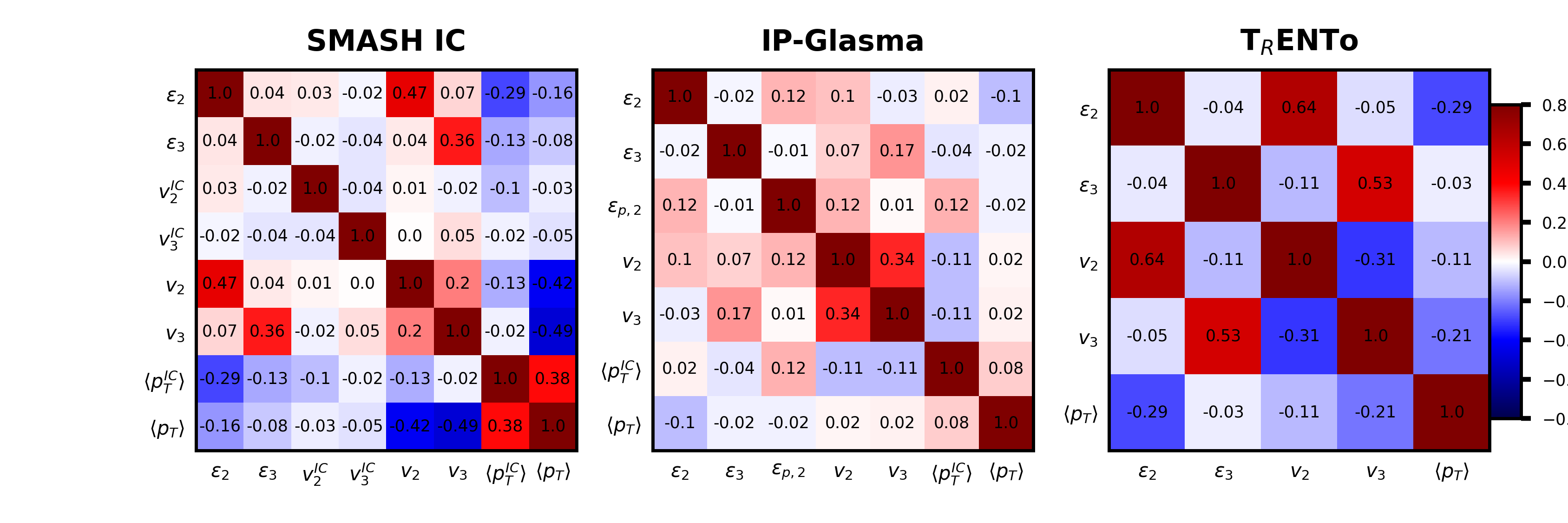}
    \includegraphics{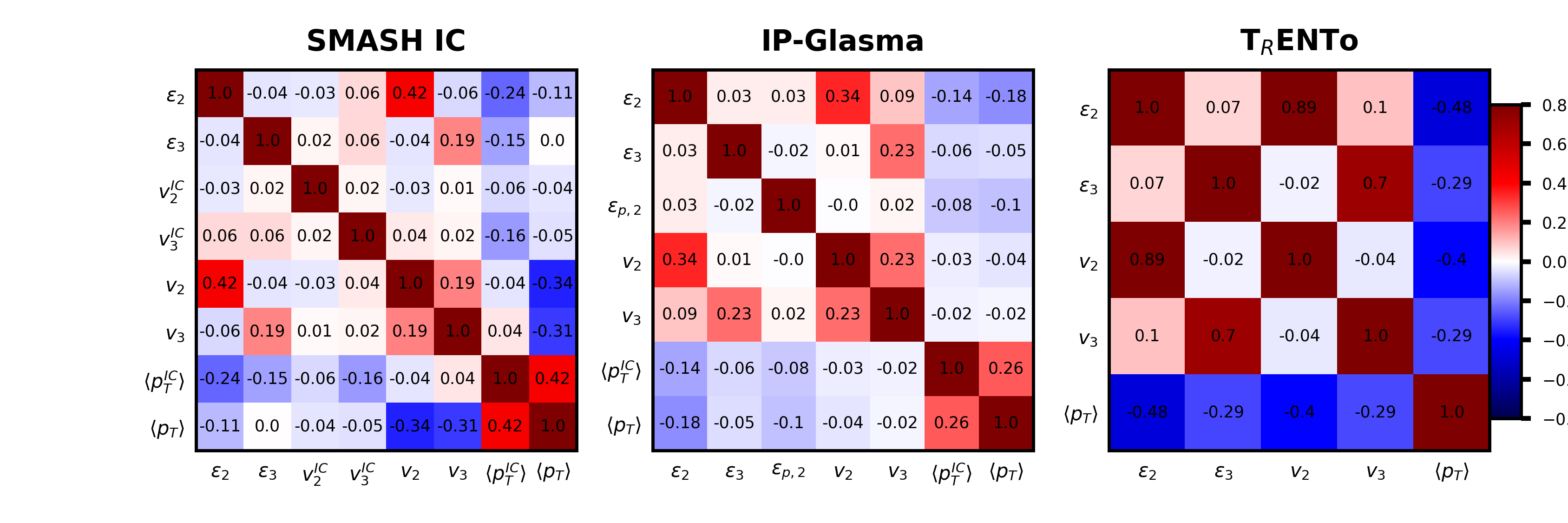}
    \caption{Pearson correlation matrix for all three initial state models, at 0-5\% centrality (top) and 20-30\% centrality (bottom).}
    \label{fig:corr}
\end{figure*}
Fig. \ref{fig:corr} shows the Pearson correlation matrix for all relevant quantities for all three models at the studied centrality classes. We find here also the observed correlations from Fig. \ref{fig:eps_distr} and \ref{fig:flow_distr}.
Regarding the $\epsilon_n$-$v_n$-correlations, we observe that these are in general the strongest for \trento, followed by SMASH. For IP-Glasma, they become especially weak for central collisions, which can be explained by the higher granularity of the IP-Glasma initial conditions and therefore the higher impact of fluctuations. Whereas they increase both for \trento and IP-Glasma when going to more off-central collisions, they decrease in the case of SMASH.
For SMASH, we see that the initial state flows do not exhibit any significant correlation with final state or initial state observables. The case is less clear for central IP-Glasma events, where the correlation between $\epsilon_p$ and $v_2$ becomes of the same magnitude as the $\epsilon_2$-$v_2$-correlation, albeit both at a very low level.
Looking at the final state transverse momentum, the picture is very different for the three models. In the case of \trento, it is slightly anticorrelated with all initial and final state properties in the central collision, and even stronger anticorrelated at off-central collisions.
In SMASH on the other hand, a strong anticorrelation is only observed towards final flows, and in the case of IP-Glasma, no significant correlation is observed. It is noteworthy that SMASH exhibits a significant correlation between radial flow and final transverse momentum, much stronger than what one finds for IP-Glasma. SMASH additionally shows a slight anticorrelation between initial state eccentricities and the radial flow. The initial ellipticity is negatively correlated to the inverse root mean square radius. Smaller, more compact sources, give larger transverse momentum, but a smaller deformation.\\

\subsection{Regression}

\begin{figure}
    \centering
    \includegraphics{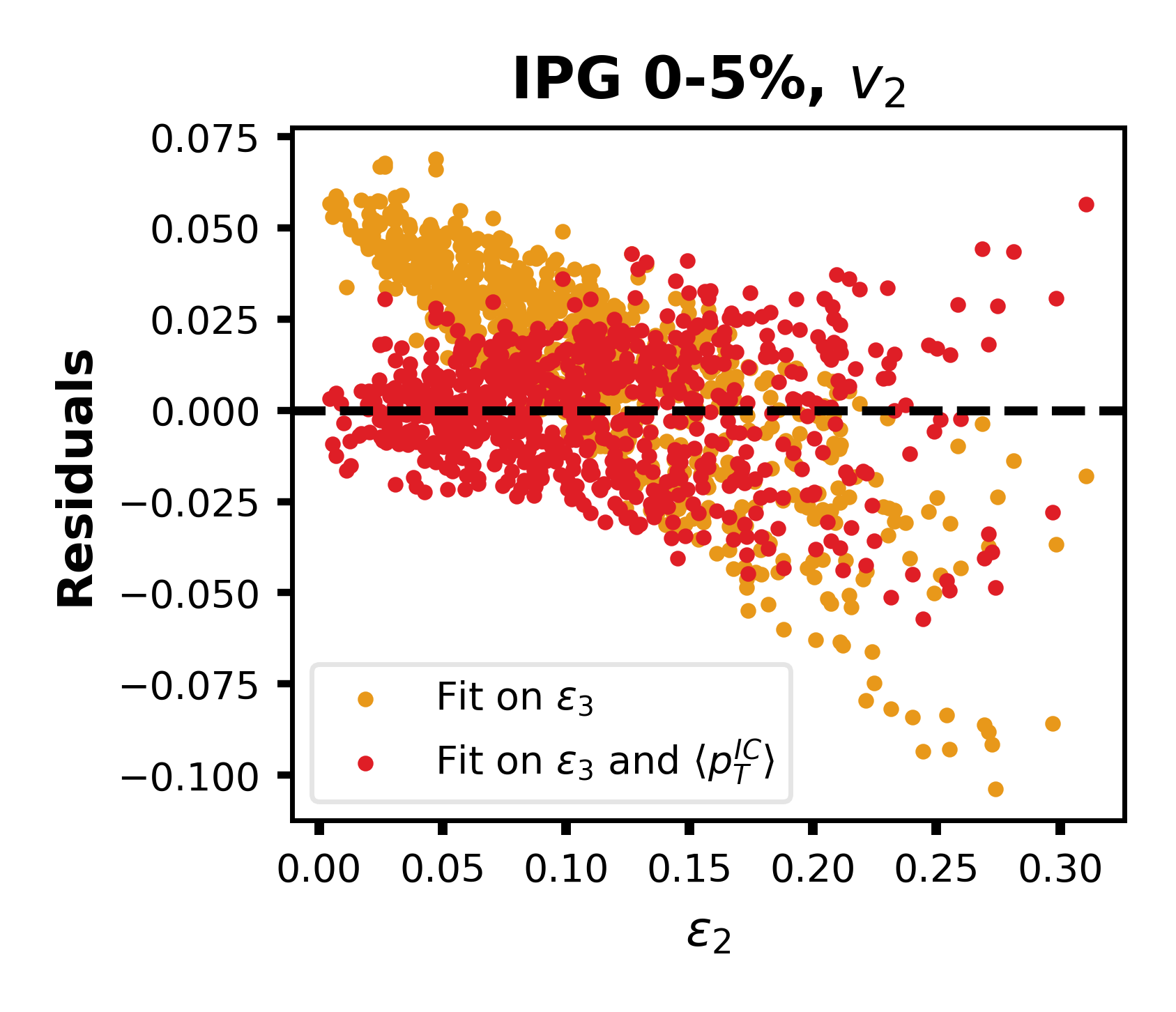}
    \includegraphics{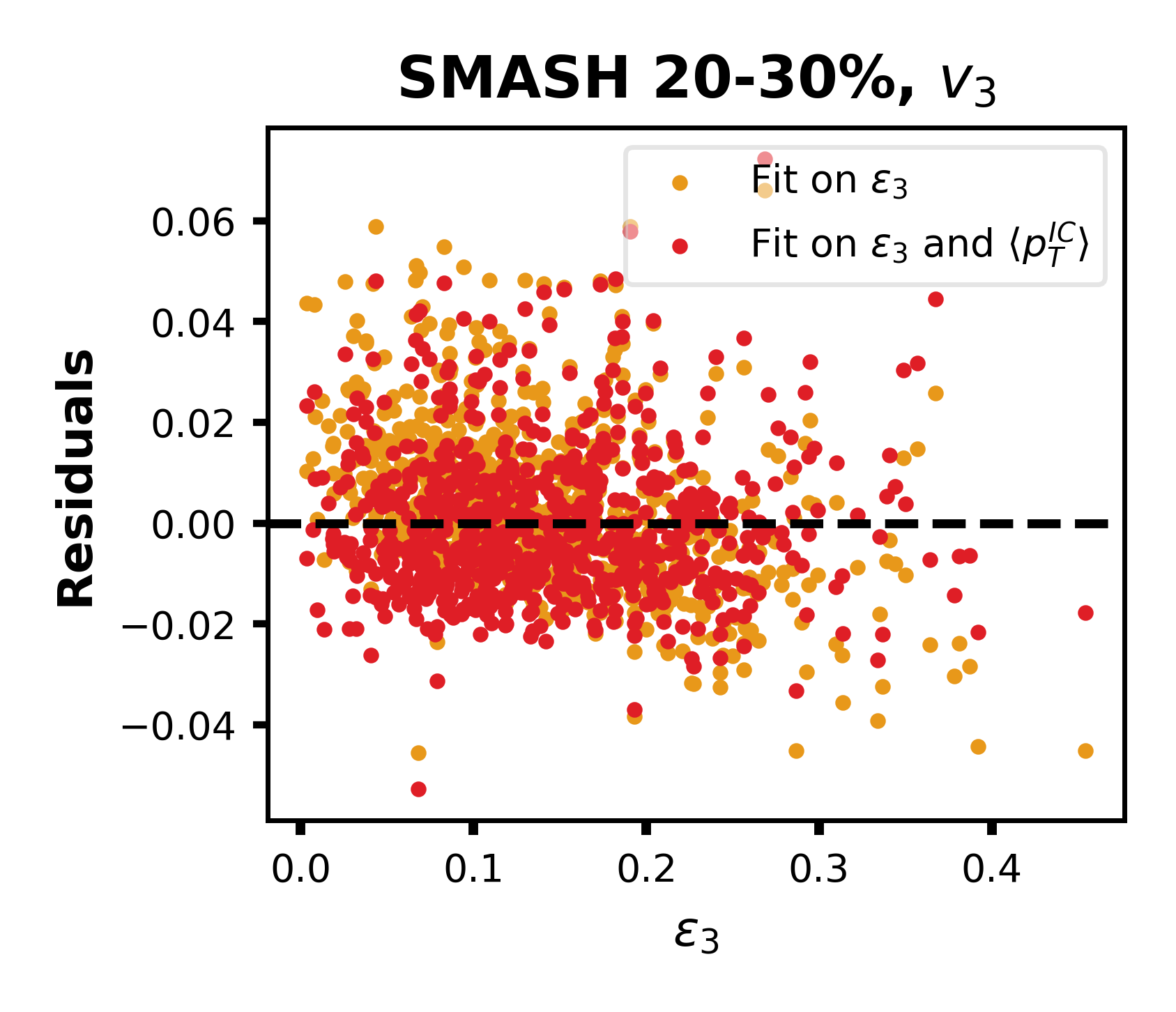}
    \caption{Residuals between observed and fitted values for a fit on $\epsilon_2$ on the one hand and $\epsilon_2$ and $\langle p_T^{IC} \rangle$ on the other hand, for a simulation with IP-Glasma at 0-5\% centrality (top). The same plot for midcentral collisions in SMASH, for the fit with $v_3$ on $\epsilon_3$ (bottom).}
    \label{fig:resid}
\end{figure}

The assumption of a linear relationship between initial eccentricities and final state flow can be seen as applying a linear regression model. Due to our access to additional initial state properties, we can test whether further initial state properties contribute to the emergence of final state flow, apart from the relevant eccentricity mode. It is important to stress that it is not sufficient to only study the correlations for this, as correlations only capture the relationship between two variables, but we are interested in a multiple linear statement. 
This can be clarified in a simple example. Assume that there are two initial state properties both contributing to $v_q$ with $v_q$ = $ 2 \epsilon_p + \epsilon_q$, and both initial state properties being perfectly anticorrelated. Although $\epsilon_q$ is also a source of $v_q$, depending on the data at hand, the Pearson correlation will be negative, as the effect is shadowed by the stronger contribution of $\epsilon_p$.
For the SMASH and IP-Glasma initial conditions at both studied centrality classes, a selection of linear regressions on different initial state properties for both $v_2$ and $v_3$ are given in the appendix. For all used explanatory data, we report the result of the coefficient as well as the p-value, which gives the probability of receiving the outcome under the null hypothesis. As a rule of thumb, a p-value smaller than 0.05 is necessary but not sufficient to consider the inclusion a dependent variable as statistically significant.  Additionally we report $r^2$, which is a common measure on how well the resulting model manages to predict the dependent variable. $r^2$ is the proportion of the variation of the dependent variable that is predictable from the independent variables on the basis of the model.
Adding further independent variables is always expected to slightly improve $r^2$. For all the observed cases, we can however determine that the inclusion of $\langle p_T^{IC} \rangle $ is statistically significant and improves $r^2$ considerably more than the inclusion of any
of the other independent variables. Due to this consistent and significant improvement, it becomes clear that the radial flow in the initial state is a next-to-leading order contribution to final state flows. A stronger transverse push yields a stronger hydrodynamic response of the spectra to the initial azimuthal deformation. The strength of the improvement in $r^2$ is generally stronger for IP-Glasma due to its higher radial flow.
This explains the differences in the response functions observed earlier: depending on the initial condition model, the presence of initial state transverse flow modifies the response to initial state eccentricites. The eccentricity alone is not the only aspect of the initial state which determines initial state flow.
The effect of including radial flow in the predictor of final state flow becomes clear in Fig. \ref{fig:resid}. Here we show the residuals of the prediction of the linear regression model with respect to the observed value in two cases where the improvement by including a second independent variable was especially significant. In both cases, this compensates to high predictions of flow for small eccentricities, and too small predictions for the flow at high eccentricities. For events with small eccentricities, radial flow becomes the main contribution to the final flow.







\subsection{SMASH at intermediate energies}
As the SMASH IC is a three-dimensional initial condition model, it can be also applied to lower collision energies. Therefore, we can extend the study of this initial condition model also to the case of PbPb @ $\sqrt{s_{NN}}$=17.3 GeV. 
\begin{figure}
    \centering
    \includegraphics{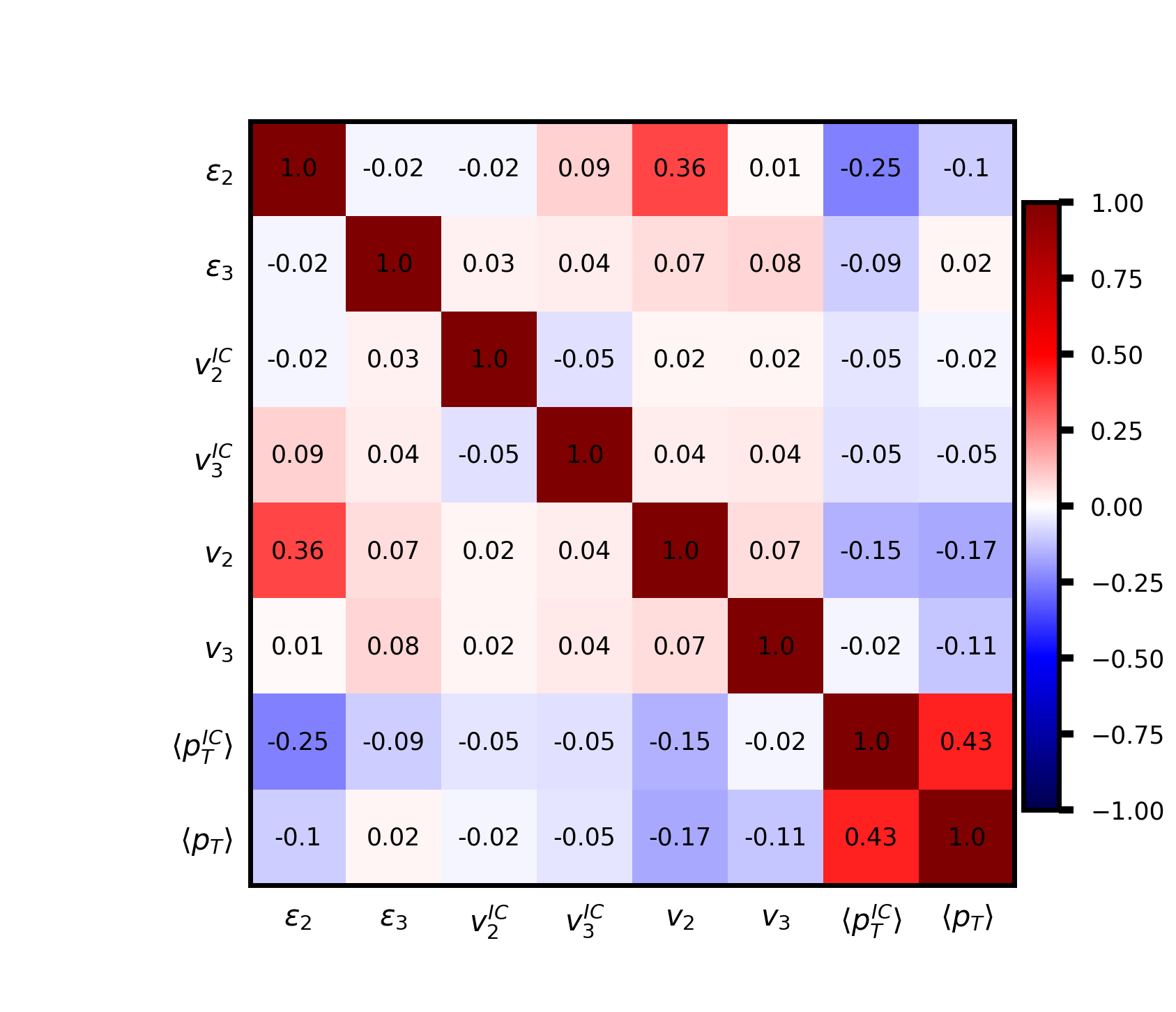}
    \includegraphics{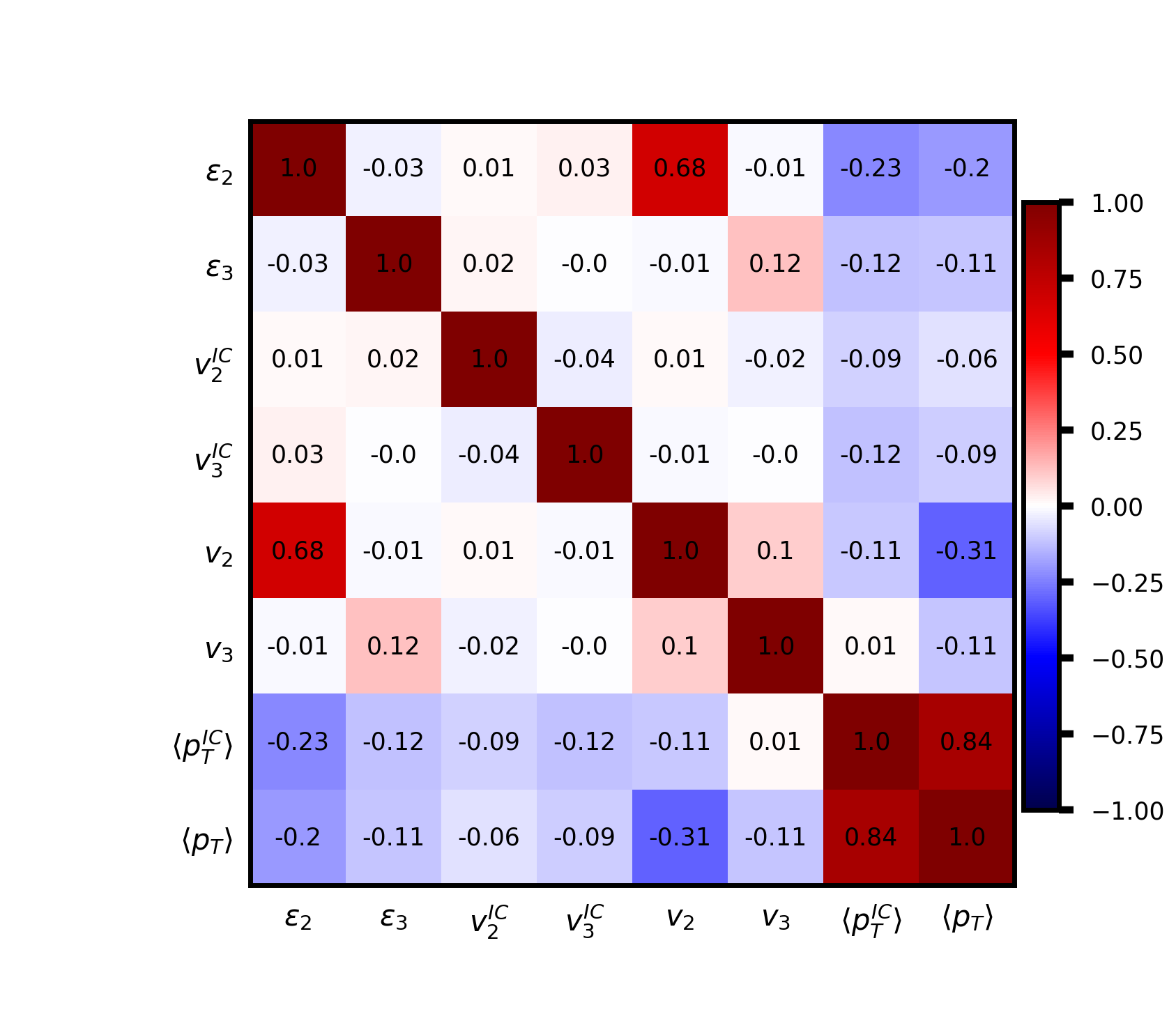}
    \caption{Pearson correlation matrix for SMASH at $\sqrt{s_{NN}}$= 17.3 GeV, at 0-5\% centrality (top) and 20-30\% centrality (bottom).}
    \label{fig:corr_173}
\end{figure}

\begin{figure}
    \centering
    \includegraphics{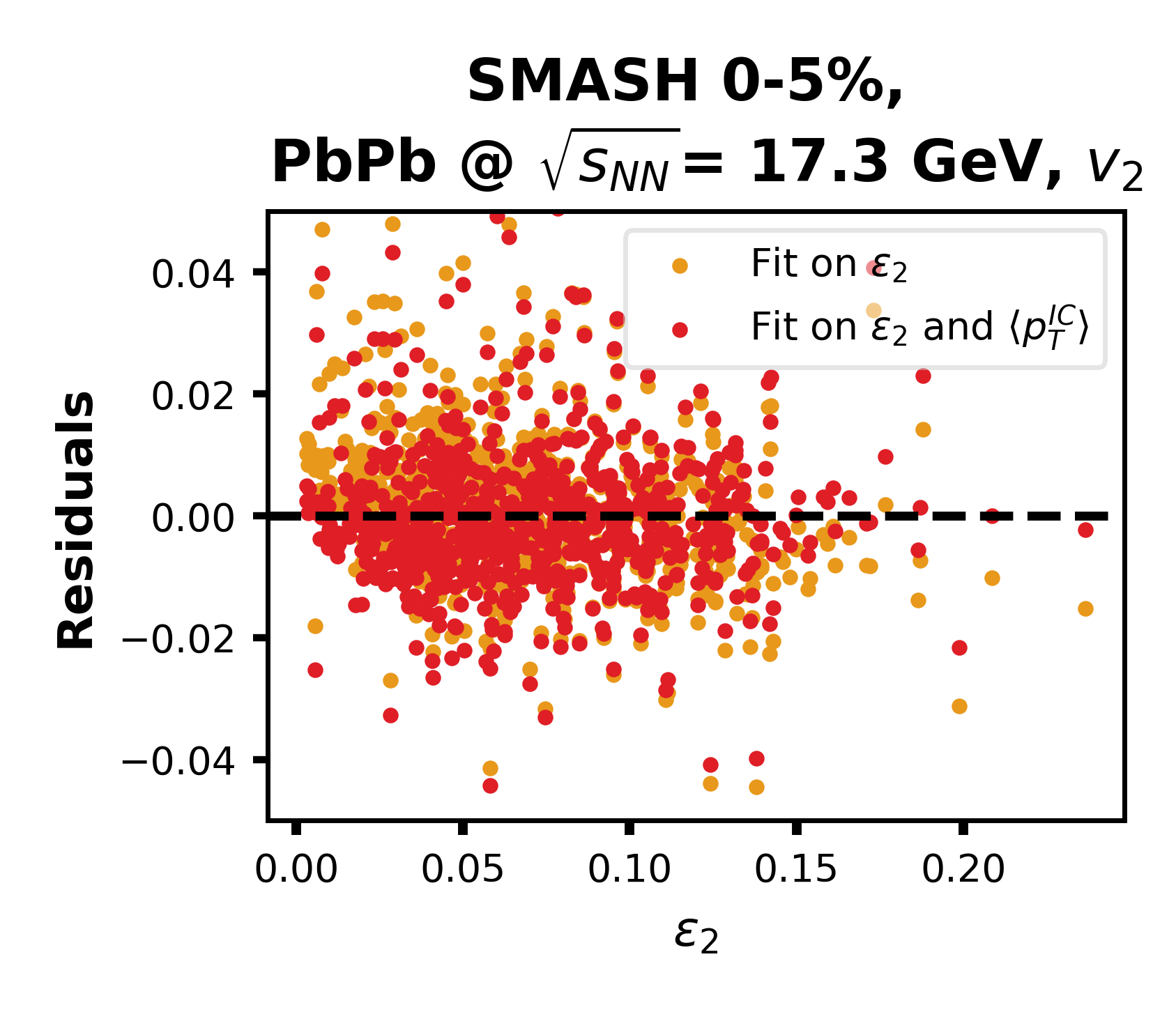}
    \caption{Residuals between observed and fitted values for a fit on $\epsilon_2$ on the one hand and $\epsilon_2$ and $\langle p_T^{IC} \rangle$ on the other hand, for a simulation with SMASH at 0-5\% centrality and $\sqrt{s_{NN}}$ = 17.3 GeV.}
    \label{fig:resid_173}
\end{figure}
The correlation matrix in Fig. \ref{fig:corr_173} shows a weaker dependence between $\epsilon_3$ and $v_3$ in comparison to the high-energy case. Additionally, the correlation between the flow modes is also reduced, hinting at a stronger role of non-flow for $v_3$. We observe a greatly increased relationship between radial flow and final transverse flow, whereas the relationship between eccentricities and radial flow remains roughly unchanged.
Again, tables of different linear regression models can be found in the appendix. Due to the weak $\epsilon_3$-$v_3$-correlation, linear regression with $v_3$ as dependent variable fails to appropriately describe the data. The including of radial flow in the linear fit again improves the regression, albeit at a smaller degree than for higher energies. The effect on the residuals remains the same than at higher energies, as can be seen in Fig. \ref{fig:resid_173}.



\section{Conclusions and Outlook}\label{sec:Conclusion}

In this work, we have presented a comparison between three different initial state models, SMASH IC, \trento and IP-Glasma, in the context of a hybrid approach for simulating heavy-ion collisions. Although averaged quantities like eccentricities and final state flows were found to be comparable across the models for large nuclei, significant differences emerged upon studying distributions and correlations on an event-by-event basis.

The distributions of eccentricities and final state flows were found to vary substantially between models, with \trento exhibiting the most peaked distributions. This demonstrates that averaged values alone are insufficient to fully characterize differences between initial state models. Additionally, the correlations between initial eccentricities and final flows were found to be noticeably model-dependent, with the strongest correlations seen for \trento and the weakest for IP-Glasma, especially in central collisions. Similarly, although the same hydrodynamic evolution was used in all cases, the response of the system to the initial eccentricities varied among the initial condition models.

This results from the observation that the transverse momentum in the initial state, as provided by SMASH IC and IP-Glasma, was found to non-negligibly affect final flows and therefore improves predictions of final flows when included alongside eccentricity in a linear regression model. This contribution from initial radial flow presents a relevant second-order effect when determining final anisotropic flows. This challenges the common assumption of a universal linear hydrodynamic response. Extending the analysis to lower energies using SMASH IC showed overall weaker correlations, however initial radial flow was still found to be beneficial for predicting final flows.

In contrast, the anisotropies of the initial state momentum were shown to not affect final state in any significant way. This means that the initial stat momentum anisotropies isotropize quickly during the hydrodynamic evolution, and do not affect the momentum space of the final state.

In summary, this comprehensive event-by-event analysis has revealed significant differences between common initial state models that are highly relevant when predicting final state observables. While averaged values appear comparable between models, the event-by-event distributions, correlations, and hydrodynamic response differ substantially. Our results highlight the importance of initial transverse momentum as a contributor to final flows, emphasizing the need to characterize initial conditions beyond eccentricity alone when correlating to final state observables.

Future studies could extend this analysis to further initial condition models. Most importantly, IP-Glasma can be combined with alternative approaches for the pre-equilibrium dynamics \cite{PhysRevC.99.034910}. Alternatively, the study could be also performed in an anisotropic hydrodynamics setup \cite{MCNELIS2021108077}, which could potentially improve model uncertainties at the point of fluidization. Additionally, it would be worthwhile to extend the study to further observables sensitive to the initial state. Next to higher order flow coefficients, comparisons for 3D initial state models can also be performed on the decorrelation length of anisotropic flow \cite{pang2016decorrelation}. Comparisons to other 3D initial state models would also allow to perform studies at lower energies, which are the stronghold of SMASH IC. 

\begin{acknowledgments}
This work was supported by the Deutsche Forschungsgemeinschaft (DFG, German Research
Foundation) – Project number 315477589 – TRR 211. N.G. acknowledges support by the Stiftung Polytechnische Gesellschaft Frankfurt am Main as well as the Studienstiftung des Deutschen Volkes.  Computational resources have been provided by the GreenCube at GSI. 
\end{acknowledgments}
\label{sec:appendix_a}
\FloatBarrier

\onecolumngrid
\appendix
\section{Regression tables}
The following section contains the regression results for the different models and dependent variables, as well as different choices for the set of independent variables. Each row is an independent regression. For each independent variable, the coefficient with error was well as the $p$-value is given. As a p-value smaller than 0.05 is seen as statistically significant, it is printed in bold. If a value is present in the column of a row, the respective independent variable was used in the regression. The final column contains the $r^2$-value for the regression.

We observe throughout all models, independent variables, and centralities for $\sqrt{s_{NN}}$ = 200 GeV that the most significant increase in predictive power, measured by $r^2$, is gained by adding $\langle p_T^{IC} \rangle$ as a second independent variable. Adding further variables is either not statistically significant, leading to greater $p^2$-values, or only negligibly improves $r^2$, which is always expected to stay constant or improve upon adding further independent variables. Using other independent variables apart from $\epsilon_n$ and $\langle p_T^{IC} \rangle$ fails to consistently reach improvements greater or equal to the combination of $\epsilon_n$ and $\langle p_T^{IC} \rangle$. 

\FloatBarrier
\begin{table*}
    \centering
    \begin{tabularx}{\textwidth}{X|X|X|X|X|X|X|X|X|X|X}
    coefficient $\epsilon_2$ & p-value $\epsilon_2$ &  coefficient $\epsilon_3$ & p-value $\epsilon_3$ & coefficient $\langle p_T^{IC} \rangle$ & p-value $\langle p_T^{IC} \rangle$ & coefficient $v_2^{IC}$ & p-value $v_2^{IC}$ &  coefficient $v_3^{IC}$ & p-value $v_3^{IC}$ & $r^2$ \\
    \hline
     2.86e-01 $\pm$ 5.8e-03 & \textbf{0.00} & - & - & - & - & - & - & - & - & 0.763 \\
     \hline
 2.26e-01 $\pm$ 9.1e-03 & \textbf{0.00} & 8.09e-02 $\pm$ 9.7e-03 & \textbf{0.00} & - & - & - & - & - & - & 0.783 \\
 \hline
 1.72e-01 $\pm$ 1.1e-02 & \textbf{0.00} & - & - & 3.65e-06 $\pm$ 2.9e-07 & \textbf{0.00} & - & - & - & - & 0.804 \\
 \hline
 2.29e-01 $\pm$ 9.1e-03 & \textbf{0.00} & - & - & - & - & 4.94e-01 $\pm$ 6.2e-02 & \textbf{0.00} & - & - & 0.781 \\
 \hline
 1.93e-01 $\pm$ 1.0e-02 & \textbf{0.00} & 4.50e-02 $\pm$ 1.1e-02 & \textbf{0.00} & - & - & 2.45e-01 $\pm$ 7.1e-02 & \textbf{0.00} & 3.14e-01 $\pm$ 7.8e-02 & \textbf{0.00} & 0.794 \\
 \hline
 1.70e-01 $\pm$ 1.1e-02 & \textbf{0.00} & 1.12e-02 $\pm$ 1.2e-02 & 3.57e-01 & 3.21e-06 $\pm$ 5.3e-07 & \textbf{0.00} & 2.10e-02 $\pm$ 7.8e-02 & 7.89e-01 & 4.77e-02 $\pm$ 8.8e-02 & 5.87e-01 & 0.804 \\
    \end{tabularx}
    \caption{Regression results for $v_2$ with SMASH IC at 0-5\% centrality, $\sqrt{s_{NN}}=$ 200 GeV.}
    \label{tab:smash_05_v2}
\end{table*}
\begin{table*}
    \centering
    \begin{tabularx}{\textwidth}{X|X|X|X|X|X|X|X|X|X|X}
    coefficient $\epsilon_2$ & p-value $\epsilon_2$ &  coefficient $\epsilon_3$ & p-value $\epsilon_3$ & coefficient $\langle p_T^{IC} \rangle$ & p-value $\langle p_T^{IC} \rangle$ & coefficient $v_2^{IC}$ & p-value $v_2^{IC}$ &  coefficient $v_3^{IC}$ & p-value $v_3^{IC}$ & $r^2$ \\
    \hline
    - & - & 1.93e-01 $\pm$ 4.7e-03 & \textbf{0.00} & - & - & - & - & - & - & 0.688 \\
\hline
 5.28e-02 $\pm$ 7.0e-03 & \textbf{0.00} & 1.48e-01 $\pm$ 7.5e-03 & \textbf{0.00} & - & - & - & - & - & - & 0.711 \\
\hline
 - & - & 1.06e-01 $\pm$ 9.4e-03 & \textbf{0.00} & 2.54e-06 $\pm$ 2.4e-07 & \textbf{0.00} & - & - & - & - & 0.728 \\
\hline
 - & - & 1.54e-01 $\pm$ 7.5e-03 & \textbf{0.00} & - & - & 3.11e-01 $\pm$ 4.8e-02 & \textbf{0.00} & - & - & 0.705 \\
\hline
 2.98e-02 $\pm$ 7.9e-03 & \textbf{0.00} & 1.23e-01 $\pm$ 8.5e-03 & \textbf{0.00} & - & - & 1.09e-01 $\pm$ 5.4e-02 & 4.50e-02 & 2.92e-01 $\pm$ 6.0e-02 & \textbf{0.00} & 0.724 \\
\hline
 1.66e-02 $\pm$ 8.4e-03 & 4.75e-02 & 1.04e-01 $\pm$ 9.4e-03 & \textbf{0.00} & 1.81e-06 $\pm$ 4.1e-07 & \textbf{0.00} & -1.79e-02 $\pm$ 6.1e-02 & 7.68e-01 & 1.41e-01 $\pm$ 6.8e-02 & 3.80e-02 & 0.731 \\
    \end{tabularx}
    \caption{Regression results for $v_3$ with SMASH IC at 0-5\% centrality, $\sqrt{s_{NN}}=$ 200 GeV.}
    \label{tab:smash_05_v3}
\end{table*}
\begin{table*}
    \centering
    \begin{tabularx}{\textwidth}{X|X|X|X|X|X|X|X|X|X|X}
    coefficient $\epsilon_2$ & p-value $\epsilon_2$ &  coefficient $\epsilon_3$ & p-value $\epsilon_3$ & coefficient $\langle p_T^{IC} \rangle$ & p-value $\langle p_T^{IC} \rangle$ & coefficient $v_2^{IC}$ & p-value $v_2^{IC}$ &  coefficient $v_3^{IC}$ & p-value $v_3^{IC}$ & $r^2$ \\
    \hline
2.53e-01 $\pm$ 3.6e-03 & \textbf{0.00} & - & - & - & - & - & - & - & - & 0.869 \\
\hline
 2.28e-01 $\pm$ 6.5e-03 & \textbf{0.00} & 5.51e-02 $\pm$ 1.2e-02 & \textbf{0.00} & - & - & - & - & - & - & 0.872 \\
\hline
 1.60e-01 $\pm$ 9.6e-03 & \textbf{0.00} & - & - & 1.91e-05 $\pm$ 1.9e-06 & \textbf{0.00} & - & - & - & - & 0.885 \\
\hline
 2.28e-01 $\pm$ 6.5e-03 & \textbf{0.00} & - & - & - & - & 4.01e-01 $\pm$ 8.8e-02 & \textbf{0.00} & - & - & 0.872 \\
\hline
 2.03e-01 $\pm$ 8.4e-03 & \textbf{0.00} & 3.37e-02 $\pm$ 1.3e-02 & \textbf{0.01} & - & - & 2.61e-01 $\pm$ 9.3e-02 & \textbf{0.00} & 3.30e-01 $\pm$ 1.1e-01 & \textbf{0.00} & 0.876 \\
\hline
 1.57e-01 $\pm$ 1.0e-02 & \textbf{0.00} & 1.31e-03 $\pm$ 1.3e-02 & 9.20e-01 & 1.81e-05 $\pm$ 2.3e-06 & \textbf{0.00} & -7.09e-03 $\pm$ 9.5e-02 & 9.41e-01 & 1.26e-01 $\pm$ 1.1e-01 & 2.38e-01 & 0.885 \\
    \end{tabularx}
    \caption{Regression results for $v_3$ with SMASH IC at 20-30\% centrality, $\sqrt{s_{NN}}=$ 200 GeV.}
    \label{tab:smash_2030_v2}
\end{table*}
\begin{table*}
    \centering
    \begin{tabularx}{\textwidth}{X|X|X|X|X|X|X|X|X|X|X}
    coefficient $\epsilon_2$ & p-value $\epsilon_2$ &  coefficient $\epsilon_3$ & p-value $\epsilon_3$ & coefficient $\langle p_T^{IC} \rangle$ & p-value $\langle p_T^{IC} \rangle$ & coefficient $v_2^{IC}$ & p-value $v_2^{IC}$ &  coefficient $v_3^{IC}$ & p-value $v_3^{IC}$ & $r^2$ \\
\hline
 - & - & 1.33e-01 $\pm$ 3.8e-03 & \textbf{0.00} & - & - & - & - & - & - & 0.617 \\
\hline
 3.76e-02 $\pm$ 3.5e-03 & \textbf{0.00} & 7.41e-02 $\pm$ 6.6e-03 & \textbf{0.00} & - & - & - & - & - & - & 0.668 \\
\hline
 - & - & 4.64e-02 $\pm$ 6.9e-03 & \textbf{0.00} & 1.03e-05 $\pm$ 7.1e-07 & \textbf{0.00} & - & - & - & - & 0.700 \\
\hline
 - & - & 9.11e-02 $\pm$ 5.9e-03 & \textbf{0.00} & - & - & 3.85e-01 $\pm$ 4.3e-02 & \textbf{0.00} & - & - & 0.654 \\
\hline
 2.31e-02 $\pm$ 4.6e-03 & \textbf{0.00} & 6.16e-02 $\pm$ 6.9e-03 & \textbf{0.00} & - & - & 1.79e-01 $\pm$ 5.0e-02 & \textbf{0.00} & 1.61e-01 $\pm$ 5.8e-02 & \textbf{0.01} & 0.678 \\
\hline
 -7.00e-04 $\pm$ 5.4e-03 & 8.97e-01 & 4.48e-02 $\pm$ 7.1e-03 & \textbf{0.00} & 9.44e-06 $\pm$ 1.3e-06 & \textbf{0.00} & 3.91e-02 $\pm$ 5.2e-02 & 4.49e-01 & 5.52e-02 $\pm$ 5.8e-02 & 3.40e-01 & 0.701 \\
    \end{tabularx}
    \caption{Regression results for $v_3$ with SMASH IC at 20-30\% centrality, $\sqrt{s_{NN}}=$ 200 GeV.}
    \label{tab:smash_2030_v3}
\end{table*}
\begin{table*}
    \centering
    \begin{tabularx}{\textwidth}{X|X|X|X|X|X|X|X|X}
    coefficient $\epsilon_2$ & p-value $\epsilon_2$ &  coefficient $\epsilon_3$ & p-value $\epsilon_3$ & coefficient $\langle p_T^{IC} \rangle$ & p-value $\langle p_T^{IC} \rangle$ & coefficient $\epsilon_p$ & p-value $\epsilon_p$ & $r^2$ \\
\hline
 4.39e-01 $\pm$ 9.4e-03 & \textbf{0.00} & - & - & - & - & - & - & 0.745 \\
\hline
 2.47e-01 $\pm$ 1.2e-02 & \textbf{0.00} & 2.68e-01 $\pm$ 1.3e-02 & \textbf{0.00} & - & - & - & - & 0.838 \\
\hline
 3.55e-02 $\pm$ 1.0e-02 & \textbf{0.00} & - & - & 5.99e-06 $\pm$ 1.3e-07 & \textbf{0.00} & - & - & 0.930 \\
\hline
 2.41e-01 $\pm$ 1.2e-02 & \textbf{0.00} & - & - & - & - & 4.02e+00 $\pm$ 2.0e-01 & \textbf{0.00} & 0.833 \\
\hline
 1.67e-01 $\pm$ 1.2e-02 & \textbf{0.00} & 1.90e-01 $\pm$ 1.3e-02 & \textbf{0.00} & - & - & 2.74e+00 $\pm$ 2.0e-01 & \textbf{0.00} & 0.872 \\
\hline
 3.36e-02 $\pm$ 1.0e-02 & \textbf{0.00} & 3.32e-02 $\pm$ 1.1e-02 & \textbf{0.00} & 5.37e-06 $\pm$ 2.1e-07 & \textbf{0.00} & 4.07e-01 $\pm$ 1.7e-01 & 1.73e-02 & 0.931 \\
    \end{tabularx}
    \caption{Regression results for $v_2$ with IP-Glasma at 0-5\% centrality, $\sqrt{s_{NN}}=$ 200 GeV.}
    \label{tab:ipg_05_v2}
\end{table*}
\begin{table*}
    \centering
    \begin{tabularx}{\textwidth}{X|X|X|X|X|X|X|X|X}
    coefficient $\epsilon_2$ & p-value $\epsilon_2$ &  coefficient $\epsilon_3$ & p-value $\epsilon_3$ & coefficient $\langle p_T^{IC} \rangle$ & p-value $\langle p_T^{IC} \rangle$ & coefficient $\epsilon_p$ & p-value $\epsilon_p$ & $r^2$ \\
\hline
 - & - & 1.39e-01 $\pm$ 3.5e-03 & \textbf{0.00} & - & - & - & - & 0.680 \\
\hline
 5.60e-02 $\pm$ 4.7e-03 & \textbf{0.00} & 9.19e-02 $\pm$ 5.1e-03 & \textbf{0.00} & - & - & - & - & 0.732 \\
\hline
 - & - & 3.53e-02 $\pm$ 6.0e-03 & \textbf{0.00} & 1.42e-06 $\pm$ 7.2e-08 & \textbf{0.00} & - & - & 0.789 \\
\hline
 - & - & 8.84e-02 $\pm$ 5.1e-03 & \textbf{0.00} & - & - & 9.64e-01 $\pm$ 7.6e-02 & \textbf{0.00} & 0.737 \\
\hline
 3.62e-02 $\pm$ 5.1e-03 & \textbf{0.00} & 7.26e-02 $\pm$ 5.4e-03 & \textbf{0.00} & - & - & 6.84e-01 $\pm$ 8.3e-02 & \textbf{0.00} & 0.754 \\
\hline
 4.33e-03 $\pm$ 5.5e-03 & 4.30e-01 & 3.52e-02 $\pm$ 6.0e-03 & \textbf{0.00} & 1.28e-06 $\pm$ 1.1e-07 & \textbf{0.00} & 1.29e-01 $\pm$ 9.1e-02 & 1.58e-01 & 0.790 \\
    \end{tabularx}
    \caption{Regression results for $v_3$ with IP-Glasma at 0-5\% centrality, $\sqrt{s_{NN}}=$ 200 GeV.}
    \label{tab:ipg_05_v3}
\end{table*}
\begin{table*}
    \centering
    \begin{tabularx}{\textwidth}{X|X|X|X|X|X|X|X|X}
    coefficient $\epsilon_2$ & p-value $\epsilon_2$ &  coefficient $\epsilon_3$ & p-value $\epsilon_3$ & coefficient $\langle p_T^{IC} \rangle$ & p-value $\langle p_T^{IC} \rangle$ & coefficient $\epsilon_p$ & p-value $\epsilon_p$ & $r^2$ \\
\hline
 2.77e-01 $\pm$ 3.8e-03 & \textbf{0.00} & - & - & - & - & - & - & 0.879 \\
\hline
 2.21e-01 $\pm$ 6.6e-03 & \textbf{0.00} & 1.08e-01 $\pm$ 1.1e-02 & \textbf{0.00} & - & - & - & - & 0.894 \\
\hline
 1.26e-01 $\pm$ 7.8e-03 & \textbf{0.00} & - & - & 1.17e-05 $\pm$ 5.6e-07 & \textbf{0.00} & - & - & 0.924 \\
\hline
 2.26e-01 $\pm$ 6.5e-03 & \textbf{0.00} & - & - & - & - & 1.84e+00 $\pm$ 2.0e-01 & \textbf{0.00} & 0.892 \\
\hline
 1.92e-01 $\pm$ 7.5e-03 & \textbf{0.00} & 8.79e-02 $\pm$ 1.1e-02 & \textbf{0.00} & - & - & 1.43e+00 $\pm$ 1.9e-01 & \textbf{0.00} & 0.901 \\
\hline
 1.19e-01 $\pm$ 8.1e-03 & \textbf{0.00} & 2.47e-02 $\pm$ 1.0e-02 & 1.57e-02 & 1.04e-05 $\pm$ 6.8e-07 & \textbf{0.00} & 3.97e-01 $\pm$ 1.8e-01 & 2.98e-02 & 0.925 \\
    \end{tabularx}
    \caption{Regression results for $v_2$ with IP-Glasma at 20-30\% centrality, $\sqrt{s_{NN}}=$ 200 GeV.}
    \label{tab:ipg_2030_v2}
\end{table*}
\begin{table*}
    \centering
    \begin{tabularx}{\textwidth}{X|X|X|X|X|X|X|X|X}
    coefficient $\epsilon_2$ & p-value $\epsilon_2$ &  coefficient $\epsilon_3$ & p-value $\epsilon_3$ & coefficient $\langle p_T^{IC} \rangle$ & p-value $\langle p_T^{IC} \rangle$ & coefficient $\epsilon_p$ & p-value $\epsilon_p$ & $r^2$ \\
\hline
 - & - & 1.45e-01 $\pm$ 3.3e-03 & \textbf{0.00} & - & - & - & - & 0.725 \\
\hline
 5.47e-02 $\pm$ 3.2e-03 & \textbf{0.00} & 7.04e-02 $\pm$ 5.2e-03 & \textbf{0.00} & - & - & - & - & 0.801 \\
\hline
 - & - & 5.06e-02 $\pm$ 5.5e-03 & \textbf{0.00} & 4.78e-06 $\pm$ 2.4e-07 & \textbf{0.00} & - & - & 0.818 \\
\hline
 - & - & 9.84e-02 $\pm$ 4.8e-03 & \textbf{0.00} & - & - & 1.12e+00 $\pm$ 8.9e-02 & \textbf{0.00} & 0.773 \\
\hline
 4.41e-02 $\pm$ 3.7e-03 & \textbf{0.00} & 6.31e-02 $\pm$ 5.3e-03 & \textbf{0.00} & - & - & 5.19e-01 $\pm$ 9.7e-02 & \textbf{0.00} & 0.808 \\
\hline
 2.19e-02 $\pm$ 4.4e-03 & \textbf{0.00} & 4.38e-02 $\pm$ 5.5e-03 & \textbf{0.00} & 3.18e-06 $\pm$ 3.7e-07 & \textbf{0.00} & 2.03e-01 $\pm$ 9.9e-02 & 4.02e-02 & 0.826 \\
    \end{tabularx}
    \caption{Regression results for $v_3$ with IP-Glasma at 20-30\% centrality, $\sqrt{s_{NN}}=$ 200 GeV.}
    \label{tab:ipg_2030_v3}
\end{table*}
\begin{table*}
    \centering
    \begin{tabularx}{\textwidth}{X|X|X|X|X|X|X|X|X|X|X}
    coefficient $\epsilon_2$ & p-value $\epsilon_2$ &  coefficient $\epsilon_3$ & p-value $\epsilon_3$ & coefficient $\langle p_T^{IC} \rangle$ & p-value $\langle p_T^{IC} \rangle$ & coefficient $v_2^{IC}$ & p-value $v_2^{IC}$ &  coefficient $v_3^{IC}$ & p-value $v_3^{IC}$ & $r^2$ \\
    \hline
 2.52e-01 $\pm$ 7.3e-03 & \textbf{0.00} & - & - & - & - & - & - & - & - & 0.616 \\
\hline
 1.96e-01 $\pm$ 1.1e-02 & \textbf{0.00} & 7.88e-02 $\pm$ 1.2e-02 & \textbf{0.00} & - & - & - & - & - & - & 0.636 \\
\hline
 1.67e-01 $\pm$ 1.4e-02 & \textbf{0.00} & - & - & 9.01e-06 $\pm$ 1.3e-06 & \textbf{0.00} & - & - & - & - & 0.638 \\
\hline
 2.04e-01 $\pm$ 1.1e-02 & \textbf{0.00} & - & - & - & - & 1.85e-01 $\pm$ 3.4e-02 & \textbf{0.00} & - & - & 0.630 \\
\hline
 1.75e-01 $\pm$ 1.3e-02 & \textbf{0.00} & 5.44e-02 $\pm$ 1.5e-02 & \textbf{0.00} & - & - & 9.14e-02 $\pm$ 4.0e-02 & 2.18e-02 & 6.02e-02 $\pm$ 4.7e-02 & 1.98e-01 & 0.640 \\
\hline
 1.65e-01 $\pm$ 1.5e-02 & \textbf{0.00} & 4.13e-02 $\pm$ 1.6e-02 & 1.22e-02 & 4.32e-06 $\pm$ 2.4e-06 & 7.07e-02 & 4.98e-02 $\pm$ 4.6e-02 & 2.79e-01 & 1.70e-02 $\pm$ 5.2e-02 & 7.46e-01 & 0.642 \\
    \end{tabularx}
    \caption{Regression results for $v_2$ with SMASH IC at 0-5\% centrality, $\sqrt{s_{NN}}=$ 17.3 GeV.}
    \label{tab:smash_173_05_v2}
\end{table*}
\begin{table*}
    \centering
    \begin{tabularx}{\textwidth}{X|X|X|X|X|X|X|X|X|X|X}
    coefficient $\epsilon_2$ & p-value $\epsilon_2$ &  coefficient $\epsilon_3$ & p-value $\epsilon_3$ & coefficient $\langle p_T^{IC} \rangle$ & p-value $\langle p_T^{IC} \rangle$ & coefficient $v_2^{IC}$ & p-value $v_2^{IC}$ &  coefficient $v_3^{IC}$ & p-value $v_3^{IC}$ & $r^2$ \\
\hline
 - & - & 1.01e-01 $\pm$ 1.1e-02 & \textbf{0.00} & - & - & - & - & - & - & 0.097 \\
\hline
 2.78e-02 $\pm$ 1.6e-02 & 8.83e-02 & 7.71e-02 $\pm$ 1.8e-02 & \textbf{0.00} & - & - & - & - & - & - & 0.101 \\
\hline
 - & - & 5.42e-02 $\pm$ 2.4e-02 & 2.29e-02 & 4.44e-06 $\pm$ 2.0e-06 & 2.73e-02 & - & - & - & - & 0.103 \\
\hline
 - & - & 7.33e-02 $\pm$ 1.8e-02 & \textbf{0.00} & - & - & 9.44e-02 $\pm$ 5.1e-02 & 6.25e-02 & - & - & 0.102 \\
\hline
 8.21e-03 $\pm$ 1.9e-02 & 6.73e-01 & 5.53e-02 $\pm$ 2.1e-02 & \textbf{0.01} & - & - & 5.11e-02 $\pm$ 5.8e-02 & 3.76e-01 & 9.06e-02 $\pm$ 6.8e-02 & 1.81e-01 & 0.105 \\
\hline
 5.26e-03 $\pm$ 2.1e-02 & 8.04e-01 & 5.16e-02 $\pm$ 2.4e-02 & 3.09e-02 & 1.22e-06 $\pm$ 3.5e-06 & 7.25e-01 & 3.93e-02 $\pm$ 6.7e-02 & 5.56e-01 & 7.85e-02 $\pm$ 7.6e-02 & 3.03e-01 & 0.105 \\
    \end{tabularx}
    \caption{Regression results for $v_3$ with SMASH IC at 0-5\% centrality, $\sqrt{s_{NN}}=$ 17.3 GeV.}
    \label{tab:smash_173_05_v3}
\end{table*}
\begin{table*}
    \centering
    \begin{tabularx}{\textwidth}{X|X|X|X|X|X|X|X|X|X|X}
    coefficient $\epsilon_2$ & p-value $\epsilon_2$ &  coefficient $\epsilon_3$ & p-value $\epsilon_3$ & coefficient $\langle p_T^{IC} \rangle$ & p-value $\langle p_T^{IC} \rangle$ & coefficient $v_2^{IC}$ & p-value $v_2^{IC}$ &  coefficient $v_3^{IC}$ & p-value $v_3^{IC}$ & $r^2$ \\
   \hline
 2.11e-01 $\pm$ 2.2e-03 & \textbf{0.00} & - & - & - & - & - & - & - & - & 0.922 \\
\hline
 1.94e-01 $\pm$ 3.9e-03 & \textbf{0.00} & 3.92e-02 $\pm$ 7.6e-03 & \textbf{0.00} & - & - & - & - & - & - & 0.925 \\
\hline
 1.63e-01 $\pm$ 5.6e-03 & \textbf{0.00} & - & - & 3.09e-05 $\pm$ 3.3e-06 & \textbf{0.00} & - & - & - & - & 0.930 \\
\hline
 1.94e-01 $\pm$ 4.0e-03 & \textbf{0.00} & - & - & - & - & 1.23e-01 $\pm$ 2.5e-02 & \textbf{0.00} & - & - & 0.925 \\
\hline
 1.82e-01 $\pm$ 5.0e-03 & \textbf{0.00} & 2.73e-02 $\pm$ 8.1e-03 & \textbf{0.00} & - & - & 8.37e-02 $\pm$ 2.6e-02 & \textbf{0.00} & 4.83e-02 $\pm$ 2.8e-02 & 8.62e-02 & 0.927 \\
\hline
 1.62e-01 $\pm$ 5.7e-03 & \textbf{0.00} & 7.30e-03 $\pm$ 8.5e-03 & 3.88e-01 & 2.96e-05 $\pm$ 4.5e-06 & \textbf{0.00} & 1.37e-02 $\pm$ 2.8e-02 & 6.20e-01 & -2.11e-02 $\pm$ 2.9e-02 & 4.73e-01 & 0.931 \\
    \end{tabularx}
    \caption{Regression results for $v_2$ with SMASH IC at 20-30\% centrality, $\sqrt{s_{NN}}=$ 17.3 GeV.}
    \label{tab:smash_173_2030_v2}
\end{table*}
\begin{table*}
    \centering
    \begin{tabularx}{\textwidth}{X|X|X|X|X|X|X|X|X|X|X}
    coefficient $\epsilon_2$ & p-value $\epsilon_2$ &  coefficient $\epsilon_3$ & p-value $\epsilon_3$ & coefficient $\langle p_T^{IC} \rangle$ & p-value $\langle p_T^{IC} \rangle$ & coefficient $v_2^{IC}$ & p-value $v_2^{IC}$ &  coefficient $v_3^{IC}$ & p-value $v_3^{IC}$ & $r^2$ \\
\hline
 - & - & 6.38e-02 $\pm$ 4.9e-03 & \textbf{0.00} & - & - & - & - & - & - & 0.184 \\
\hline
 1.09e-02 $\pm$ 4.4e-03 & 1.34e-02 & 4.62e-02 $\pm$ 8.6e-03 & \textbf{0.00} & - & - & - & - & - & - & 0.190 \\
\hline
 - & - & 3.57e-02 $\pm$ 9.8e-03 & \textbf{0.00} & 9.79e-06 $\pm$ 3.0e-06 & \textbf{0.00} & - & - & - & - & 0.195 \\
\hline
 - & - & 5.36e-02 $\pm$ 8.0e-03 & \textbf{0.00} & - & - & 4.03e-02 $\pm$ 2.5e-02 & 1.06e-01 & - & - & 0.187 \\
\hline
 7.93e-03 $\pm$ 5.8e-03 & 1.70e-01 & 4.34e-02 $\pm$ 9.3e-03 & \textbf{0.00} & - & - & 4.53e-03 $\pm$ 3.0e-02 & 8.81e-01 & 2.80e-02 $\pm$ 3.2e-02 & 3.86e-01 & 0.191 \\
\hline
 8.43e-04 $\pm$ 6.7e-03 & 9.01e-01 & 3.62e-02 $\pm$ 1.0e-02 & \textbf{0.00} & 1.08e-05 $\pm$ 5.3e-06 & 4.19e-02 & -2.09e-02 $\pm$ 3.3e-02 & 5.22e-01 & 2.79e-03 $\pm$ 3.5e-02 & 9.36e-01 & 0.196 \\
    \end{tabularx}
    \caption{Regression results for $v_3$ with SMASH IC at 20-30\% centrality, $\sqrt{s_{NN}}=$ 17.3 GeV.}
    \label{tab:smash_173_2030_v3}
\end{table*}
\FloatBarrier
\twocolumngrid 
\bibliography{2023_ic}

\begin{thebibliography}{74}%
\makeatletter
\providecommand \@ifxundefined [1]{%
 \@ifx{#1\undefined}
}%
\providecommand \@ifnum [1]{%
 \ifnum #1\expandafter \@firstoftwo
 \else \expandafter \@secondoftwo
 \fi
}%
\providecommand \@ifx [1]{%
 \ifx #1\expandafter \@firstoftwo
 \else \expandafter \@secondoftwo
 \fi
}%
\providecommand \natexlab [1]{#1}%
\providecommand \enquote  [1]{``#1''}%
\providecommand \bibnamefont  [1]{#1}%
\providecommand \bibfnamefont [1]{#1}%
\providecommand \citenamefont [1]{#1}%
\providecommand \href@noop [0]{\@secondoftwo}%
\providecommand \href [0]{\begingroup \@sanitize@url \@href}%
\providecommand \@href[1]{\@@startlink{#1}\@@href}%
\providecommand \@@href[1]{\endgroup#1\@@endlink}%
\providecommand \@sanitize@url [0]{\catcode `\\12\catcode `\$12\catcode
  `\&12\catcode `\#12\catcode `\^12\catcode `\_12\catcode `\%12\relax}%
\providecommand \@@startlink[1]{}%
\providecommand \@@endlink[0]{}%
\providecommand \url  [0]{\begingroup\@sanitize@url \@url }%
\providecommand \@url [1]{\endgroup\@href {#1}{\urlprefix }}%
\providecommand \urlprefix  [0]{URL }%
\providecommand \Eprint [0]{\href }%
\providecommand \doibase [0]{http://dx.doi.org/}%
\providecommand \selectlanguage [0]{\@gobble}%
\providecommand \bibinfo  [0]{\@secondoftwo}%
\providecommand \bibfield  [0]{\@secondoftwo}%
\providecommand \translation [1]{[#1]}%
\providecommand \BibitemOpen [0]{}%
\providecommand \bibitemStop [0]{}%
\providecommand \bibitemNoStop [0]{.\EOS\space}%
\providecommand \EOS [0]{\spacefactor3000\relax}%
\providecommand \BibitemShut  [1]{\csname bibitem#1\endcsname}%
\let\auto@bib@innerbib\@empty
\bibitem [{\citenamefont {Adare}\ \emph {et~al.}(2011)\citenamefont {Adare}
  \emph {et~al.}}]{PHENIX:2011yyh}%
  \BibitemOpen
  \bibfield  {author} {\bibinfo {author} {\bibfnamefont {A.}~\bibnamefont
  {Adare}} \emph {et~al.} (\bibinfo {collaboration} {PHENIX}),\ }\href
  {\doibase 10.1103/PhysRevLett.107.252301} {\bibfield  {journal} {\bibinfo
  {journal} {Phys. Rev. Lett.}\ }\textbf {\bibinfo {volume} {107}},\ \bibinfo
  {pages} {252301} (\bibinfo {year} {2011})},\ \Eprint
  {http://arxiv.org/abs/1105.3928} {arXiv:1105.3928 [nucl-ex]} \BibitemShut
  {NoStop}%
\bibitem [{\citenamefont {Adams}\ \emph {et~al.}(2005)\citenamefont {Adams}
  \emph {et~al.}}]{STAR:2004jwm}%
  \BibitemOpen
  \bibfield  {author} {\bibinfo {author} {\bibfnamefont {J.}~\bibnamefont
  {Adams}} \emph {et~al.} (\bibinfo {collaboration} {STAR}),\ }\href {\doibase
  10.1103/PhysRevC.72.014904} {\bibfield  {journal} {\bibinfo  {journal} {Phys.
  Rev. C}\ }\textbf {\bibinfo {volume} {72}},\ \bibinfo {pages} {014904}
  (\bibinfo {year} {2005})},\ \Eprint {http://arxiv.org/abs/nucl-ex/0409033}
  {arXiv:nucl-ex/0409033} \BibitemShut {NoStop}%
\bibitem [{\citenamefont {Elfner}\ and\ \citenamefont
  {M\"uller}(2022)}]{Elfner:2022iae}%
  \BibitemOpen
  \bibfield  {author} {\bibinfo {author} {\bibfnamefont {H.}~\bibnamefont
  {Elfner}}\ and\ \bibinfo {author} {\bibfnamefont {B.}~\bibnamefont
  {M\"uller}},\ }\href@noop {} {\  (\bibinfo {year} {2022})},\ \Eprint
  {http://arxiv.org/abs/2210.12056} {arXiv:2210.12056 [nucl-th]} \BibitemShut
  {NoStop}%
\bibitem [{\citenamefont {Noronha-Hostler}\ \emph {et~al.}(2016)\citenamefont
  {Noronha-Hostler}, \citenamefont {Yan}, \citenamefont {Gardim},\ and\
  \citenamefont {Ollitrault}}]{Noronha-Hostler:2015dbi}%
  \BibitemOpen
  \bibfield  {author} {\bibinfo {author} {\bibfnamefont {J.}~\bibnamefont
  {Noronha-Hostler}}, \bibinfo {author} {\bibfnamefont {L.}~\bibnamefont
  {Yan}}, \bibinfo {author} {\bibfnamefont {F.~G.}\ \bibnamefont {Gardim}}, \
  and\ \bibinfo {author} {\bibfnamefont {J.-Y.}\ \bibnamefont {Ollitrault}},\
  }\href {\doibase 10.1103/PhysRevC.93.014909} {\bibfield  {journal} {\bibinfo
  {journal} {Phys. Rev. C}\ }\textbf {\bibinfo {volume} {93}},\ \bibinfo
  {pages} {014909} (\bibinfo {year} {2016})},\ \Eprint
  {http://arxiv.org/abs/1511.03896} {arXiv:1511.03896 [nucl-th]} \BibitemShut
  {NoStop}%
\bibitem [{\citenamefont {Sch\"afer}\ \emph {et~al.}(2022)\citenamefont
  {Sch\"afer}, \citenamefont {Karpenko}, \citenamefont {Wu}, \citenamefont
  {Hammelmann},\ and\ \citenamefont {Elfner}}]{Schafer:2021csj}%
  \BibitemOpen
  \bibfield  {author} {\bibinfo {author} {\bibfnamefont {A.}~\bibnamefont
  {Sch\"afer}}, \bibinfo {author} {\bibfnamefont {I.}~\bibnamefont {Karpenko}},
  \bibinfo {author} {\bibfnamefont {X.-Y.}\ \bibnamefont {Wu}}, \bibinfo
  {author} {\bibfnamefont {J.}~\bibnamefont {Hammelmann}}, \ and\ \bibinfo
  {author} {\bibfnamefont {H.}~\bibnamefont {Elfner}} (\bibinfo {collaboration}
  {SMASH}),\ }\href {\doibase 10.1140/epja/s10050-022-00872-x} {\bibfield
  {journal} {\bibinfo  {journal} {Eur. Phys. J. A}\ }\textbf {\bibinfo {volume}
  {58}},\ \bibinfo {pages} {230} (\bibinfo {year} {2022})},\ \Eprint
  {http://arxiv.org/abs/2112.08724} {arXiv:2112.08724 [hep-ph]} \BibitemShut
  {NoStop}%
\bibitem [{\citenamefont {Karpenko}\ \emph {et~al.}(2015)\citenamefont
  {Karpenko}, \citenamefont {Huovinen}, \citenamefont {Petersen},\ and\
  \citenamefont {Bleicher}}]{Karpenko_2015}%
  \BibitemOpen
  \bibfield  {author} {\bibinfo {author} {\bibfnamefont {I.~A.}\ \bibnamefont
  {Karpenko}}, \bibinfo {author} {\bibfnamefont {P.}~\bibnamefont {Huovinen}},
  \bibinfo {author} {\bibfnamefont {H.}~\bibnamefont {Petersen}}, \ and\
  \bibinfo {author} {\bibfnamefont {M.}~\bibnamefont {Bleicher}},\ }\href@noop
  {} {\bibfield  {journal} {\bibinfo  {journal} {Phys. Rev. C}\ }\textbf
  {\bibinfo {volume} {91}},\ \bibinfo {pages} {064901} (\bibinfo {year}
  {2015})}\BibitemShut {NoStop}%
\bibitem [{\citenamefont {Petersen}\ \emph
  {et~al.}(2008{\natexlab{a}})\citenamefont {Petersen}, \citenamefont
  {Steinheimer}, \citenamefont {Burau}, \citenamefont {Bleicher},\ and\
  \citenamefont {St\"ocker}}]{Petersen_2008}%
  \BibitemOpen
  \bibfield  {author} {\bibinfo {author} {\bibfnamefont {H.}~\bibnamefont
  {Petersen}}, \bibinfo {author} {\bibfnamefont {J.}~\bibnamefont
  {Steinheimer}}, \bibinfo {author} {\bibfnamefont {G.}~\bibnamefont {Burau}},
  \bibinfo {author} {\bibfnamefont {M.}~\bibnamefont {Bleicher}}, \ and\
  \bibinfo {author} {\bibfnamefont {H.}~\bibnamefont {St\"ocker}},\ }\href
  {\doibase 10.1103/physrevc.78.044901} {\bibfield  {journal} {\bibinfo
  {journal} {Phys. Rev. C}\ }\textbf {\bibinfo {volume} {78}},\ \bibinfo
  {pages} {044901} (\bibinfo {year} {2008}{\natexlab{a}})}\BibitemShut
  {NoStop}%
\bibitem [{\citenamefont {Wu}\ \emph {et~al.}(2022)\citenamefont {Wu},
  \citenamefont {Qin}, \citenamefont {Pang},\ and\ \citenamefont
  {Wang}}]{Wu_2022}%
  \BibitemOpen
  \bibfield  {author} {\bibinfo {author} {\bibfnamefont {X.-Y.}\ \bibnamefont
  {Wu}}, \bibinfo {author} {\bibfnamefont {G.-Y.}\ \bibnamefont {Qin}},
  \bibinfo {author} {\bibfnamefont {L.-G.}\ \bibnamefont {Pang}}, \ and\
  \bibinfo {author} {\bibfnamefont {X.-N.}\ \bibnamefont {Wang}},\ }\href@noop
  {} {\bibfield  {journal} {\bibinfo  {journal} {Phys. Rev. C}\ }\textbf
  {\bibinfo {volume} {105}},\ \bibinfo {pages} {034909} (\bibinfo {year}
  {2022})}\BibitemShut {NoStop}%
\bibitem [{\citenamefont {Shen}\ \emph {et~al.}(2017)\citenamefont {Shen},
  \citenamefont {Denicol}, \citenamefont {Gale}, \citenamefont {Jeon},
  \citenamefont {Monnai},\ and\ \citenamefont {Schenke}}]{Shen_2017}%
  \BibitemOpen
  \bibfield  {author} {\bibinfo {author} {\bibfnamefont {C.}~\bibnamefont
  {Shen}}, \bibinfo {author} {\bibfnamefont {G.}~\bibnamefont {Denicol}},
  \bibinfo {author} {\bibfnamefont {C.}~\bibnamefont {Gale}}, \bibinfo {author}
  {\bibfnamefont {S.}~\bibnamefont {Jeon}}, \bibinfo {author} {\bibfnamefont
  {A.}~\bibnamefont {Monnai}}, \ and\ \bibinfo {author} {\bibfnamefont
  {B.}~\bibnamefont {Schenke}},\ }\href {\doibase
  10.1016/j.nuclphysa.2017.06.008} {\bibfield  {journal} {\bibinfo  {journal}
  {Nucl. Phys. A}\ }\textbf {\bibinfo {volume} {967}},\ \bibinfo {pages} {796}
  (\bibinfo {year} {2017})}\BibitemShut {NoStop}%
\bibitem [{\citenamefont {Akamatsu}\ \emph {et~al.}(2018)\citenamefont
  {Akamatsu}, \citenamefont {Asakawa}, \citenamefont {Hirano}, \citenamefont
  {Kitazawa}, \citenamefont {Morita}, \citenamefont {Murase}, \citenamefont
  {Nara}, \citenamefont {Nonaka},\ and\ \citenamefont
  {Ohnishi}}]{akamatsu2018dynamically}%
  \BibitemOpen
  \bibfield  {author} {\bibinfo {author} {\bibfnamefont {Y.}~\bibnamefont
  {Akamatsu}}, \bibinfo {author} {\bibfnamefont {M.}~\bibnamefont {Asakawa}},
  \bibinfo {author} {\bibfnamefont {T.}~\bibnamefont {Hirano}}, \bibinfo
  {author} {\bibfnamefont {M.}~\bibnamefont {Kitazawa}}, \bibinfo {author}
  {\bibfnamefont {K.}~\bibnamefont {Morita}}, \bibinfo {author} {\bibfnamefont
  {K.}~\bibnamefont {Murase}}, \bibinfo {author} {\bibfnamefont
  {Y.}~\bibnamefont {Nara}}, \bibinfo {author} {\bibfnamefont {C.}~\bibnamefont
  {Nonaka}}, \ and\ \bibinfo {author} {\bibfnamefont {A.}~\bibnamefont
  {Ohnishi}},\ }\href@noop {} {\bibfield  {journal} {\bibinfo  {journal} {Phys.
  Rev. C}\ }\textbf {\bibinfo {volume} {98}},\ \bibinfo {pages} {024909}
  (\bibinfo {year} {2018})}\BibitemShut {NoStop}%
\bibitem [{\citenamefont {Du}\ and\ \citenamefont {Heinz}(2020)}]{du20203+}%
  \BibitemOpen
  \bibfield  {author} {\bibinfo {author} {\bibfnamefont {L.}~\bibnamefont
  {Du}}\ and\ \bibinfo {author} {\bibfnamefont {U.}~\bibnamefont {Heinz}},\
  }\href@noop {} {\bibfield  {journal} {\bibinfo  {journal} {Comput. Phys.
  Commun.}\ }\textbf {\bibinfo {volume} {251}},\ \bibinfo {pages} {107090}
  (\bibinfo {year} {2020})}\BibitemShut {NoStop}%
\bibitem [{\citenamefont {Nandi}\ \emph {et~al.}(2020)\citenamefont {Nandi},
  \citenamefont {Kumar},\ and\ \citenamefont {Sharma}}]{nandi2020constraining}%
  \BibitemOpen
  \bibfield  {author} {\bibinfo {author} {\bibfnamefont {A.}~\bibnamefont
  {Nandi}}, \bibinfo {author} {\bibfnamefont {L.}~\bibnamefont {Kumar}}, \ and\
  \bibinfo {author} {\bibfnamefont {N.}~\bibnamefont {Sharma}},\ }\href@noop {}
  {\bibfield  {journal} {\bibinfo  {journal} {Phys. Rev. C}\ }\textbf {\bibinfo
  {volume} {102}},\ \bibinfo {pages} {024902} (\bibinfo {year}
  {2020})}\BibitemShut {NoStop}%
\bibitem [{\citenamefont {Petersen}(2014)}]{Petersen_2014}%
  \BibitemOpen
  \bibfield  {author} {\bibinfo {author} {\bibfnamefont {H.}~\bibnamefont
  {Petersen}},\ }\href {\doibase 10.1088/0954-3899/41/12/124005} {\bibfield
  {journal} {\bibinfo  {journal} {J. Phys. G Nucl. Part. Phys.}\ }\textbf
  {\bibinfo {volume} {41}},\ \bibinfo {pages} {124005} (\bibinfo {year}
  {2014})}\BibitemShut {NoStop}%
\bibitem [{\citenamefont {Everett}\ \emph
  {et~al.}(2021{\natexlab{a}})\citenamefont {Everett}, \citenamefont {Ke},
  \citenamefont {Paquet}, \citenamefont {Vujanovic}, \citenamefont {Bass},
  \citenamefont {Du}, \citenamefont {Gale}, \citenamefont {Heffernan},
  \citenamefont {Heinz}, \citenamefont {Liyanage}, \citenamefont {Luzum},
  \citenamefont {Majumder}, \citenamefont {McNelis}, \citenamefont {Shen},
  \citenamefont {Xu}, \citenamefont {Angerami}, \citenamefont {Cao},
  \citenamefont {Chen}, \citenamefont {Coleman}, \citenamefont {Cunqueiro},
  \citenamefont {Dai}, \citenamefont {Ehlers}, \citenamefont {Elfner},
  \citenamefont {Fan}, \citenamefont {Fries}, \citenamefont {Garza},
  \citenamefont {He}, \citenamefont {Jacak}, \citenamefont {Jacobs},
  \citenamefont {Jeon}, \citenamefont {Kim}, \citenamefont {Kordell},
  \citenamefont {Kumar}, \citenamefont {Mak}, \citenamefont {Mulligan},
  \citenamefont {Nattrass}, \citenamefont {Oliinychenko}, \citenamefont {Park},
  \citenamefont {Putschke}, \citenamefont {Roland}, \citenamefont {Schenke},
  \citenamefont {Schwiebert}, \citenamefont {Silva}, \citenamefont {Sirimanna},
  \citenamefont {Soltz}, \citenamefont {Tachibana}, \citenamefont {Wang},\ and\
  \citenamefont {Wolpert}}]{JETSCAPE:2020mzn}%
  \BibitemOpen
  \bibfield  {author} {\bibinfo {author} {\bibfnamefont {D.}~\bibnamefont
  {Everett}}, \bibinfo {author} {\bibfnamefont {W.}~\bibnamefont {Ke}},
  \bibinfo {author} {\bibfnamefont {J.-F.}\ \bibnamefont {Paquet}}, \bibinfo
  {author} {\bibfnamefont {G.}~\bibnamefont {Vujanovic}}, \bibinfo {author}
  {\bibfnamefont {S.~A.}\ \bibnamefont {Bass}}, \bibinfo {author}
  {\bibfnamefont {L.}~\bibnamefont {Du}}, \bibinfo {author} {\bibfnamefont
  {C.}~\bibnamefont {Gale}}, \bibinfo {author} {\bibfnamefont {M.}~\bibnamefont
  {Heffernan}}, \bibinfo {author} {\bibfnamefont {U.}~\bibnamefont {Heinz}},
  \bibinfo {author} {\bibfnamefont {D.}~\bibnamefont {Liyanage}}, \bibinfo
  {author} {\bibfnamefont {M.}~\bibnamefont {Luzum}}, \bibinfo {author}
  {\bibfnamefont {A.}~\bibnamefont {Majumder}}, \bibinfo {author}
  {\bibfnamefont {M.}~\bibnamefont {McNelis}}, \bibinfo {author} {\bibfnamefont
  {C.}~\bibnamefont {Shen}}, \bibinfo {author} {\bibfnamefont {Y.}~\bibnamefont
  {Xu}}, \bibinfo {author} {\bibfnamefont {A.}~\bibnamefont {Angerami}},
  \bibinfo {author} {\bibfnamefont {S.}~\bibnamefont {Cao}}, \bibinfo {author}
  {\bibfnamefont {Y.}~\bibnamefont {Chen}}, \bibinfo {author} {\bibfnamefont
  {J.}~\bibnamefont {Coleman}}, \bibinfo {author} {\bibfnamefont
  {L.}~\bibnamefont {Cunqueiro}}, \bibinfo {author} {\bibfnamefont
  {T.}~\bibnamefont {Dai}}, \bibinfo {author} {\bibfnamefont {R.}~\bibnamefont
  {Ehlers}}, \bibinfo {author} {\bibfnamefont {H.}~\bibnamefont {Elfner}},
  \bibinfo {author} {\bibfnamefont {W.}~\bibnamefont {Fan}}, \bibinfo {author}
  {\bibfnamefont {R.~J.}\ \bibnamefont {Fries}}, \bibinfo {author}
  {\bibfnamefont {F.}~\bibnamefont {Garza}}, \bibinfo {author} {\bibfnamefont
  {Y.}~\bibnamefont {He}}, \bibinfo {author} {\bibfnamefont {B.~V.}\
  \bibnamefont {Jacak}}, \bibinfo {author} {\bibfnamefont {P.~M.}\ \bibnamefont
  {Jacobs}}, \bibinfo {author} {\bibfnamefont {S.}~\bibnamefont {Jeon}},
  \bibinfo {author} {\bibfnamefont {B.}~\bibnamefont {Kim}}, \bibinfo {author}
  {\bibfnamefont {M.}~\bibnamefont {Kordell}}, \bibinfo {author} {\bibfnamefont
  {A.}~\bibnamefont {Kumar}}, \bibinfo {author} {\bibfnamefont
  {S.}~\bibnamefont {Mak}}, \bibinfo {author} {\bibfnamefont {J.}~\bibnamefont
  {Mulligan}}, \bibinfo {author} {\bibfnamefont {C.}~\bibnamefont {Nattrass}},
  \bibinfo {author} {\bibfnamefont {D.}~\bibnamefont {Oliinychenko}}, \bibinfo
  {author} {\bibfnamefont {C.}~\bibnamefont {Park}}, \bibinfo {author}
  {\bibfnamefont {J.~H.}\ \bibnamefont {Putschke}}, \bibinfo {author}
  {\bibfnamefont {G.}~\bibnamefont {Roland}}, \bibinfo {author} {\bibfnamefont
  {B.}~\bibnamefont {Schenke}}, \bibinfo {author} {\bibfnamefont
  {L.}~\bibnamefont {Schwiebert}}, \bibinfo {author} {\bibfnamefont
  {A.}~\bibnamefont {Silva}}, \bibinfo {author} {\bibfnamefont
  {C.}~\bibnamefont {Sirimanna}}, \bibinfo {author} {\bibfnamefont {R.~A.}\
  \bibnamefont {Soltz}}, \bibinfo {author} {\bibfnamefont {Y.}~\bibnamefont
  {Tachibana}}, \bibinfo {author} {\bibfnamefont {X.-N.}\ \bibnamefont {Wang}},
  \ and\ \bibinfo {author} {\bibfnamefont {R.~L.}\ \bibnamefont {Wolpert}}
  (\bibinfo {collaboration} {{JETSCAPE Collaboration}}),\ }\href {\doibase
  10.1103/PhysRevC.103.054904} {\bibfield  {journal} {\bibinfo  {journal}
  {Phys. Rev. C}\ }\textbf {\bibinfo {volume} {103}},\ \bibinfo {pages}
  {054904} (\bibinfo {year} {2021}{\natexlab{a}})},\ \Eprint
  {http://arxiv.org/abs/2011.01430} {arXiv:2011.01430 [hep-ph]} \BibitemShut
  {NoStop}%
\bibitem [{\citenamefont {Ghiglieri}\ \emph {et~al.}(2018)\citenamefont
  {Ghiglieri}, \citenamefont {Moore},\ and\ \citenamefont
  {Teaney}}]{Ghiglieri:2018dib}%
  \BibitemOpen
  \bibfield  {author} {\bibinfo {author} {\bibfnamefont {J.}~\bibnamefont
  {Ghiglieri}}, \bibinfo {author} {\bibfnamefont {G.~D.}\ \bibnamefont
  {Moore}}, \ and\ \bibinfo {author} {\bibfnamefont {D.}~\bibnamefont
  {Teaney}},\ }\href {\doibase 10.1007/JHEP03(2018)179} {\bibfield  {journal}
  {\bibinfo  {journal} {JHEP}\ }\textbf {\bibinfo {volume} {03}},\ \bibinfo
  {pages} {179} (\bibinfo {year} {2018})},\ \Eprint
  {http://arxiv.org/abs/1802.09535} {arXiv:1802.09535 [hep-ph]} \BibitemShut
  {NoStop}%
\bibitem [{\citenamefont {Auvinen}\ \emph {et~al.}(2020)\citenamefont
  {Auvinen}, \citenamefont {Eskola}, \citenamefont {Huovinen}, \citenamefont
  {Niemi}, \citenamefont {Paatelainen},\ and\ \citenamefont
  {Petreczky}}]{Auvinen:2020mpc}%
  \BibitemOpen
  \bibfield  {author} {\bibinfo {author} {\bibfnamefont {J.}~\bibnamefont
  {Auvinen}}, \bibinfo {author} {\bibfnamefont {K.~J.}\ \bibnamefont {Eskola}},
  \bibinfo {author} {\bibfnamefont {P.}~\bibnamefont {Huovinen}}, \bibinfo
  {author} {\bibfnamefont {H.}~\bibnamefont {Niemi}}, \bibinfo {author}
  {\bibfnamefont {R.}~\bibnamefont {Paatelainen}}, \ and\ \bibinfo {author}
  {\bibfnamefont {P.}~\bibnamefont {Petreczky}},\ }\href {\doibase
  10.1103/PhysRevC.102.044911} {\bibfield  {journal} {\bibinfo  {journal}
  {Phys. Rev. C}\ }\textbf {\bibinfo {volume} {102}},\ \bibinfo {pages}
  {044911} (\bibinfo {year} {2020})},\ \Eprint
  {http://arxiv.org/abs/2006.12499} {arXiv:2006.12499 [nucl-th]} \BibitemShut
  {NoStop}%
\bibitem [{\citenamefont {Nijs}\ \emph {et~al.}(2021)\citenamefont {Nijs},
  \citenamefont {van~der Schee}, \citenamefont {G\"ursoy},\ and\ \citenamefont
  {Snellings}}]{Nijs:2020ors}%
  \BibitemOpen
  \bibfield  {author} {\bibinfo {author} {\bibfnamefont {G.}~\bibnamefont
  {Nijs}}, \bibinfo {author} {\bibfnamefont {W.}~\bibnamefont {van~der Schee}},
  \bibinfo {author} {\bibfnamefont {U.}~\bibnamefont {G\"ursoy}}, \ and\
  \bibinfo {author} {\bibfnamefont {R.}~\bibnamefont {Snellings}},\ }\href
  {\doibase 10.1103/PhysRevLett.126.202301} {\bibfield  {journal} {\bibinfo
  {journal} {Phys. Rev. Lett.}\ }\textbf {\bibinfo {volume} {126}},\ \bibinfo
  {pages} {202301} (\bibinfo {year} {2021})},\ \Eprint
  {http://arxiv.org/abs/2010.15130} {arXiv:2010.15130 [nucl-th]} \BibitemShut
  {NoStop}%
\bibitem [{\citenamefont {{G. Nijs, W. van der Schee, U. G\"ursoy, and R.
  Snellings}}(2021)}]{Nijs:2020roc}%
  \BibitemOpen
  \bibfield  {author} {\bibinfo {author} {\bibnamefont {{G. Nijs, W. van der
  Schee, U. G\"ursoy, and R. Snellings}}},\ }\href {\doibase
  10.1103/PhysRevC.103.054909} {\bibfield  {journal} {\bibinfo  {journal}
  {Phys. Rev. C}\ }\textbf {\bibinfo {volume} {103}},\ \bibinfo {pages}
  {054909} (\bibinfo {year} {2021})},\ \Eprint
  {http://arxiv.org/abs/2010.15134} {arXiv:2010.15134 [nucl-th]} \BibitemShut
  {NoStop}%
\bibitem [{\citenamefont {Greiner}\ \emph {et~al.}(2011)\citenamefont
  {Greiner}, \citenamefont {Noronha-Hostler},\ and\ \citenamefont
  {Noronha}}]{greiner2011hagedorn}%
  \BibitemOpen
  \bibfield  {author} {\bibinfo {author} {\bibfnamefont {C.}~\bibnamefont
  {Greiner}}, \bibinfo {author} {\bibfnamefont {J.}~\bibnamefont
  {Noronha-Hostler}}, \ and\ \bibinfo {author} {\bibfnamefont {J.}~\bibnamefont
  {Noronha}},\ }\href@noop {} {\enquote {\bibinfo {title} {Hagedorn states and
  thermalization in heavy ion collisions},}\ } (\bibinfo {year} {2011}),\
  \Eprint {http://arxiv.org/abs/1105.1756} {arXiv:1105.1756 [nucl-th]}
  \BibitemShut {NoStop}%
\bibitem [{\citenamefont {Gorenstein}\ \emph {et~al.}(2008)\citenamefont
  {Gorenstein}, \citenamefont {Hauer},\ and\ \citenamefont
  {Moroz}}]{Gorenstein_2008}%
  \BibitemOpen
  \bibfield  {author} {\bibinfo {author} {\bibfnamefont {M.~I.}\ \bibnamefont
  {Gorenstein}}, \bibinfo {author} {\bibfnamefont {M.}~\bibnamefont {Hauer}}, \
  and\ \bibinfo {author} {\bibfnamefont {O.~N.}\ \bibnamefont {Moroz}},\ }\href
  {\doibase 10.1103/physrevc.77.024911} {\bibfield  {journal} {\bibinfo
  {journal} {Phys. Rev. C}\ }\textbf {\bibinfo {volume} {77}},\ \bibinfo
  {pages} {024911} (\bibinfo {year} {2008})}\BibitemShut {NoStop}%
\bibitem [{\citenamefont {Csernai}\ \emph {et~al.}(2006)\citenamefont
  {Csernai}, \citenamefont {Kapusta},\ and\ \citenamefont
  {McLerran}}]{csernai2006strongly}%
  \BibitemOpen
  \bibfield  {author} {\bibinfo {author} {\bibfnamefont {L.~P.}\ \bibnamefont
  {Csernai}}, \bibinfo {author} {\bibfnamefont {J.~I.}\ \bibnamefont
  {Kapusta}}, \ and\ \bibinfo {author} {\bibfnamefont {L.~D.}\ \bibnamefont
  {McLerran}},\ }\href@noop {} {\bibfield  {journal} {\bibinfo  {journal}
  {Phys. Rev. Lett.}\ }\textbf {\bibinfo {volume} {97}},\ \bibinfo {pages}
  {152303} (\bibinfo {year} {2006})}\BibitemShut {NoStop}%
\bibitem [{\citenamefont {Liyanage}\ \emph {et~al.}(2023)\citenamefont
  {Liyanage}, \citenamefont {S\"urer}, \citenamefont {Plumlee}, \citenamefont
  {Wild},\ and\ \citenamefont {Heinz}}]{Liyanage:2023nds}%
  \BibitemOpen
  \bibfield  {author} {\bibinfo {author} {\bibfnamefont {D.}~\bibnamefont
  {Liyanage}}, \bibinfo {author} {\bibfnamefont {O.}~\bibnamefont {S\"urer}},
  \bibinfo {author} {\bibfnamefont {M.}~\bibnamefont {Plumlee}}, \bibinfo
  {author} {\bibfnamefont {S.~M.}\ \bibnamefont {Wild}}, \ and\ \bibinfo
  {author} {\bibfnamefont {U.}~\bibnamefont {Heinz}},\ }\href@noop {} {\
  (\bibinfo {year} {2023})},\ \Eprint {http://arxiv.org/abs/2302.14184}
  {arXiv:2302.14184 [nucl-th]} \BibitemShut {NoStop}%
\bibitem [{\citenamefont {Everett}\ \emph
  {et~al.}(2021{\natexlab{b}})\citenamefont {Everett}, \citenamefont {Ke},
  \citenamefont {Paquet}, \citenamefont {Vujanovic}, \citenamefont {Bass},
  \citenamefont {Du}, \citenamefont {Gale}, \citenamefont {Heffernan},
  \citenamefont {Heinz}, \citenamefont {Liyanage}, \citenamefont {Luzum},
  \citenamefont {Majumder}, \citenamefont {McNelis}, \citenamefont {Shen},
  \citenamefont {Xu}, \citenamefont {Angerami}, \citenamefont {Cao},
  \citenamefont {Chen}, \citenamefont {Coleman}, \citenamefont {Cunqueiro},
  \citenamefont {Dai}, \citenamefont {Ehlers}, \citenamefont {Elfner},
  \citenamefont {Fan}, \citenamefont {Fries}, \citenamefont {Garza},
  \citenamefont {He}, \citenamefont {Jacak}, \citenamefont {Jacobs},
  \citenamefont {Jeon}, \citenamefont {Kim}, \citenamefont {Kordell},
  \citenamefont {Kumar}, \citenamefont {Mak}, \citenamefont {Mulligan},
  \citenamefont {Nattrass}, \citenamefont {Oliinychenko}, \citenamefont {Park},
  \citenamefont {Putschke}, \citenamefont {Roland}, \citenamefont {Schenke},
  \citenamefont {Schwiebert}, \citenamefont {Silva}, \citenamefont {Sirimanna},
  \citenamefont {Soltz}, \citenamefont {Tachibana}, \citenamefont {Wang},\ and\
  \citenamefont {Wolpert}}]{JETSCAPE:2020shq}%
  \BibitemOpen
  \bibfield  {author} {\bibinfo {author} {\bibfnamefont {D.}~\bibnamefont
  {Everett}}, \bibinfo {author} {\bibfnamefont {W.}~\bibnamefont {Ke}},
  \bibinfo {author} {\bibfnamefont {J.-F.}\ \bibnamefont {Paquet}}, \bibinfo
  {author} {\bibfnamefont {G.}~\bibnamefont {Vujanovic}}, \bibinfo {author}
  {\bibfnamefont {S.~A.}\ \bibnamefont {Bass}}, \bibinfo {author}
  {\bibfnamefont {L.}~\bibnamefont {Du}}, \bibinfo {author} {\bibfnamefont
  {C.}~\bibnamefont {Gale}}, \bibinfo {author} {\bibfnamefont {M.}~\bibnamefont
  {Heffernan}}, \bibinfo {author} {\bibfnamefont {U.}~\bibnamefont {Heinz}},
  \bibinfo {author} {\bibfnamefont {D.}~\bibnamefont {Liyanage}}, \bibinfo
  {author} {\bibfnamefont {M.}~\bibnamefont {Luzum}}, \bibinfo {author}
  {\bibfnamefont {A.}~\bibnamefont {Majumder}}, \bibinfo {author}
  {\bibfnamefont {M.}~\bibnamefont {McNelis}}, \bibinfo {author} {\bibfnamefont
  {C.}~\bibnamefont {Shen}}, \bibinfo {author} {\bibfnamefont {Y.}~\bibnamefont
  {Xu}}, \bibinfo {author} {\bibfnamefont {A.}~\bibnamefont {Angerami}},
  \bibinfo {author} {\bibfnamefont {S.}~\bibnamefont {Cao}}, \bibinfo {author}
  {\bibfnamefont {Y.}~\bibnamefont {Chen}}, \bibinfo {author} {\bibfnamefont
  {J.}~\bibnamefont {Coleman}}, \bibinfo {author} {\bibfnamefont
  {L.}~\bibnamefont {Cunqueiro}}, \bibinfo {author} {\bibfnamefont
  {T.}~\bibnamefont {Dai}}, \bibinfo {author} {\bibfnamefont {R.}~\bibnamefont
  {Ehlers}}, \bibinfo {author} {\bibfnamefont {H.}~\bibnamefont {Elfner}},
  \bibinfo {author} {\bibfnamefont {W.}~\bibnamefont {Fan}}, \bibinfo {author}
  {\bibfnamefont {R.~J.}\ \bibnamefont {Fries}}, \bibinfo {author}
  {\bibfnamefont {F.}~\bibnamefont {Garza}}, \bibinfo {author} {\bibfnamefont
  {Y.}~\bibnamefont {He}}, \bibinfo {author} {\bibfnamefont {B.~V.}\
  \bibnamefont {Jacak}}, \bibinfo {author} {\bibfnamefont {P.~M.}\ \bibnamefont
  {Jacobs}}, \bibinfo {author} {\bibfnamefont {S.}~\bibnamefont {Jeon}},
  \bibinfo {author} {\bibfnamefont {B.}~\bibnamefont {Kim}}, \bibinfo {author}
  {\bibfnamefont {M.}~\bibnamefont {Kordell}}, \bibinfo {author} {\bibfnamefont
  {A.}~\bibnamefont {Kumar}}, \bibinfo {author} {\bibfnamefont
  {S.}~\bibnamefont {Mak}}, \bibinfo {author} {\bibfnamefont {J.}~\bibnamefont
  {Mulligan}}, \bibinfo {author} {\bibfnamefont {C.}~\bibnamefont {Nattrass}},
  \bibinfo {author} {\bibfnamefont {D.}~\bibnamefont {Oliinychenko}}, \bibinfo
  {author} {\bibfnamefont {C.}~\bibnamefont {Park}}, \bibinfo {author}
  {\bibfnamefont {J.~H.}\ \bibnamefont {Putschke}}, \bibinfo {author}
  {\bibfnamefont {G.}~\bibnamefont {Roland}}, \bibinfo {author} {\bibfnamefont
  {B.}~\bibnamefont {Schenke}}, \bibinfo {author} {\bibfnamefont
  {L.}~\bibnamefont {Schwiebert}}, \bibinfo {author} {\bibfnamefont
  {A.}~\bibnamefont {Silva}}, \bibinfo {author} {\bibfnamefont
  {C.}~\bibnamefont {Sirimanna}}, \bibinfo {author} {\bibfnamefont {R.~A.}\
  \bibnamefont {Soltz}}, \bibinfo {author} {\bibfnamefont {Y.}~\bibnamefont
  {Tachibana}}, \bibinfo {author} {\bibfnamefont {X.-N.}\ \bibnamefont {Wang}},
  \ and\ \bibinfo {author} {\bibfnamefont {R.~L.}\ \bibnamefont {Wolpert}}
  (\bibinfo {collaboration} {JETSCAPE Collaboration}),\ }\href {\doibase
  10.1103/PhysRevLett.126.242301} {\bibfield  {journal} {\bibinfo  {journal}
  {Phys. Rev. Lett.}\ }\textbf {\bibinfo {volume} {126}},\ \bibinfo {pages}
  {242301} (\bibinfo {year} {2021}{\natexlab{b}})},\ \Eprint
  {http://arxiv.org/abs/2010.03928} {arXiv:2010.03928 [hep-ph]} \BibitemShut
  {NoStop}%
\bibitem [{\citenamefont {Kolb}\ \emph {et~al.}(2001)\citenamefont {Kolb},
  \citenamefont {Heinz}, \citenamefont {Huovinen}, \citenamefont {Eskola},\
  and\ \citenamefont {Tuominen}}]{Kolb:2001qz}%
  \BibitemOpen
  \bibfield  {author} {\bibinfo {author} {\bibfnamefont {P.~F.}\ \bibnamefont
  {Kolb}}, \bibinfo {author} {\bibfnamefont {U.~W.}\ \bibnamefont {Heinz}},
  \bibinfo {author} {\bibfnamefont {P.}~\bibnamefont {Huovinen}}, \bibinfo
  {author} {\bibfnamefont {K.~J.}\ \bibnamefont {Eskola}}, \ and\ \bibinfo
  {author} {\bibfnamefont {K.}~\bibnamefont {Tuominen}},\ }\href {\doibase
  10.1016/S0375-9474(01)01114-9} {\bibfield  {journal} {\bibinfo  {journal}
  {Nucl. Phys. A}\ }\textbf {\bibinfo {volume} {696}},\ \bibinfo {pages} {197}
  (\bibinfo {year} {2001})},\ \Eprint {http://arxiv.org/abs/hep-ph/0103234}
  {arXiv:hep-ph/0103234} \BibitemShut {NoStop}%
\bibitem [{\citenamefont {Bartels}\ \emph {et~al.}(2002)\citenamefont
  {Bartels}, \citenamefont {Golec-Biernat},\ and\ \citenamefont
  {Kowalski}}]{Bartels:2002cj}%
  \BibitemOpen
  \bibfield  {author} {\bibinfo {author} {\bibfnamefont {J.}~\bibnamefont
  {Bartels}}, \bibinfo {author} {\bibfnamefont {K.~J.}\ \bibnamefont
  {Golec-Biernat}}, \ and\ \bibinfo {author} {\bibfnamefont {H.}~\bibnamefont
  {Kowalski}},\ }\href {\doibase 10.1103/PhysRevD.66.014001} {\bibfield
  {journal} {\bibinfo  {journal} {Phys. Rev. D}\ }\textbf {\bibinfo {volume}
  {66}},\ \bibinfo {pages} {014001} (\bibinfo {year} {2002})},\ \Eprint
  {http://arxiv.org/abs/hep-ph/0203258} {arXiv:hep-ph/0203258} \BibitemShut
  {NoStop}%
\bibitem [{\citenamefont {Kowalski}\ and\ \citenamefont
  {Teaney}(2003)}]{Kowalski:2003hm}%
  \BibitemOpen
  \bibfield  {author} {\bibinfo {author} {\bibfnamefont {H.}~\bibnamefont
  {Kowalski}}\ and\ \bibinfo {author} {\bibfnamefont {D.}~\bibnamefont
  {Teaney}},\ }\href {\doibase 10.1103/PhysRevD.68.114005} {\bibfield
  {journal} {\bibinfo  {journal} {Phys. Rev. D}\ }\textbf {\bibinfo {volume}
  {68}},\ \bibinfo {pages} {114005} (\bibinfo {year} {2003})},\ \Eprint
  {http://arxiv.org/abs/hep-ph/0304189} {arXiv:hep-ph/0304189} \BibitemShut
  {NoStop}%
\bibitem [{\citenamefont {Lin}\ \emph {et~al.}(2005)\citenamefont {Lin},
  \citenamefont {Ko}, \citenamefont {Li}, \citenamefont {Zhang},\ and\
  \citenamefont {Pal}}]{Lin:2004en}%
  \BibitemOpen
  \bibfield  {author} {\bibinfo {author} {\bibfnamefont {Z.-W.}\ \bibnamefont
  {Lin}}, \bibinfo {author} {\bibfnamefont {C.~M.}\ \bibnamefont {Ko}},
  \bibinfo {author} {\bibfnamefont {B.-A.}\ \bibnamefont {Li}}, \bibinfo
  {author} {\bibfnamefont {B.}~\bibnamefont {Zhang}}, \ and\ \bibinfo {author}
  {\bibfnamefont {S.}~\bibnamefont {Pal}},\ }\href {\doibase
  10.1103/PhysRevC.72.064901} {\bibfield  {journal} {\bibinfo  {journal} {Phys.
  Rev. C}\ }\textbf {\bibinfo {volume} {72}},\ \bibinfo {pages} {064901}
  (\bibinfo {year} {2005})},\ \Eprint {http://arxiv.org/abs/nucl-th/0411110}
  {arXiv:nucl-th/0411110} \BibitemShut {NoStop}%
\bibitem [{\citenamefont {Hirano}\ \emph {et~al.}(2006)\citenamefont {Hirano},
  \citenamefont {Heinz}, \citenamefont {Kharzeev}, \citenamefont {Lacey},\ and\
  \citenamefont {Nara}}]{Hirano:2005xf}%
  \BibitemOpen
  \bibfield  {author} {\bibinfo {author} {\bibfnamefont {T.}~\bibnamefont
  {Hirano}}, \bibinfo {author} {\bibfnamefont {U.~W.}\ \bibnamefont {Heinz}},
  \bibinfo {author} {\bibfnamefont {D.}~\bibnamefont {Kharzeev}}, \bibinfo
  {author} {\bibfnamefont {R.}~\bibnamefont {Lacey}}, \ and\ \bibinfo {author}
  {\bibfnamefont {Y.}~\bibnamefont {Nara}},\ }\href {\doibase
  10.1016/j.physletb.2006.03.060} {\bibfield  {journal} {\bibinfo  {journal}
  {Phys. Lett. B}\ }\textbf {\bibinfo {volume} {636}},\ \bibinfo {pages} {299}
  (\bibinfo {year} {2006})},\ \Eprint {http://arxiv.org/abs/nucl-th/0511046}
  {arXiv:nucl-th/0511046} \BibitemShut {NoStop}%
\bibitem [{\citenamefont {Drescher}\ \emph {et~al.}(2006)\citenamefont
  {Drescher}, \citenamefont {Dumitru}, \citenamefont {Hayashigaki},\ and\
  \citenamefont {Nara}}]{Drescher:2006pi}%
  \BibitemOpen
  \bibfield  {author} {\bibinfo {author} {\bibfnamefont {H.-J.}\ \bibnamefont
  {Drescher}}, \bibinfo {author} {\bibfnamefont {A.}~\bibnamefont {Dumitru}},
  \bibinfo {author} {\bibfnamefont {A.}~\bibnamefont {Hayashigaki}}, \ and\
  \bibinfo {author} {\bibfnamefont {Y.}~\bibnamefont {Nara}},\ }\href {\doibase
  10.1103/PhysRevC.74.044905} {\bibfield  {journal} {\bibinfo  {journal} {Phys.
  Rev. C}\ }\textbf {\bibinfo {volume} {74}},\ \bibinfo {pages} {044905}
  (\bibinfo {year} {2006})},\ \Eprint {http://arxiv.org/abs/nucl-th/0605012}
  {arXiv:nucl-th/0605012} \BibitemShut {NoStop}%
\bibitem [{\citenamefont {Drescher}\ and\ \citenamefont
  {Nara}(2007)}]{Drescher_2007}%
  \BibitemOpen
  \bibfield  {author} {\bibinfo {author} {\bibfnamefont {H.-J.}\ \bibnamefont
  {Drescher}}\ and\ \bibinfo {author} {\bibfnamefont {Y.}~\bibnamefont
  {Nara}},\ }\href {\doibase 10.1103/physrevc.75.034905} {\bibfield  {journal}
  {\bibinfo  {journal} {Physical Review C}\ }\textbf {\bibinfo {volume} {75}}
  (\bibinfo {year} {2007}),\ 10.1103/physrevc.75.034905}\BibitemShut {NoStop}%
\bibitem [{\citenamefont {Petersen}\ \emph
  {et~al.}(2008{\natexlab{b}})\citenamefont {Petersen}, \citenamefont
  {Steinheimer}, \citenamefont {Burau}, \citenamefont {Bleicher},\ and\
  \citenamefont {St\"ocker}}]{Petersen:2008dd}%
  \BibitemOpen
  \bibfield  {author} {\bibinfo {author} {\bibfnamefont {H.}~\bibnamefont
  {Petersen}}, \bibinfo {author} {\bibfnamefont {J.}~\bibnamefont
  {Steinheimer}}, \bibinfo {author} {\bibfnamefont {G.}~\bibnamefont {Burau}},
  \bibinfo {author} {\bibfnamefont {M.}~\bibnamefont {Bleicher}}, \ and\
  \bibinfo {author} {\bibfnamefont {H.}~\bibnamefont {St\"ocker}},\ }\href
  {\doibase 10.1103/PhysRevC.78.044901} {\bibfield  {journal} {\bibinfo
  {journal} {Phys. Rev. C}\ }\textbf {\bibinfo {volume} {78}},\ \bibinfo
  {pages} {044901} (\bibinfo {year} {2008}{\natexlab{b}})},\ \Eprint
  {http://arxiv.org/abs/0806.1695} {arXiv:0806.1695 [nucl-th]} \BibitemShut
  {NoStop}%
\bibitem [{\citenamefont {Werner}\ \emph {et~al.}(2010)\citenamefont {Werner},
  \citenamefont {Karpenko}, \citenamefont {Pierog}, \citenamefont {Bleicher},\
  and\ \citenamefont {Mikhailov}}]{Werner:2010aa}%
  \BibitemOpen
  \bibfield  {author} {\bibinfo {author} {\bibfnamefont {K.}~\bibnamefont
  {Werner}}, \bibinfo {author} {\bibfnamefont {I.}~\bibnamefont {Karpenko}},
  \bibinfo {author} {\bibfnamefont {T.}~\bibnamefont {Pierog}}, \bibinfo
  {author} {\bibfnamefont {M.}~\bibnamefont {Bleicher}}, \ and\ \bibinfo
  {author} {\bibfnamefont {K.}~\bibnamefont {Mikhailov}},\ }\href {\doibase
  10.1103/PhysRevC.82.044904} {\bibfield  {journal} {\bibinfo  {journal} {Phys.
  Rev. C}\ }\textbf {\bibinfo {volume} {82}},\ \bibinfo {pages} {044904}
  (\bibinfo {year} {2010})},\ \Eprint {http://arxiv.org/abs/1004.0805}
  {arXiv:1004.0805 [nucl-th]} \BibitemShut {NoStop}%
\bibitem [{\citenamefont {Accardi}\ \emph {et~al.}(2016)\citenamefont {Accardi}
  \emph {et~al.}}]{Accardi:2012qut}%
  \BibitemOpen
  \bibfield  {author} {\bibinfo {author} {\bibfnamefont {A.}~\bibnamefont
  {Accardi}} \emph {et~al.},\ }\href {\doibase 10.1140/epja/i2016-16268-9}
  {\bibfield  {journal} {\bibinfo  {journal} {Eur. Phys. J. A}\ }\textbf
  {\bibinfo {volume} {52}},\ \bibinfo {pages} {268} (\bibinfo {year} {2016})},\
  \Eprint {http://arxiv.org/abs/1212.1701} {arXiv:1212.1701 [nucl-ex]}
  \BibitemShut {NoStop}%
\bibitem [{\citenamefont {Schenke}\ \emph
  {et~al.}(2012{\natexlab{a}})\citenamefont {Schenke}, \citenamefont
  {Tribedy},\ and\ \citenamefont {Venugopalan}}]{Schenke:2012wb}%
  \BibitemOpen
  \bibfield  {author} {\bibinfo {author} {\bibfnamefont {B.}~\bibnamefont
  {Schenke}}, \bibinfo {author} {\bibfnamefont {P.}~\bibnamefont {Tribedy}}, \
  and\ \bibinfo {author} {\bibfnamefont {R.}~\bibnamefont {Venugopalan}},\
  }\href {\doibase 10.1103/PhysRevLett.108.252301} {\bibfield  {journal}
  {\bibinfo  {journal} {Phys. Rev. Lett.}\ }\textbf {\bibinfo {volume} {108}},\
  \bibinfo {pages} {252301} (\bibinfo {year} {2012}{\natexlab{a}})},\ \Eprint
  {http://arxiv.org/abs/1202.6646} {arXiv:1202.6646 [nucl-th]} \BibitemShut
  {NoStop}%
\bibitem [{\citenamefont {Rybczynski}\ \emph {et~al.}(2014)\citenamefont
  {Rybczynski}, \citenamefont {Stefanek}, \citenamefont {Broniowski},\ and\
  \citenamefont {Bozek}}]{Rybczynski:2013yba}%
  \BibitemOpen
  \bibfield  {author} {\bibinfo {author} {\bibfnamefont {M.}~\bibnamefont
  {Rybczynski}}, \bibinfo {author} {\bibfnamefont {G.}~\bibnamefont
  {Stefanek}}, \bibinfo {author} {\bibfnamefont {W.}~\bibnamefont
  {Broniowski}}, \ and\ \bibinfo {author} {\bibfnamefont {P.}~\bibnamefont
  {Bozek}},\ }\href {\doibase 10.1016/j.cpc.2014.02.016} {\bibfield  {journal}
  {\bibinfo  {journal} {Comput. Phys. Commun.}\ }\textbf {\bibinfo {volume}
  {185}},\ \bibinfo {pages} {1759} (\bibinfo {year} {2014})},\ \Eprint
  {http://arxiv.org/abs/1310.5475} {arXiv:1310.5475 [nucl-th]} \BibitemShut
  {NoStop}%
\bibitem [{\citenamefont {Moreland}\ \emph {et~al.}(2015)\citenamefont
  {Moreland}, \citenamefont {Bernhard},\ and\ \citenamefont
  {Bass}}]{Moreland:2014oya}%
  \BibitemOpen
  \bibfield  {author} {\bibinfo {author} {\bibfnamefont {J.~S.}\ \bibnamefont
  {Moreland}}, \bibinfo {author} {\bibfnamefont {J.~E.}\ \bibnamefont
  {Bernhard}}, \ and\ \bibinfo {author} {\bibfnamefont {S.~A.}\ \bibnamefont
  {Bass}},\ }\href {\doibase 10.1103/PhysRevC.92.011901} {\bibfield  {journal}
  {\bibinfo  {journal} {Phys. Rev. C}\ }\textbf {\bibinfo {volume} {92}},\
  \bibinfo {pages} {011901} (\bibinfo {year} {2015})},\ \Eprint
  {http://arxiv.org/abs/1412.4708} {arXiv:1412.4708 [nucl-th]} \BibitemShut
  {NoStop}%
\bibitem [{\citenamefont {van~der Schee}\ and\ \citenamefont
  {Schenke}(2015)}]{vanderSchee:2015rta}%
  \BibitemOpen
  \bibfield  {author} {\bibinfo {author} {\bibfnamefont {W.}~\bibnamefont
  {van~der Schee}}\ and\ \bibinfo {author} {\bibfnamefont {B.}~\bibnamefont
  {Schenke}},\ }\href {\doibase 10.1103/PhysRevC.92.064907} {\bibfield
  {journal} {\bibinfo  {journal} {Phys. Rev. C}\ }\textbf {\bibinfo {volume}
  {92}},\ \bibinfo {pages} {064907} (\bibinfo {year} {2015})},\ \Eprint
  {http://arxiv.org/abs/1507.08195} {arXiv:1507.08195 [nucl-th]} \BibitemShut
  {NoStop}%
\bibitem [{\citenamefont {Garcia-Montero}\ \emph {et~al.}(2023)\citenamefont
  {Garcia-Montero}, \citenamefont {Elfner},\ and\ \citenamefont
  {Schlichting}}]{Garcia-Montero:2023gex}%
  \BibitemOpen
  \bibfield  {author} {\bibinfo {author} {\bibfnamefont {O.}~\bibnamefont
  {Garcia-Montero}}, \bibinfo {author} {\bibfnamefont {H.}~\bibnamefont
  {Elfner}}, \ and\ \bibinfo {author} {\bibfnamefont {S.}~\bibnamefont
  {Schlichting}},\ }\href@noop {} {\  (\bibinfo {year} {2023})},\ \Eprint
  {http://arxiv.org/abs/2308.11713} {arXiv:2308.11713 [hep-ph]} \BibitemShut
  {NoStop}%
\bibitem [{\citenamefont {Holopainen}\ \emph {et~al.}(2011)\citenamefont
  {Holopainen}, \citenamefont {Niemi},\ and\ \citenamefont
  {Eskola}}]{Holopainen:2010gz}%
  \BibitemOpen
  \bibfield  {author} {\bibinfo {author} {\bibfnamefont {H.}~\bibnamefont
  {Holopainen}}, \bibinfo {author} {\bibfnamefont {H.}~\bibnamefont {Niemi}}, \
  and\ \bibinfo {author} {\bibfnamefont {K.~J.}\ \bibnamefont {Eskola}},\
  }\href {\doibase 10.1103/PhysRevC.83.034901} {\bibfield  {journal} {\bibinfo
  {journal} {Phys. Rev. C}\ }\textbf {\bibinfo {volume} {83}},\ \bibinfo
  {pages} {034901} (\bibinfo {year} {2011})},\ \Eprint
  {http://arxiv.org/abs/1007.0368} {arXiv:1007.0368 [hep-ph]} \BibitemShut
  {NoStop}%
\bibitem [{\citenamefont {Song}\ \emph {et~al.}(2011)\citenamefont {Song},
  \citenamefont {Bass}, \citenamefont {Heinz}, \citenamefont {Hirano},\ and\
  \citenamefont {Shen}}]{Song:2010mg}%
  \BibitemOpen
  \bibfield  {author} {\bibinfo {author} {\bibfnamefont {H.}~\bibnamefont
  {Song}}, \bibinfo {author} {\bibfnamefont {S.~A.}\ \bibnamefont {Bass}},
  \bibinfo {author} {\bibfnamefont {U.}~\bibnamefont {Heinz}}, \bibinfo
  {author} {\bibfnamefont {T.}~\bibnamefont {Hirano}}, \ and\ \bibinfo {author}
  {\bibfnamefont {C.}~\bibnamefont {Shen}},\ }\href {\doibase
  10.1103/PhysRevLett.106.192301} {\bibfield  {journal} {\bibinfo  {journal}
  {Phys. Rev. Lett.}\ }\textbf {\bibinfo {volume} {106}},\ \bibinfo {pages}
  {192301} (\bibinfo {year} {2011})},\ \bibinfo {note} {[Erratum:
  Phys.Rev.Lett. 109, 139904 (2012)]},\ \Eprint
  {http://arxiv.org/abs/1011.2783} {arXiv:1011.2783 [nucl-th]} \BibitemShut
  {NoStop}%
\bibitem [{hyb()}]{hybridurl}%
  \BibitemOpen
  \href {https://github.com/smash-transport/smash-vhlle-hybrid} {}\bibinfo
  {howpublished}
  {\url{https://github.com/smash-transport/smash-vhlle-hybrid}}\BibitemShut
  {NoStop}%
\bibitem [{\citenamefont {Weil}\ \emph {et~al.}(2016)\citenamefont {Weil},
  \citenamefont {Steinberg}, \citenamefont {Staudenmaier}, \citenamefont
  {Pang}, \citenamefont {Oliinychenko}, \citenamefont {Mohs}, \citenamefont
  {Kretz}, \citenamefont {Kehrenberg}, \citenamefont {Goldschmidt},
  \citenamefont {B\"auchle}, \citenamefont {Auvinen}, \citenamefont {Attems},\
  and\ \citenamefont {Petersen}}]{Weil_2016}%
  \BibitemOpen
  \bibfield  {author} {\bibinfo {author} {\bibfnamefont {J.}~\bibnamefont
  {Weil}}, \bibinfo {author} {\bibfnamefont {V.}~\bibnamefont {Steinberg}},
  \bibinfo {author} {\bibfnamefont {J.}~\bibnamefont {Staudenmaier}}, \bibinfo
  {author} {\bibfnamefont {L.~G.}\ \bibnamefont {Pang}}, \bibinfo {author}
  {\bibfnamefont {D.}~\bibnamefont {Oliinychenko}}, \bibinfo {author}
  {\bibfnamefont {J.}~\bibnamefont {Mohs}}, \bibinfo {author} {\bibfnamefont
  {M.}~\bibnamefont {Kretz}}, \bibinfo {author} {\bibfnamefont
  {T.}~\bibnamefont {Kehrenberg}}, \bibinfo {author} {\bibfnamefont
  {A.}~\bibnamefont {Goldschmidt}}, \bibinfo {author} {\bibfnamefont
  {B.}~\bibnamefont {B\"auchle}}, \bibinfo {author} {\bibfnamefont
  {J.}~\bibnamefont {Auvinen}}, \bibinfo {author} {\bibfnamefont
  {M.}~\bibnamefont {Attems}}, \ and\ \bibinfo {author} {\bibfnamefont
  {H.}~\bibnamefont {Petersen}},\ }\href {\doibase 10.1103/physrevc.94.054905}
  {\bibfield  {journal} {\bibinfo  {journal} {Phys. Rev. C}\ }\textbf {\bibinfo
  {volume} {94}},\ \bibinfo {pages} {054905} (\bibinfo {year}
  {2016})}\BibitemShut {NoStop}%
\bibitem [{\citenamefont {Oliinychenko}\ \emph {et~al.}(2021)\citenamefont
  {Oliinychenko}, \citenamefont {Steinberg}, \citenamefont {Staudenmaier},
  \citenamefont {Weil}, \citenamefont {Sch\"afer}, \citenamefont {(Petersen)},
  \citenamefont {Ryu}, \citenamefont {Mohs}, \citenamefont {Li}, \citenamefont
  {Sorensen}, \citenamefont {Mitrovic}, \citenamefont {Pang}, \citenamefont
  {void 0ne}, \citenamefont {Sciarra}, \citenamefont {Garcia-Montero},
  \citenamefont {Mayer}, \citenamefont {K\"ubler},\ and\ \citenamefont
  {Nikita}}]{dmytro_oliinychenko_2021_5796168}%
  \BibitemOpen
  \bibfield  {author} {\bibinfo {author} {\bibfnamefont {D.}~\bibnamefont
  {Oliinychenko}}, \bibinfo {author} {\bibfnamefont {V.}~\bibnamefont
  {Steinberg}}, \bibinfo {author} {\bibfnamefont {J.}~\bibnamefont
  {Staudenmaier}}, \bibinfo {author} {\bibfnamefont {J.}~\bibnamefont {Weil}},
  \bibinfo {author} {\bibfnamefont {A.}~\bibnamefont {Sch\"afer}}, \bibinfo
  {author} {\bibfnamefont {H.~E.}\ \bibnamefont {(Petersen)}}, \bibinfo
  {author} {\bibfnamefont {S.}~\bibnamefont {Ryu}}, \bibinfo {author}
  {\bibfnamefont {J.}~\bibnamefont {Mohs}}, \bibinfo {author} {\bibfnamefont
  {F.}~\bibnamefont {Li}}, \bibinfo {author} {\bibfnamefont {A.}~\bibnamefont
  {Sorensen}}, \bibinfo {author} {\bibfnamefont {D.}~\bibnamefont {Mitrovic}},
  \bibinfo {author} {\bibfnamefont {L.}~\bibnamefont {Pang}}, \bibinfo {author}
  {\bibnamefont {void 0ne}}, \bibinfo {author} {\bibfnamefont {A.}~\bibnamefont
  {Sciarra}}, \bibinfo {author} {\bibfnamefont {O.}~\bibnamefont
  {Garcia-Montero}}, \bibinfo {author} {\bibfnamefont {M.}~\bibnamefont
  {Mayer}}, \bibinfo {author} {\bibfnamefont {N.}~\bibnamefont {K\"ubler}}, \
  and\ \bibinfo {author} {\bibnamefont {Nikita}},\ }\href {\doibase
  10.5281/zenodo.5796168} {\enquote {\bibinfo {title} {{smash-transport/smash:
  SMASH-2.1}},}\ } (\bibinfo {year} {2021})\BibitemShut {NoStop}%
\bibitem [{sma()}]{smashurl}%
  \BibitemOpen
  \href {https://smash-transport.github.io/} {}\bibinfo {howpublished}
  {\url{https://smash-transport.github.io/}}\BibitemShut {NoStop}%
\bibitem [{\citenamefont {Tanabashi}\ \emph {et~al.}(2018)\citenamefont
  {Tanabashi} \emph {et~al.}}]{ParticleDataGroup:2018ovx}%
  \BibitemOpen
  \bibfield  {author} {\bibinfo {author} {\bibfnamefont {M.}~\bibnamefont
  {Tanabashi}} \emph {et~al.} (\bibinfo {collaboration} {Particle Data
  Group}),\ }\href {\doibase 10.1103/PhysRevD.98.030001} {\bibfield  {journal}
  {\bibinfo  {journal} {Phys. Rev. D}\ }\textbf {\bibinfo {volume} {98}},\
  \bibinfo {pages} {030001} (\bibinfo {year} {2018})}\BibitemShut {NoStop}%
\bibitem [{\citenamefont {Sj\"ostrand}\ \emph {et~al.}(2008)\citenamefont
  {Sj\"ostrand}, \citenamefont {Mrenna},\ and\ \citenamefont
  {Skands}}]{Sj_strand_2008}%
  \BibitemOpen
  \bibfield  {author} {\bibinfo {author} {\bibfnamefont {T.}~\bibnamefont
  {Sj\"ostrand}}, \bibinfo {author} {\bibfnamefont {S.}~\bibnamefont {Mrenna}},
  \ and\ \bibinfo {author} {\bibfnamefont {P.}~\bibnamefont {Skands}},\ }\href
  {\doibase 10.1016/j.cpc.2008.01.036} {\ \textbf {\bibinfo {volume} {178}},\
  \bibinfo {pages} {852} (\bibinfo {year} {2008})}\BibitemShut {NoStop}%
\bibitem [{\citenamefont {Karpenko}\ \emph {et~al.}(2014)\citenamefont
  {Karpenko}, \citenamefont {Huovinen},\ and\ \citenamefont
  {Bleicher}}]{Karpenko_2014}%
  \BibitemOpen
  \bibfield  {author} {\bibinfo {author} {\bibfnamefont {I.}~\bibnamefont
  {Karpenko}}, \bibinfo {author} {\bibfnamefont {P.}~\bibnamefont {Huovinen}},
  \ and\ \bibinfo {author} {\bibfnamefont {M.}~\bibnamefont {Bleicher}},\
  }\href@noop {} {\bibfield  {journal} {\bibinfo  {journal} {Comput. Phys.
  Commun.}\ }\textbf {\bibinfo {volume} {185}},\ \bibinfo {pages} {3016}
  (\bibinfo {year} {2014})}\BibitemShut {NoStop}%
\bibitem [{\citenamefont {Denicol}\ \emph {et~al.}(2014)\citenamefont
  {Denicol}, \citenamefont {Jeon},\ and\ \citenamefont {Gale}}]{Denicol_2014}%
  \BibitemOpen
  \bibfield  {author} {\bibinfo {author} {\bibfnamefont {G.~S.}\ \bibnamefont
  {Denicol}}, \bibinfo {author} {\bibfnamefont {S.}~\bibnamefont {Jeon}}, \
  and\ \bibinfo {author} {\bibfnamefont {C.}~\bibnamefont {Gale}},\ }\href
  {\doibase 10.1103/physrevc.90.024912} {\bibfield  {journal} {\bibinfo
  {journal} {Phys. Rev. C}\ }\textbf {\bibinfo {volume} {90}},\ \bibinfo
  {pages} {024912} (\bibinfo {year} {2014})}\BibitemShut {NoStop}%
\bibitem [{\citenamefont {Ryu}\ \emph {et~al.}(2015)\citenamefont {Ryu},
  \citenamefont {Paquet}, \citenamefont {Shen}, \citenamefont {Denicol},
  \citenamefont {Schenke}, \citenamefont {Jeon},\ and\ \citenamefont
  {Gale}}]{Ryu_2015}%
  \BibitemOpen
  \bibfield  {author} {\bibinfo {author} {\bibfnamefont {S.}~\bibnamefont
  {Ryu}}, \bibinfo {author} {\bibfnamefont {J.-F.}\ \bibnamefont {Paquet}},
  \bibinfo {author} {\bibfnamefont {C.}~\bibnamefont {Shen}}, \bibinfo {author}
  {\bibfnamefont {G.~S.}\ \bibnamefont {Denicol}}, \bibinfo {author}
  {\bibfnamefont {B.}~\bibnamefont {Schenke}}, \bibinfo {author} {\bibfnamefont
  {S.}~\bibnamefont {Jeon}}, \ and\ \bibinfo {author} {\bibfnamefont
  {C.}~\bibnamefont {Gale}},\ }\href {\doibase 10.1103/physrevlett.115.132301}
  {\bibfield  {journal} {\bibinfo  {journal} {Phys. Rev. Lett.}\ }\textbf
  {\bibinfo {volume} {115}},\ \bibinfo {pages} {132301} (\bibinfo {year}
  {2015})}\BibitemShut {NoStop}%
\bibitem [{\citenamefont {Steinheimer}\ \emph {et~al.}(2011)\citenamefont
  {Steinheimer}, \citenamefont {Schramm},\ and\ \citenamefont
  {Stocker}}]{Steinheimer:2011ea}%
  \BibitemOpen
  \bibfield  {author} {\bibinfo {author} {\bibfnamefont {J.}~\bibnamefont
  {Steinheimer}}, \bibinfo {author} {\bibfnamefont {S.}~\bibnamefont
  {Schramm}}, \ and\ \bibinfo {author} {\bibfnamefont {H.}~\bibnamefont
  {Stocker}},\ }\href {\doibase 10.1103/PhysRevC.84.045208} {\bibfield
  {journal} {\bibinfo  {journal} {Phys. Rev. C}\ }\textbf {\bibinfo {volume}
  {84}},\ \bibinfo {pages} {045208} (\bibinfo {year} {2011})},\ \Eprint
  {http://arxiv.org/abs/1108.2596} {arXiv:1108.2596 [hep-ph]} \BibitemShut
  {NoStop}%
\bibitem [{\citenamefont {Motornenko}\ \emph {et~al.}(2020)\citenamefont
  {Motornenko}, \citenamefont {Steinheimer}, \citenamefont {Vovchenko},
  \citenamefont {Schramm},\ and\ \citenamefont
  {Stoecker}}]{Motornenko:2019arp}%
  \BibitemOpen
  \bibfield  {author} {\bibinfo {author} {\bibfnamefont {A.}~\bibnamefont
  {Motornenko}}, \bibinfo {author} {\bibfnamefont {J.}~\bibnamefont
  {Steinheimer}}, \bibinfo {author} {\bibfnamefont {V.}~\bibnamefont
  {Vovchenko}}, \bibinfo {author} {\bibfnamefont {S.}~\bibnamefont {Schramm}},
  \ and\ \bibinfo {author} {\bibfnamefont {H.}~\bibnamefont {Stoecker}},\
  }\href {\doibase 10.1103/PhysRevC.101.034904} {\bibfield  {journal} {\bibinfo
   {journal} {Phys. Rev. C}\ }\textbf {\bibinfo {volume} {101}},\ \bibinfo
  {pages} {034904} (\bibinfo {year} {2020})},\ \Eprint
  {http://arxiv.org/abs/1905.00866} {arXiv:1905.00866 [hep-ph]} \BibitemShut
  {NoStop}%
\bibitem [{\citenamefont {Most}\ \emph {et~al.}(2023)\citenamefont {Most},
  \citenamefont {Motornenko}, \citenamefont {Steinheimer}, \citenamefont
  {Dexheimer}, \citenamefont {Hanauske}, \citenamefont {Rezzolla},\ and\
  \citenamefont {Stoecker}}]{Most:2022wgo}%
  \BibitemOpen
  \bibfield  {author} {\bibinfo {author} {\bibfnamefont {E.~R.}\ \bibnamefont
  {Most}}, \bibinfo {author} {\bibfnamefont {A.}~\bibnamefont {Motornenko}},
  \bibinfo {author} {\bibfnamefont {J.}~\bibnamefont {Steinheimer}}, \bibinfo
  {author} {\bibfnamefont {V.}~\bibnamefont {Dexheimer}}, \bibinfo {author}
  {\bibfnamefont {M.}~\bibnamefont {Hanauske}}, \bibinfo {author}
  {\bibfnamefont {L.}~\bibnamefont {Rezzolla}}, \ and\ \bibinfo {author}
  {\bibfnamefont {H.}~\bibnamefont {Stoecker}},\ }\href {\doibase
  10.1103/PhysRevD.107.043034} {\bibfield  {journal} {\bibinfo  {journal}
  {Phys. Rev. D}\ }\textbf {\bibinfo {volume} {107}},\ \bibinfo {pages}
  {043034} (\bibinfo {year} {2023})},\ \Eprint
  {http://arxiv.org/abs/2201.13150} {arXiv:2201.13150 [nucl-th]} \BibitemShut
  {NoStop}%
\bibitem [{\citenamefont {Huovinen}\ and\ \citenamefont
  {Petersen}(2012)}]{Huovinen_2012}%
  \BibitemOpen
  \bibfield  {author} {\bibinfo {author} {\bibfnamefont {P.}~\bibnamefont
  {Huovinen}}\ and\ \bibinfo {author} {\bibfnamefont {H.}~\bibnamefont
  {Petersen}},\ }\href {\doibase 10.1140/epja/i2012-12171-9} {\bibfield
  {journal} {\bibinfo  {journal} {Eur. Phys. J. A}\ }\textbf {\bibinfo {volume}
  {48}} (\bibinfo {year} {2012}),\ 10.1140/epja/i2012-12171-9}\BibitemShut
  {NoStop}%
\bibitem [{\citenamefont {G\"otz}\ and\ \citenamefont
  {Elfner}(2022)}]{Gotz:2022naz}%
  \BibitemOpen
  \bibfield  {author} {\bibinfo {author} {\bibfnamefont {N.}~\bibnamefont
  {G\"otz}}\ and\ \bibinfo {author} {\bibfnamefont {H.}~\bibnamefont
  {Elfner}},\ }\href {\doibase 10.1103/PhysRevC.106.054904} {\bibfield
  {journal} {\bibinfo  {journal} {Phys. Rev. C}\ }\textbf {\bibinfo {volume}
  {106}},\ \bibinfo {pages} {054904} (\bibinfo {year} {2022})},\ \Eprint
  {http://arxiv.org/abs/2207.05778} {arXiv:2207.05778 [hep-ph]} \BibitemShut
  {NoStop}%
\bibitem [{\citenamefont {Cimerman}\ \emph {et~al.}(2021)\citenamefont
  {Cimerman}, \citenamefont {Karpenko}, \citenamefont {Tom\'a\v{s}ik},\ and\
  \citenamefont {Trzeciak}}]{Cimerman:2020iny}%
  \BibitemOpen
  \bibfield  {author} {\bibinfo {author} {\bibfnamefont {J.}~\bibnamefont
  {Cimerman}}, \bibinfo {author} {\bibfnamefont {I.}~\bibnamefont {Karpenko}},
  \bibinfo {author} {\bibfnamefont {B.}~\bibnamefont {Tom\'a\v{s}ik}}, \ and\
  \bibinfo {author} {\bibfnamefont {B.~A.}\ \bibnamefont {Trzeciak}},\ }\href
  {\doibase 10.1103/PhysRevC.103.034902} {\bibfield  {journal} {\bibinfo
  {journal} {Phys. Rev. C}\ }\textbf {\bibinfo {volume} {103}},\ \bibinfo
  {pages} {034902} (\bibinfo {year} {2021})},\ \Eprint
  {http://arxiv.org/abs/2012.10266} {arXiv:2012.10266 [nucl-th]} \BibitemShut
  {NoStop}%
\bibitem [{sam()}]{samplerurl}%
  \BibitemOpen
  \href {https://github.com/smash-transport/smash-hadron-sampler} {}\bibinfo
  {howpublished}
  {\url{https://github.com/smash-transport/smash-hadron-sampler}}\BibitemShut
  {NoStop}%
\bibitem [{\citenamefont {Cooper}\ \emph {et~al.}(1975)\citenamefont {Cooper},
  \citenamefont {Frye},\ and\ \citenamefont {Schonberg}}]{cooper1975landau}%
  \BibitemOpen
  \bibfield  {author} {\bibinfo {author} {\bibfnamefont {F.}~\bibnamefont
  {Cooper}}, \bibinfo {author} {\bibfnamefont {G.}~\bibnamefont {Frye}}, \ and\
  \bibinfo {author} {\bibfnamefont {E.}~\bibnamefont {Schonberg}},\ }\href@noop
  {} {\bibfield  {journal} {\bibinfo  {journal} {Phys. Rev. D}\ }\textbf
  {\bibinfo {volume} {11}},\ \bibinfo {pages} {192} (\bibinfo {year}
  {1975})}\BibitemShut {NoStop}%
\bibitem [{\citenamefont {Oliinychenko}\ and\ \citenamefont
  {Petersen}(2016)}]{Oliinychenko:2015lva}%
  \BibitemOpen
  \bibfield  {author} {\bibinfo {author} {\bibfnamefont {D.}~\bibnamefont
  {Oliinychenko}}\ and\ \bibinfo {author} {\bibfnamefont {H.}~\bibnamefont
  {Petersen}},\ }\href {\doibase 10.1103/PhysRevC.93.034905} {\bibfield
  {journal} {\bibinfo  {journal} {Phys. Rev. C}\ }\textbf {\bibinfo {volume}
  {93}},\ \bibinfo {pages} {034905} (\bibinfo {year} {2016})},\ \Eprint
  {http://arxiv.org/abs/1508.04378} {arXiv:1508.04378 [nucl-th]} \BibitemShut
  {NoStop}%
\bibitem [{\citenamefont {Inghirami}\ and\ \citenamefont
  {Elfner}(2022)}]{Inghirami:2022afu}%
  \BibitemOpen
  \bibfield  {author} {\bibinfo {author} {\bibfnamefont {G.}~\bibnamefont
  {Inghirami}}\ and\ \bibinfo {author} {\bibfnamefont {H.}~\bibnamefont
  {Elfner}},\ }\href@noop {} {\  (\bibinfo {year} {2022})},\ \Eprint
  {http://arxiv.org/abs/2201.05934} {arXiv:2201.05934 [hep-ph]} \BibitemShut
  {NoStop}%
\bibitem [{\citenamefont {Auvinen}\ and\ \citenamefont
  {Petersen}(2013)}]{Auvinen:2013sba}%
  \BibitemOpen
  \bibfield  {author} {\bibinfo {author} {\bibfnamefont {J.}~\bibnamefont
  {Auvinen}}\ and\ \bibinfo {author} {\bibfnamefont {H.}~\bibnamefont
  {Petersen}},\ }\href {\doibase 10.1103/PhysRevC.88.064908} {\bibfield
  {journal} {\bibinfo  {journal} {Phys. Rev. C}\ }\textbf {\bibinfo {volume}
  {88}},\ \bibinfo {pages} {064908} (\bibinfo {year} {2013})},\ \Eprint
  {http://arxiv.org/abs/1310.1764} {arXiv:1310.1764 [nucl-th]} \BibitemShut
  {NoStop}%
\bibitem [{\citenamefont {Everett}\ \emph {et~al.}(2022)\citenamefont {Everett}
  \emph {et~al.}}]{JETSCAPE:2022cob}%
  \BibitemOpen
  \bibfield  {author} {\bibinfo {author} {\bibfnamefont {D.}~\bibnamefont
  {Everett}} \emph {et~al.} (\bibinfo {collaboration} {JETSCAPE}),\ }\href
  {\doibase 10.1103/PhysRevC.106.064901} {\bibfield  {journal} {\bibinfo
  {journal} {Phys. Rev. C}\ }\textbf {\bibinfo {volume} {106}},\ \bibinfo
  {pages} {064901} (\bibinfo {year} {2022})},\ \Eprint
  {http://arxiv.org/abs/2203.08286} {arXiv:2203.08286 [hep-ph]} \BibitemShut
  {NoStop}%
\bibitem [{\citenamefont {Bernhard}\ \emph {et~al.}(2019)\citenamefont
  {Bernhard}, \citenamefont {Moreland},\ and\ \citenamefont
  {Bass}}]{Bernhard:2019bmu}%
  \BibitemOpen
  \bibfield  {author} {\bibinfo {author} {\bibfnamefont {J.~E.}\ \bibnamefont
  {Bernhard}}, \bibinfo {author} {\bibfnamefont {J.~S.}\ \bibnamefont
  {Moreland}}, \ and\ \bibinfo {author} {\bibfnamefont {S.~A.}\ \bibnamefont
  {Bass}},\ }\href {\doibase 10.1038/s41567-019-0611-8} {\bibfield  {journal}
  {\bibinfo  {journal} {Nature Phys.}\ }\textbf {\bibinfo {volume} {15}},\
  \bibinfo {pages} {1113} (\bibinfo {year} {2019})}\BibitemShut {NoStop}%
\bibitem [{\citenamefont {Bernhard}\ \emph {et~al.}(2016)\citenamefont
  {Bernhard}, \citenamefont {Moreland}, \citenamefont {Bass}, \citenamefont
  {Liu},\ and\ \citenamefont {Heinz}}]{Bernhard:2016tnd}%
  \BibitemOpen
  \bibfield  {author} {\bibinfo {author} {\bibfnamefont {J.~E.}\ \bibnamefont
  {Bernhard}}, \bibinfo {author} {\bibfnamefont {J.~S.}\ \bibnamefont
  {Moreland}}, \bibinfo {author} {\bibfnamefont {S.~A.}\ \bibnamefont {Bass}},
  \bibinfo {author} {\bibfnamefont {J.}~\bibnamefont {Liu}}, \ and\ \bibinfo
  {author} {\bibfnamefont {U.}~\bibnamefont {Heinz}},\ }\href {\doibase
  10.1103/PhysRevC.94.024907} {\bibfield  {journal} {\bibinfo  {journal} {Phys.
  Rev. C}\ }\textbf {\bibinfo {volume} {94}},\ \bibinfo {pages} {024907}
  (\bibinfo {year} {2016})},\ \Eprint {http://arxiv.org/abs/1605.03954}
  {arXiv:1605.03954 [nucl-th]} \BibitemShut {NoStop}%
\bibitem [{\citenamefont {Gelis}\ \emph {et~al.}(2010)\citenamefont {Gelis},
  \citenamefont {Iancu}, \citenamefont {Jalilian-Marian},\ and\ \citenamefont
  {Venugopalan}}]{Gelis:2010nm}%
  \BibitemOpen
  \bibfield  {author} {\bibinfo {author} {\bibfnamefont {F.}~\bibnamefont
  {Gelis}}, \bibinfo {author} {\bibfnamefont {E.}~\bibnamefont {Iancu}},
  \bibinfo {author} {\bibfnamefont {J.}~\bibnamefont {Jalilian-Marian}}, \ and\
  \bibinfo {author} {\bibfnamefont {R.}~\bibnamefont {Venugopalan}},\ }\href
  {\doibase 10.1146/annurev.nucl.010909.083629} {\bibfield  {journal} {\bibinfo
   {journal} {Ann. Rev. Nucl. Part. Sci.}\ }\textbf {\bibinfo {volume} {60}},\
  \bibinfo {pages} {463} (\bibinfo {year} {2010})},\ \Eprint
  {http://arxiv.org/abs/1002.0333} {arXiv:1002.0333 [hep-ph]} \BibitemShut
  {NoStop}%
\bibitem [{\citenamefont {Krasnitz}\ and\ \citenamefont
  {Venugopalan}(1999)}]{Krasnitz:1998ns}%
  \BibitemOpen
  \bibfield  {author} {\bibinfo {author} {\bibfnamefont {A.}~\bibnamefont
  {Krasnitz}}\ and\ \bibinfo {author} {\bibfnamefont {R.}~\bibnamefont
  {Venugopalan}},\ }\href {\doibase 10.1016/S0550-3213(99)00366-1} {\bibfield
  {journal} {\bibinfo  {journal} {Nucl. Phys. B}\ }\textbf {\bibinfo {volume}
  {557}},\ \bibinfo {pages} {237} (\bibinfo {year} {1999})},\ \Eprint
  {http://arxiv.org/abs/hep-ph/9809433} {arXiv:hep-ph/9809433} \BibitemShut
  {NoStop}%
\bibitem [{\citenamefont {Gale}\ \emph {et~al.}(2013)\citenamefont {Gale},
  \citenamefont {Jeon},\ and\ \citenamefont {Schenke}}]{Gale:2013da}%
  \BibitemOpen
  \bibfield  {author} {\bibinfo {author} {\bibfnamefont {C.}~\bibnamefont
  {Gale}}, \bibinfo {author} {\bibfnamefont {S.}~\bibnamefont {Jeon}}, \ and\
  \bibinfo {author} {\bibfnamefont {B.}~\bibnamefont {Schenke}},\ }\href
  {\doibase 10.1142/S0217751X13400113} {\bibfield  {journal} {\bibinfo
  {journal} {Int. J. Mod. Phys. A}\ }\textbf {\bibinfo {volume} {28}},\
  \bibinfo {pages} {1340011} (\bibinfo {year} {2013})},\ \Eprint
  {http://arxiv.org/abs/1301.5893} {arXiv:1301.5893 [nucl-th]} \BibitemShut
  {NoStop}%
\bibitem [{\citenamefont {Schenke}\ \emph
  {et~al.}(2012{\natexlab{b}})\citenamefont {Schenke}, \citenamefont
  {Tribedy},\ and\ \citenamefont {Venugopalan}}]{Schenke:2012hg}%
  \BibitemOpen
  \bibfield  {author} {\bibinfo {author} {\bibfnamefont {B.}~\bibnamefont
  {Schenke}}, \bibinfo {author} {\bibfnamefont {P.}~\bibnamefont {Tribedy}}, \
  and\ \bibinfo {author} {\bibfnamefont {R.}~\bibnamefont {Venugopalan}},\
  }\href {\doibase 10.1103/PhysRevC.86.034908} {\bibfield  {journal} {\bibinfo
  {journal} {Phys. Rev. C}\ }\textbf {\bibinfo {volume} {86}},\ \bibinfo
  {pages} {034908} (\bibinfo {year} {2012}{\natexlab{b}})},\ \Eprint
  {http://arxiv.org/abs/1206.6805} {arXiv:1206.6805 [hep-ph]} \BibitemShut
  {NoStop}%
\bibitem [{\citenamefont {Adler}\ \emph {et~al.}(2002)\citenamefont {Adler}
  \emph {et~al.}}]{STAR:2002hbo}%
  \BibitemOpen
  \bibfield  {author} {\bibinfo {author} {\bibfnamefont {C.}~\bibnamefont
  {Adler}} \emph {et~al.} (\bibinfo {collaboration} {STAR}),\ }\href {\doibase
  10.1103/PhysRevC.66.034904} {\bibfield  {journal} {\bibinfo  {journal} {Phys.
  Rev. C}\ }\textbf {\bibinfo {volume} {66}},\ \bibinfo {pages} {034904}
  (\bibinfo {year} {2002})},\ \Eprint {http://arxiv.org/abs/nucl-ex/0206001}
  {arXiv:nucl-ex/0206001} \BibitemShut {NoStop}%
\bibitem [{\citenamefont {Schenke}\ \emph {et~al.}(2020)\citenamefont
  {Schenke}, \citenamefont {Shen},\ and\ \citenamefont
  {Tribedy}}]{Schenke_2020}%
  \BibitemOpen
  \bibfield  {author} {\bibinfo {author} {\bibfnamefont {B.}~\bibnamefont
  {Schenke}}, \bibinfo {author} {\bibfnamefont {C.}~\bibnamefont {Shen}}, \
  and\ \bibinfo {author} {\bibfnamefont {P.}~\bibnamefont {Tribedy}},\ }\href
  {\doibase 10.1016/j.physletb.2020.135322} {\bibfield  {journal} {\bibinfo
  {journal} {Physics Letters B}\ }\textbf {\bibinfo {volume} {803}},\ \bibinfo
  {pages} {135322} (\bibinfo {year} {2020})}\BibitemShut {NoStop}%
\bibitem [{\citenamefont {Roch}\ \emph {et~al.}(2023)\citenamefont {Roch},
  \citenamefont {Saß},\ and\ \citenamefont
  {Götz}}]{hendrik_roch_2023_10288639}%
  \BibitemOpen
  \bibfield  {author} {\bibinfo {author} {\bibfnamefont {H.}~\bibnamefont
  {Roch}}, \bibinfo {author} {\bibfnamefont {N.}~\bibnamefont {Saß}}, \ and\
  \bibinfo {author} {\bibfnamefont {N.}~\bibnamefont {Götz}},\ }\href
  {\doibase 10.5281/zenodo.10288639} {\enquote {\bibinfo {title}
  {smash-transport/sparkx: v1.1.0-newton},}\ } (\bibinfo {year}
  {2023})\BibitemShut {NoStop}%
\bibitem [{\citenamefont {Alvioli}\ \emph {et~al.}(2009)\citenamefont
  {Alvioli}, \citenamefont {Drescher},\ and\ \citenamefont
  {Strikman}}]{Alvioli:2009ab}%
  \BibitemOpen
  \bibfield  {author} {\bibinfo {author} {\bibfnamefont {M.}~\bibnamefont
  {Alvioli}}, \bibinfo {author} {\bibfnamefont {H.~J.}\ \bibnamefont
  {Drescher}}, \ and\ \bibinfo {author} {\bibfnamefont {M.}~\bibnamefont
  {Strikman}},\ }\href {\doibase 10.1016/j.physletb.2009.08.067} {\bibfield
  {journal} {\bibinfo  {journal} {Phys. Lett. B}\ }\textbf {\bibinfo {volume}
  {680}},\ \bibinfo {pages} {225} (\bibinfo {year} {2009})},\ \Eprint
  {http://arxiv.org/abs/0905.2670} {arXiv:0905.2670 [nucl-th]} \BibitemShut
  {NoStop}%
\bibitem [{\citenamefont {Kurkela}\ \emph {et~al.}(2019)\citenamefont
  {Kurkela}, \citenamefont {Mazeliauskas}, \citenamefont {Paquet},
  \citenamefont {Schlichting},\ and\ \citenamefont
  {Teaney}}]{PhysRevC.99.034910}%
  \BibitemOpen
  \bibfield  {author} {\bibinfo {author} {\bibfnamefont {A.}~\bibnamefont
  {Kurkela}}, \bibinfo {author} {\bibfnamefont {A.}~\bibnamefont
  {Mazeliauskas}}, \bibinfo {author} {\bibfnamefont {J.-F. m.~c.}\ \bibnamefont
  {Paquet}}, \bibinfo {author} {\bibfnamefont {S.}~\bibnamefont {Schlichting}},
  \ and\ \bibinfo {author} {\bibfnamefont {D.}~\bibnamefont {Teaney}},\ }\href
  {\doibase 10.1103/PhysRevC.99.034910} {\bibfield  {journal} {\bibinfo
  {journal} {Phys. Rev. C}\ }\textbf {\bibinfo {volume} {99}},\ \bibinfo
  {pages} {034910} (\bibinfo {year} {2019})}\BibitemShut {NoStop}%
\bibitem [{\citenamefont {McNelis}\ \emph {et~al.}(2021)\citenamefont
  {McNelis}, \citenamefont {Bazow},\ and\ \citenamefont
  {Heinz}}]{MCNELIS2021108077}%
  \BibitemOpen
  \bibfield  {author} {\bibinfo {author} {\bibfnamefont {M.}~\bibnamefont
  {McNelis}}, \bibinfo {author} {\bibfnamefont {D.}~\bibnamefont {Bazow}}, \
  and\ \bibinfo {author} {\bibfnamefont {U.}~\bibnamefont {Heinz}},\ }\href
  {\doibase https://doi.org/10.1016/j.cpc.2021.108077} {\bibfield  {journal}
  {\bibinfo  {journal} {Computer Physics Communications}\ }\textbf {\bibinfo
  {volume} {267}},\ \bibinfo {pages} {108077} (\bibinfo {year}
  {2021})}\BibitemShut {NoStop}%
\bibitem [{\citenamefont {Pang}\ \emph {et~al.}(2016)\citenamefont {Pang},
  \citenamefont {Petersen}, \citenamefont {Qin}, \citenamefont {Roy},\ and\
  \citenamefont {Wang}}]{pang2016decorrelation}%
  \BibitemOpen
  \bibfield  {author} {\bibinfo {author} {\bibfnamefont {L.-G.}\ \bibnamefont
  {Pang}}, \bibinfo {author} {\bibfnamefont {H.}~\bibnamefont {Petersen}},
  \bibinfo {author} {\bibfnamefont {G.-Y.}\ \bibnamefont {Qin}}, \bibinfo
  {author} {\bibfnamefont {V.}~\bibnamefont {Roy}}, \ and\ \bibinfo {author}
  {\bibfnamefont {X.-N.}\ \bibnamefont {Wang}},\ }\href@noop {} {\bibfield
  {journal} {\bibinfo  {journal} {The European Physical Journal A}\ }\textbf
  {\bibinfo {volume} {52}},\ \bibinfo {pages} {97} (\bibinfo {year}
  {2016})}\BibitemShut {NoStop}%
\end{thebibliography}%
\end{document}